\documentclass[reprint,twocolumn,showpacs,nofootinbib,superscriptaddress,notitlepage,floatfix]{revtex4-2}
\usepackage{hyperref}
\usepackage{graphicx}
\usepackage{amsmath}
\usepackage{amssymb}
\usepackage{slashed}
\usepackage{amsfonts}
\usepackage{xcolor}
\usepackage{multirow}
\usepackage[caption=false]{subfig}
\usepackage{array}
\usepackage{mwe}

\usepackage[utf8]{inputenc}
\usepackage[normalem]{ulem}

\DeclareRobustCommand{\eq}[1]{Eq.~\eqref{eq:#1}}

\DeclareRobustCommand{\fig}[1]{Fig.~\ref{fig:#1}}

\DeclareRobustCommand{\app}[1]{App.~\ref{app:#1}}

\DeclareRobustCommand{\tbl}[1]{Table~\ref{tbl:#1}}

\DeclareRobustCommand{\eq}[1]{Eq.~(\ref{eq:#1})}

\newcommand{\MS}{{\overline{\mathrm{MS}}}}

\newcommand{\eps}{\epsilon}

\newcommand{\nn}{\nonumber}

\newcommand{\Tr}{\mathrm{Tr}}

\newcommand\bets{\begin{table*}}
\newcommand\eets[1]{\label{tb:#1}\end{table*}}

\begin{document}

\title{Lattice QCD Determination of the Bjorken-$x$ Dependence of Parton Distribution Functions at Next-to-next-to-leading Order}


\author{Xiang Gao}
\email{xgao@bnl.gov}
\affiliation{Key Laboratory of Quark \& Lepton Physics (MOE) and Institute of Particle Physics,
Central China Normal University, Wuhan 430079, China}

\author{Andrew D. Hanlon}
\affiliation{Physics Department, Brookhaven National Laboratory, Bldg. 510A, Upton, New York 11973, USA}

\author{Swagato Mukherjee}
\affiliation{Physics Department, Brookhaven National Laboratory, Bldg. 510A, Upton, New York 11973, USA}

\author{Peter Petreczky}
\affiliation{Physics Department, Brookhaven National Laboratory, Bldg. 510A, Upton, New York 11973, USA}

\author{Philipp Scior}
\affiliation{Physics Department, Brookhaven National Laboratory, Bldg. 510A, Upton, New York 11973, USA}

\author{Sergey Syritsyn}
\affiliation{RIKEN-BNL Research Center, Brookhaven National Laboratory, Upton, New York 11973}
\affiliation{Department of Physics and Astronomy, Stony Brook University, Stony Brook, New York 11790}

\author{Yong Zhao}
\email{yong.zhao@anl.gov}
\affiliation{Physics Division, Argonne National Laboratory, Lemont, IL 60439, USA}

\begin{abstract}
We report the first lattice QCD calculation of pion valence quark distribution with next-to-next-to-leading order perturbative matching correction, which is done using two fine lattices with spacings $a=0.04$ fm and $0.06$ fm and valence pion mass $m_\pi=300$ MeV, at boost momentum as large as $2.42$ GeV. As a crucial step to control the systematics, we renormalize the pion valence quasi distribution in the recently proposed hybrid scheme, which features a Wilson-line mass subtraction at large distances in coordinate space, and develop a procedure to match it to the $\MS$ scheme. We demonstrate that the renormalization and the perturbative matching in Bjorken-$x$ space yield a reliable determination of the valence quark distribution for $0.03\lesssim x \lesssim 0.80$ with 5--20\% uncertainties.

\end{abstract}

\maketitle

Understanding the hadron inner structure remains one of the top fundamental questions in nuclear and particle physics. As the lightest hadrons in nature, pions are the Nambu-Goldstone bosons of quantum chromodynamics (QCD), and their quark and gluon structures can help understand the origins of hadron mass and dynamical chiral symmetry breaking.
The parton distribution functions (PDFs), which describe 1D momentum densities of quarks and gluons in a hadron,
are the simplest and most important quantities that have been extensively studied from global high-energy scattering experiments and will be probed at unprecedented precision at the future Electron-Ion Collider~\cite{Accardi:2012qut,AbdulKhalek:2021gbh}.
Besides the experimental efforts, the first-principles calculations of PDFs using lattice QCD are also expected to provide useful predictions. 

Computation of the PDFs on a Euclidean lattice has been extremely difficult because they are defined from light-cone correlations with real-time dependence in Minkowski space. For a long time, only the lowest moments of the PDFs were calculable as they are matrix elements of local gauge-invariant operators. For reviews see Refs.~\cite{Lin:2017snn,Constantinou:2020hdm}. Less than a decade ago, a breakthrough was made by large-momentum effective theory (LaMET)~\cite{Ji:2013dva,Ji:2014gla,Ji:2020ect}, which starts from a Euclidean ``quasi-PDF'' (qPDF) in a boosted hadron and obtains the PDF through a large-momentum expansion and perturbative matching of the qPDF in Bjorken-$x$ (longitudinal momentum fraction) space. Over the years, LaMET has led to much progress in the calculation of PDFs and other parton physics~\cite{Ji:2020ect,Constantinou:2020hdm}, which reinvigorated the field as other proposals~\cite{Liu:1993cv,Detmold:2005gg,Braun:2007wv,Chambers:2017dov,Radyushkin:2017cyf,Ma:2017pxb} are also being studied and implemented.

Despite substantial progress, lattice calculation of the PDF $x$-dependence has yet to achieve essential control of the systematic uncertainties~\cite{Alexandrou:2019lfo}.
In the LaMET approach, lattice renormalization is one of the most important sources of error. The nonlocal quark bilinear operator $O_\Gamma(z)\equiv \bar{\psi}(z)\Gamma W(z,0) \psi(0)$, where $\Gamma$ is a Dirac matrix and $z^\mu=(0,0,0,z)$, which defines the qPDF, suffers from a linear power divergence in the Wilson line $W(z,0)$ that must be subtracted before taking the continuum limit.
The most popular methods so far are the regularization independent momentum subtraction scheme~\cite{Constantinou:2017sej,Stewart:2017tvs,Alexandrou:2017huk,Chen:2017mzz} and other ratio schemes~\cite{Orginos:2017kos,Braun:2018brg,Li:2020xml,Fan:2020nzz}, which use the matrix element of $O_\Gamma(z)$ in an off-shell quark~\cite{Constantinou:2017sej,Stewart:2017tvs,Alexandrou:2017huk,Chen:2017mzz}, a static/boosted hadron~\cite{Orginos:2017kos,Fan:2020nzz} or the vacuum state~\cite{Braun:2018brg,Li:2020xml} as the renormalization factor.
At small $z$ the matrix elements in these schemes satisfy a factorization relation to the light-cone correlation~\cite{Ji:2017rah,Radyushkin:2017lvu,Ma:2017pxb,Izubuchi:2018srq}.
However, at large $z$ they introduce nonperturbative effects~\cite{Gao:2020ito} that propagate to the qPDF via Fourier transform (FT) of the matrix elements, which contaminates the LaMET matching in $x$-space.
To overcome this limitation, the hybrid scheme~\cite{Ji:2020brr} was proposed to subtract the linear divergence at large $z$ and match the result to the $\MS$ scheme, thus preserving the LaMET matching after FT. To date, the hybrid scheme has not been used in calculating the PDFs, except for a recent work on meson distribution amplitudes~\cite{Hua:2020gnw}.
Apart from renormalization, the accuracy of perturbative matching also controls the precision of the calculation. In all the existing lattice calculations, the matching was done at only next-to-leading order (NLO), and it is not until recently that the next-to-next-to-leading order (NNLO) matching was derived for the non-singlet quark qPDF in the $\MS$ scheme~\cite{Chen:2020ody,Li:2020xml}.

In this Letter we present a state-of-the-art calculation of pion valence quark PDF using high-statistics, superfine-spacing, and large-momentum lattice data~\cite{Gao:2020ito}, with an adapted hybrid-scheme renormalization and the first-time implementation of NNLO matching. The pion valence PDF has been extracted from global fits~\cite{Gluck:1991ey,Novikov:2020snp,Barry:2021osv,Aicher:2010cb} and studied in lattice QCD~\cite{Zhang:2018nsy,Sufian:2019bol,Izubuchi:2019lyk,Joo:2019bzr,Sufian:2020vzb,Lin:2020ssv,Gao:2020ito,Gao:2021hxl}, with both at NLO accuracy.
In this work, we subtract the linear divergence in $O_\Gamma(z)$ with sub-percent precision, and develop a procedure to match the lattice subtraction scheme to $\MS$, a crucial step in the hybrid scheme to reduce the power corrections~\cite{Ji:2020brr}.
We derive the NNLO hybrid-scheme matching and apply it to the qPDF, showing good perturbative convergence and reduced scale-variation uncertainty compared to NLO matching. Finally, we demonstrate that our analysis yields a reliable determination of the PDF for $0.03\lesssim x \lesssim 0.80$ with 5--20\% uncertainties.

Our lattice data was produced using gauge ensembles in 2+1 flavor QCD 
generated by the HotQCD collaboration~\cite{HotQCD:2014kol} with
Highly Improved Staggered Quarks~\cite{Follana:2006rc},
including two lattice
spacings $a=0.04$ and $0.06$ fm, and volumes $L_s^3\times L_t=64^4$ and
$48^3 \times 64$, respectively. We use tadpole-improved clover Wilson valence fermions 
on the hypercubic (HYP) smeared~\cite{Hasenfratz:2001hp} gauge background, with a valence pion mass $m_\pi=300$ MeV. Furthermore, the Wilson line
in $O_\Gamma(z)$ is constructed from HYP-smeared gauge links.
We use pion momenta $P^z=(2\pi n_z)/(L_sa)$ with $0\le n_z \le 5$, resulting in $P^z$ as large as $2.42$~GeV.

The qPDF $\tilde f_v(x,P^z,\mu)$ is defined in a boosted pion state $|P\rangle$ with four-momentum $P^\mu=(P^t,0,0,P^z)$:
\begin{align}
    \tilde f_v(x,P^z,\mu) &= \int{dz \over 2\pi}e^{ixP^zz}\ \tilde h(z,P^z,\mu)\,,
\end{align}
where $\tilde h(z,P^z,\mu)\equiv \langle P| O_{\gamma^t}(z) |P\rangle/(2P^t)$, and $\mu$ is the $\MS$ scale.
The operator $O_\Gamma(z)$ can be renormalized under lattice regularization as~\cite{Ji:2017oey,Ishikawa:2017faj,Green:2017xeu}
\begin{align}\label{eq:renorm}
    O^B_\Gamma(z,a) = e^{-\delta m(a) |z|}Z_O(a) O^R_\Gamma(z)\,,
\end{align}
where ``$B$'' and ``$R$'' denote bare and renormalized quantities. The factor $Z_O(a)$ includes all the logarithmic ultraviolet (UV) divergences which are independent of $z$, while the Wilson-line mass correction $\delta m(a)$ includes the linear UV divergence $\propto 1/a$ and can be expressed as
\begin{align} \label{eq:mass}
    \delta m(a) = {m_{-1}(a)\over a} +  m_0\,,
\end{align}
where $m_{-1}(a)$ is a series in the strong coupling $\alpha_s(1/a)$, and $m_0$ is an ${\cal O}(\Lambda_{\rm QCD})$ constant originating from the renormalon ambiguity in $m_{-1}(a)$~\cite{Bauer:2011ws}.

The hybrid scheme is implemented as follows:
For $0\le z\le z_S$ with $a\ll z_S \ll 1/\Lambda_{\rm QCD}$, we form the ratio $\tilde h(z,P^z,a)/\tilde h(z,0,a)$ to cancel the UV divergences and the cutoff effects from $z\sim a$~\cite{Orginos:2017kos}; at $z> z_S$ we subtract $\delta m(a)$ and determine $Z_O(a)$ by imposing a continuity condition of the renormalized matrix elements at $z=z_S$.
There are different ways to calculate $\delta m(a)$~\cite{Ji:2020brr,Zhang:2017bzy,Green:2017xeu,Green:2020xco,Alexandrou:2020qtt,LatticePartonCollaborationLPC:2021xdx}. We determine $\delta m(a)$ from the combination of the static quark-antiquark potential, $V^{\rm lat}(r)$~\cite{HotQCD:2014kol,Bazavov:2017dsy}, and the free energy of a static quark at non-zero temperature \cite{Bazavov:2016uvm,Bazavov:2018wmo,Petreczky:2021mef}, with the following normalization scheme,
\begin{align}\label{eq:pot}
    V^{\rm lat}(a,r=r_0)+ 2\delta m(a) = 0.95/r_0\,,
\end{align}
where $r_0 =0.469$ fm is the Sommer scale for 2+1 flavor QCD \cite{HotQCD:2014kol}, and the constant 0.95 defines the scheme. The linear divergence $m_{-1}(a)/a$ does not depend on the scheme, while $m_0$ does.
The results are $a\delta m=0.1586(8)$ and $0.1508(12)$ for $a=0.06$ and $0.04$ fm, respectively.

Since $m_0$ is scheme dependent, a factor of $e^{\overline{m}_0|z|}$ with $\overline{m}_0\sim {\cal O}(\Lambda_{\rm QCD})$ is needed to match the lattice scheme to $\MS$, otherwise the LaMET expansion of the qPDF will include a power correction $\propto \bar{m}_0/P^z$~\cite{Ji:2020brr}, which slows down convergence to the PDF as $P^z$ grows.
It was proposed that $\overline{m}_0$ can be obtained by comparing the subtracted matrix elements of $O_\Gamma(z)$~\cite{LatticePartonCollaborationLPC:2021xdx} or $W(z,0)$~\cite{Green:2020xco} with their $\MS$ operator product expansion (OPE), whose accuracy requires $z\lesssim 0.2$ fm~\cite{Ji:2020brr}. But due to discretization effects, the window of $z$ that can be used is actually narrow. 

Our new procedure for the hybrid scheme is distinct by the determination of $\overline{m}_0$. 
In order to use larger $z$, we construct the following ratio and compare it to a form motivated by the OPE of $\tilde h(z,0,\mu)$,
\begin{align}\label{eq:mfit}
    \lim_{a\to0}e^{\delta m(a) (z-z_0)} {\tilde h(z,0,a)\over \tilde h(z_0,0,a)} & \!=\! e^{- {\bar{m}_0}(z-z_0)}  { C_0( \mu^2z^2) + \Lambda z^2\over C_0(\mu^2z^2_0) + \Lambda z^2_0}\,,
\end{align}
where $z,z_0\gg a$, and the parameter $\Lambda\sim{\cal O}(\Lambda_{\rm QCD}^2)$.
The Wilson coefficient $C_0$ is known to NNLO~\cite{Izubuchi:2018srq,Chen:2020ody,Li:2020xml}, and $\bar{m}_0$ and $\Lambda z^2$ originate from the leading UV and infrared renormalons in $C_0$~\cite{Braun:2018brg}.
According to \eq{renorm}, 
the l.h.s. of \eq{mfit} must have a continuum limit if $\delta m(a)$ includes all the linear divergences, which is renormalization group (RG) invariant. We choose $z\ge z_0=0.24$ fm and find agreement between the $a=0.04$ fm and $a=0.06$ fm ratios at sub-percent level up to $z\sim 1$ fm (see \app{ren}).
Then we extrapolate the lattice ratios to the continuum with $a^2$-dependence~\cite{Gao:2020ito}, and fit the result to the r.h.s of \eq{mfit}.
For $z_0\le z \le 0.4$ fm, we obtain decent plateaus and $\chi^2$ values for both $\bar{m}_0$ and $\Lambda$ with the NNLO $C_0$.
By definition $\bar{m}_0$ cancels the lattice scheme dependence of $\delta m(a)$, as changing the scheme only shifts $\delta m(a)$ by a constant, but $\bar{m}_0$ will inherit the ${\cal O}(\Lambda_{\rm QCD})$ ambiguity in the $\MS$ scheme.
Since $C_0$ is at fixed order, both $\bar{m}_0$ and $\Lambda$ depend on $\mu$, which we vary to estimate the related uncertainty in the final result.
At $\mu=2.0$ GeV, $\bar{m}_0=0.151(1)$ GeV and $\Lambda=0.041(6)$ GeV$^2$, so the power correction is not negligible. Therefore, we modify the hybrid scheme by correcting the $\Lambda z^2$ term in $\tilde h(z,0,\mu)$ at short $z$ as
\begin{align}\label{eq:hybren}
&\!\tilde h(z, z_S,P^z,\mu,a)\!=\! N{\tilde h(z,P^z,a)\over \tilde h(z,0,a)}\! {C_0(z^2\mu^2) \!+\! \Lambda z^2\over C_0( z^2\mu^2)} \theta(z_S\!-\!z) \nn\\
    &+\! Ne^{\delta m'(z - z_S)} {\tilde h(z, P^z, a)\over \tilde h(z_S,0, a)} {C_0(z^2_S\mu^2) \!+\! \Lambda z^2_S \over C_0( z^2_S\mu^2) } \theta(z\!-\!z_S)\,,\!
\end{align}
where $\delta m'=\delta m + \bar{m}_0$, and $N=\tilde h(0,0,a)/\tilde h(0,P^z,a)$ normalizes $\tilde h(z, z_S,P^z,\mu,a)$ to one at $z=0$. Since $C_0$ is at fixed order, $\tilde h(z, z_S,P^z,\mu,a)$ depends on $\mu$ despite the fact that it should be RG invariant.
Such a renormalization is performed through bootstrap loops so that the correlation between different $P^z$ and $z$ is taken care of.

\begin{figure}
    \centering
    \includegraphics[width=\columnwidth]{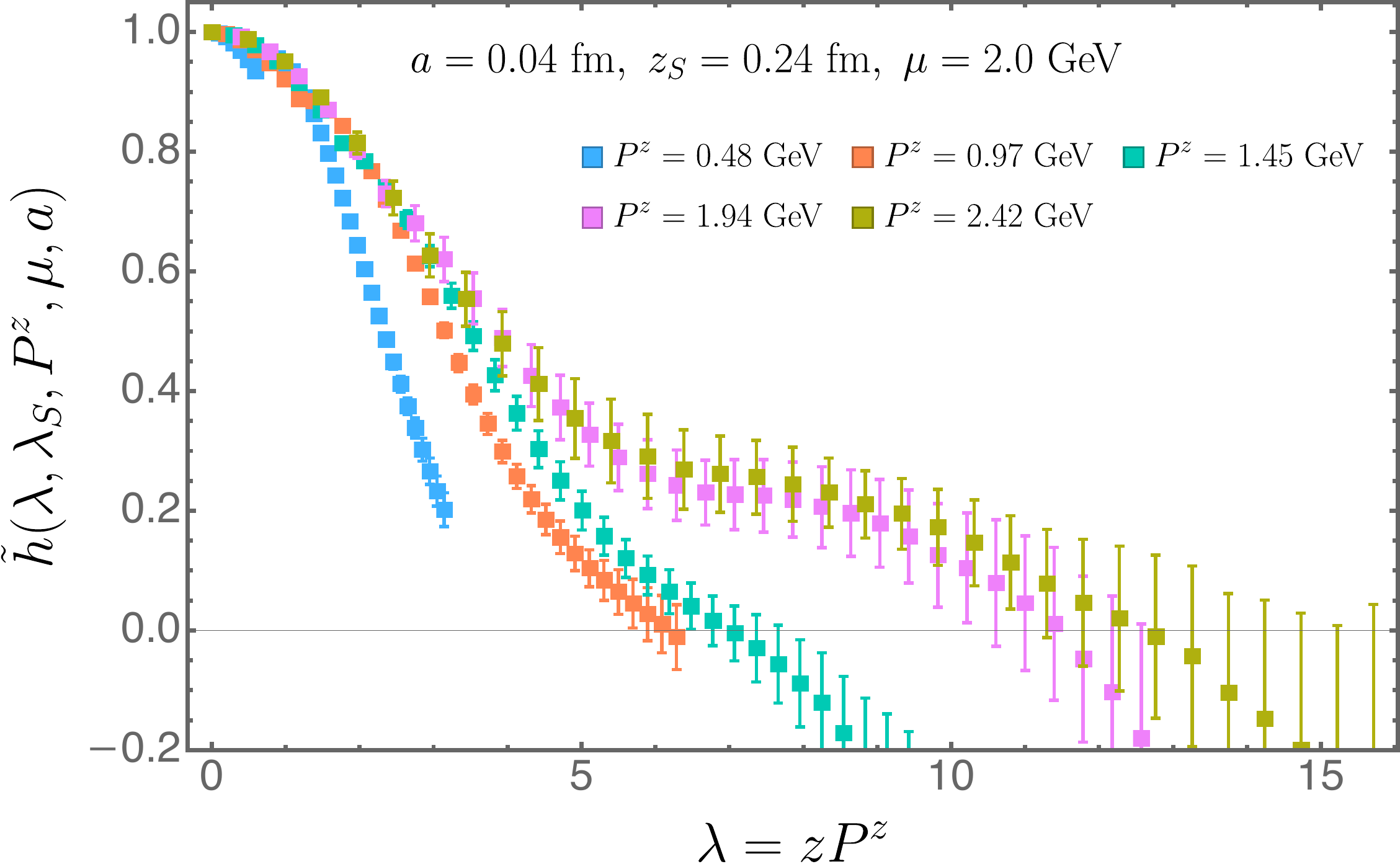}
    \caption{Renormalized matrix elements in the hybrid scheme.}
    \label{fig:hybme}
\end{figure}

The hybrid-scheme matrix elements are shown in \fig{hybme}. 
At small $z$, $\tilde h(z,P^z)$ is dominated by the leading-twist contribution.
At large $z$, the spacelike correlator for pion valence quarks will exhibit an exponential decay $\propto e^{-m_{\rm eff}|z|}$ where $m_{\rm eff}$ is an effective mass related to the system~\cite{Burkardt:1994pw}. 
When plotted as a function of $\lambda=zP^z$, $\tilde h(\lambda,P^z)$ should scale in $P^z$ at small $\lambda$, with slight violation due to QCD evolution.
Its exponential decay will emerge at a larger $\lambda$ with greater $P^z$ and with decay rate $m_{\rm eff}/P^z$. In the $P^z\to\infty$ limit, the exponential decay vanishes at finite $\lambda$ ($z\to0$), and only the leading-twist contribution remains, which almost scales in $P^z$ and features a power-law decay at large $\lambda$ that corresponds to small-$x$ PDF~\cite{Ji:2020brr}. This picture is consistent with \fig{hybme}.

The next step is a FT. 
We truncate the matrix elements at $z_L$ or $\lambda_L=z_LP^z$ where $\tilde h(\lambda_L)\sim0$, and extrapolate to $\infty$ to remove the unphysical oscillations from a truncated FT~\cite{Ji:2020brr}. The extrapolation form is $Ae^{-m_{\rm eff}|z|}/|\lambda|^d$, where $A$, $m_{\rm eff}$ and $d$ are the parameters.
Since $m_{\rm eff}$ is independent of $P^z$, by fitting to the $P^z=0$ matrix elements we find that it is around $0.1$ GeV, which is not far from the phenomenological estimate of 0.2--0.5 GeV in HQET~\cite{Beneke:1994sw}. Therefore, we impose $m_{\rm eff}>0.1$ GeV, as well as $A>0$ and $d>0$, to ensure a convergent FT on each bootstrap sample.
Since the FT converges fast with the exponential decay, the extrapolation mainly affects the small-$x$ region apart from removing the unphysical oscillations.
To verify this we vary $z_L$, which turns out to have little impact, and use different $m_{\rm eff}$ bounds and extrapolation forms, which lead to consistent qPDFs down to $x\sim 0.05$. (See \app{ft}).

Then, we match the qPDF $\tilde f_v(x,\lambda_S,P^z,\mu)$ to the $\MS$ PDF $f_v(x,\mu)$ through LaMET~\cite{Xiong:2013bka,Ma:2014jla,Izubuchi:2018srq,Ji:2020brr}:
\begin{align} \label{eq:fact}
f_v(x, \mu)&= \int_{-\infty}^{\infty} \frac{dy}{|y|} \ C^{-1}\!\left(\frac{x}{y}, \frac{\mu}{yP^z},|y|\lambda_S\right) \tilde f_v(y,\lambda_S,P^z,\mu) \nn\\
&\qquad + {\cal O}\Big(\frac{\Lambda_{\text{QCD}}^2}{(xP^z)^2},\frac{\Lambda_{\text{QCD}}^2}{((1-x)P^z)^2}\Big)\,,
\end{align}
where $\lambda_S=z_SP^z$, $z_S=0.24$ fm, and the power corrections are controlled by the parton and spectator momenta $xP^z$ and $(1-x)P^z$~\cite{Ji:2020brr}.
Here $C^{-1}$ is the inverse of the hybrid-scheme matching coefficient $C$, which we derive at NNLO~\cite{Zhao:2021xxx} by conversion from the $\MS$ result~\cite{Chen:2020ody,Li:2020xml}. Based on \eq{fact}, we can directly calculate the PDF with $P^z$-controlled power corrections for $x\in[x_{\rm min},x_{\rm max}]$.

\begin{figure}
    \centering
    \includegraphics[width=\columnwidth]{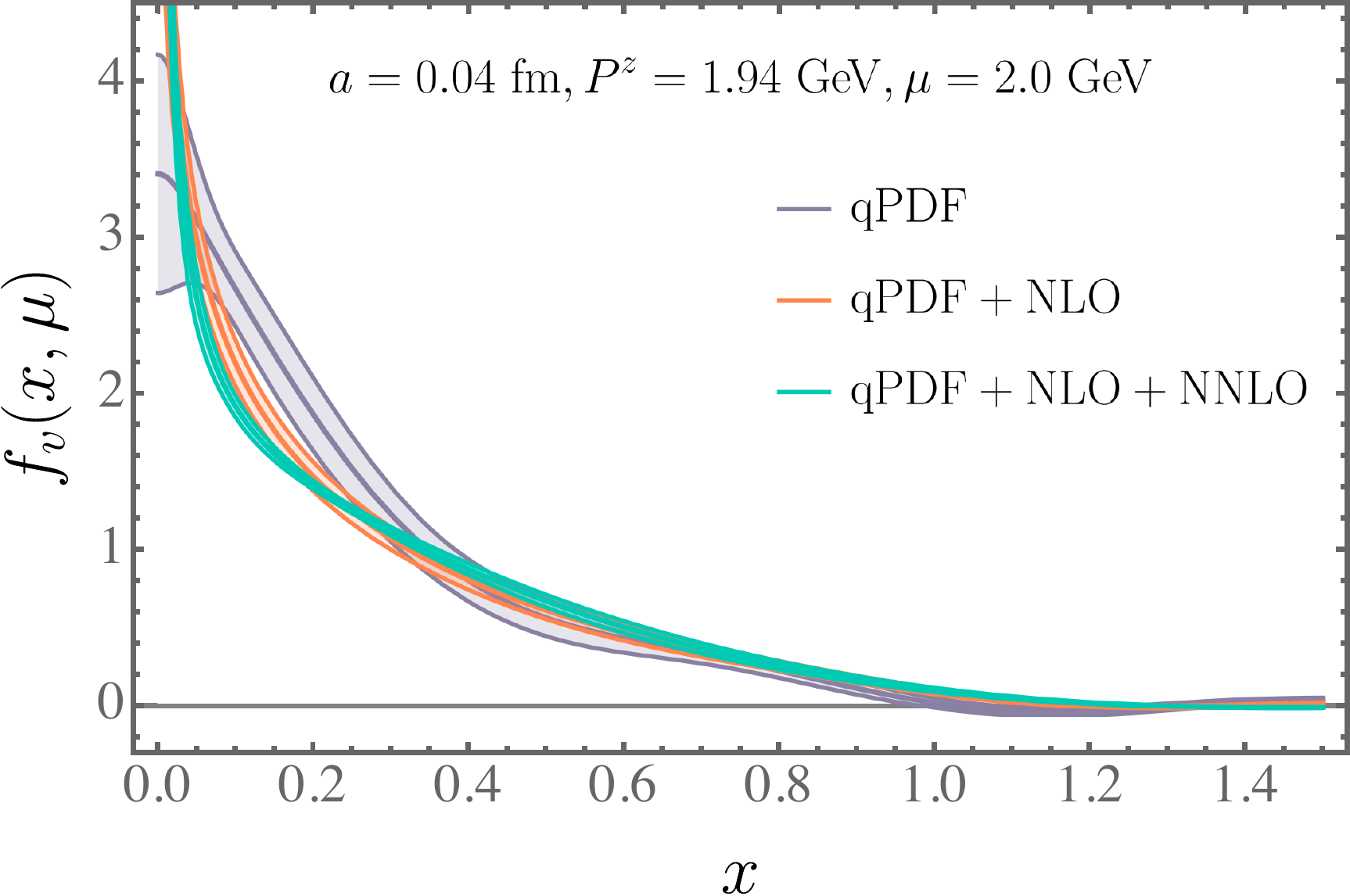}
    \caption{Comparison of PDFs obtained from the qPDF with NLO and NNLO matching corrections.}
    \label{fig:match}
\end{figure}

In \fig{match} we show the results of perturbative matching.
The matching drives the qPDF to smaller $x$ and reduces the statistical errors at moderate $x$, because matching effectively relates the qPDF from finite $P^z$ to infinity, and the qPDF evolves to smaller $x$ as $P^z$ increases.
The NNLO correction is generally smaller than the NLO correction,
which indicates good perturbative convergence, a crucial criterion for precision calculation. 
Besides, by varying $\mu$ and evolving the matched results to the same $\mu$, 
we find that the scale-variation uncertainty is reduced at NNLO, which is further evidence of improved precision.
The matching correction diverges as $x\to0$, implying that resummation of small-$x$ logarithms is needed. A resummation is also necessary as $x\to1$~\cite{Gao:2021hxl}, but these resummations are not needed for moderate $x$.

\begin{figure}
    \centering
    \includegraphics[width=\columnwidth]{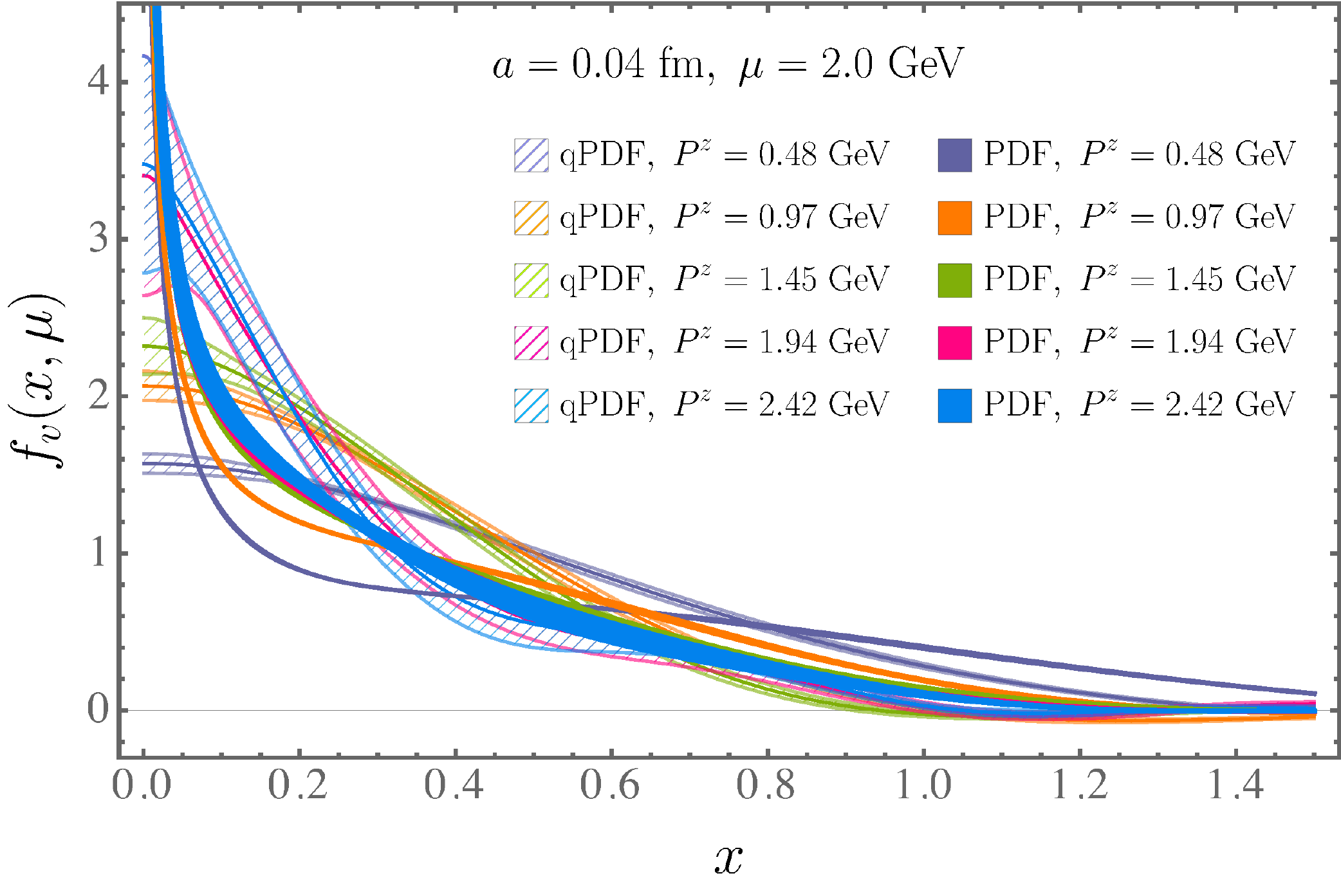}
    \caption{The PDFs obtained from the qPDFs with NNLO matching at different $P^z$.}
    \label{fig:mom}
\end{figure}

We compare the PDFs obtained at different $P^z$ with NNLO matching in \fig{mom}.
At moderate $x$, the $P^z$-dependence is remarkably reduced, and the results appear to converge for $P^z\ge 1.45$ GeV, which strongly indicates the effectiveness of LaMET matching.
At $x\gtrsim1$, each PDF curve has a small non-vanishing tail due to the power corrections in \eq{fact}, but they decrease with larger $P^z$ (see also \app{pzdependence}). To estimate the size of the power corrections, we fit the PDFs obtained at $a=0.04$ fm, $P^z=\{1.45,1.94,2.42\}$ GeV and $a=0.06$ fm, $P^z=\{1.72,2.15\}$ GeV to the \textit{ansatz} $f_v(x) + \alpha(x) / P_z^2$ for each fixed $x$, where we ignore the $a$-dependence as it has been found that the matrix elements have ${\cal O}(a^2P_z^2)$ effects that are less than 1\%~\cite{Gao:2020ito}. 
Since this fit is mainly affected by the data sets at lower $P^z$ with smaller statistical errors, which have larger power corrections, we use the result at $P^z=2.42$ GeV instead of the fitted $f_v(x)$ as our final prediction.
The power correction at $P^z=2.42$ GeV is estimated to be $ \alpha(x) / [P_z^2 f_v(x)]<0.10$ for $0.01<x<0.80$. It is surprising that the results are insensitive to $P^z$ for $x$ as small as $0.01$, nor do they show dependence on the extrapolation form in the FT as we have checked. This can be explained by that, under matching, the qPDF contributes to the PDF at larger $x$ which has less dependence on $P^z$ or the extrapolation. 
Nevertheless, it must be pointed out that the smallness here is only relative, as $\alpha(x) / P_z^2$ still diverges as $x\to0$.

\begin{figure}
\centering
\includegraphics[width=\columnwidth]{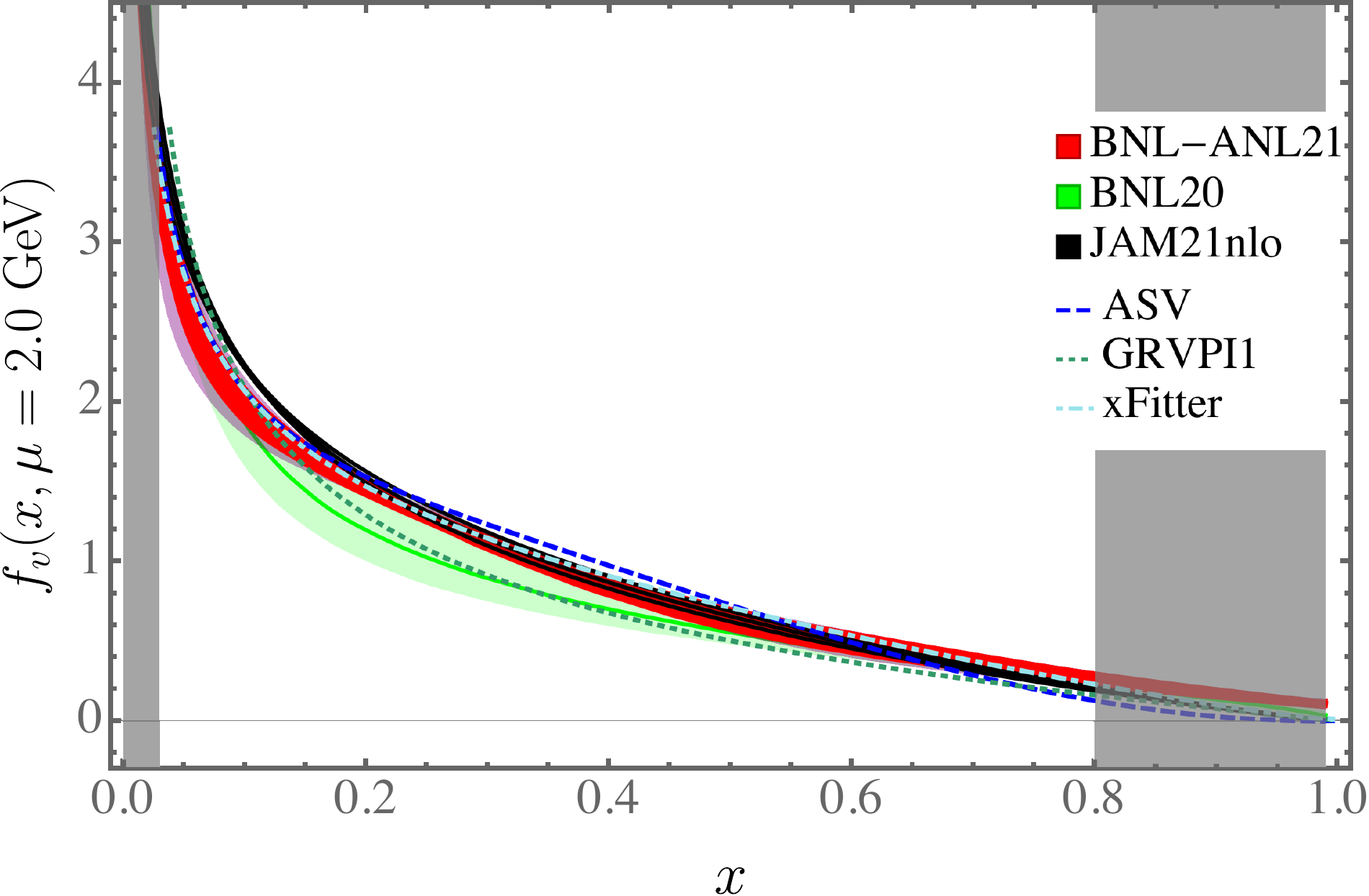}
\begin{picture}(0,0)
\put(-84,79){ \includegraphics[scale=0.48]{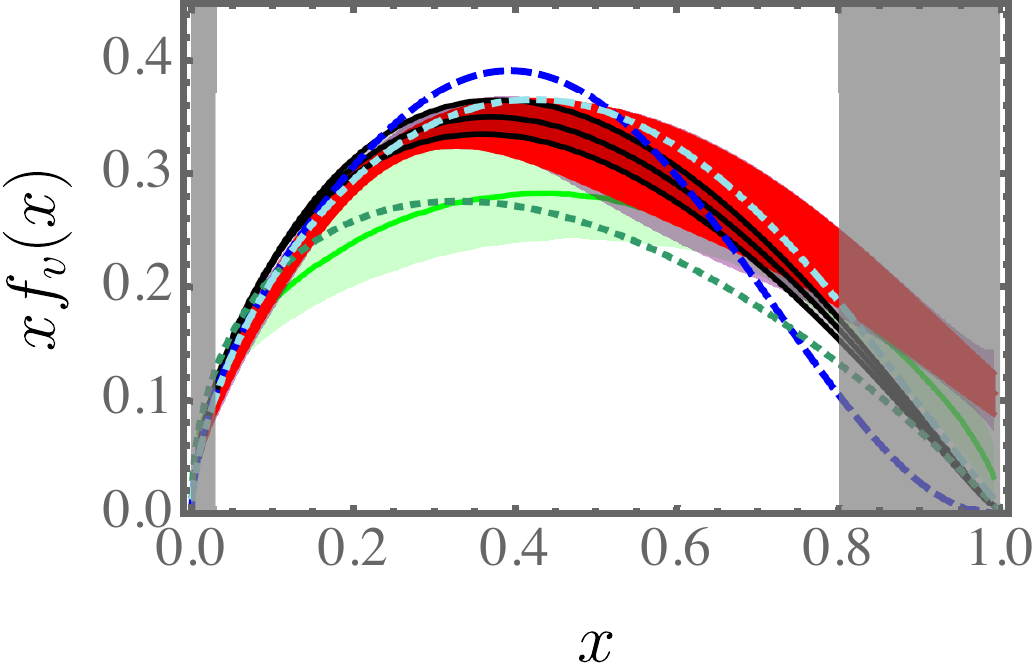}}
\end{picture}
\vspace*{-3mm}
\caption{Comparison of our prediction of $f_v(x)$, BNL-ANL21, to global fits and BNL20. The shaded regions $x<0.03$ and $x>0.8$ are excluded by requiring that estimates of ${\cal O}(\alpha_s^3)$ and power corrections be smaller than 5\% and 10\%, respectively.}
\label{fig:comp}
\end{figure}

Our final prediction for the pion valence quark PDF (BNL-ANL21) is shown in \fig{comp}, which is obtained from the qPDF at $a=0.04$ fm, $z_S=0.24$ fm, $z_L=0.92$ fm, $\mu=2.0$ GeV and $P^z=2.42$ GeV with exponential extrapolation and NNLO matching. The red band represents the statistical error, and the light purple band includes the error from scale variations, which is obtained by repeating the same analysis for $\mu=1.4$ GeV and $2.8$ GeV and evolving the PDFs to $\mu=2.0$ GeV with the NLO DGLAP kernel.
Since the hybrid-scheme parameter $\bar{m}_0$ depends on $\mu$, the small scale variation in the final result shows that the renormalization uncertainty is well under control.
We require that the ${\cal O}(\alpha_s^3)$ matching correction at $\mu=2.0$ GeV be smaller than 5\%, which propagates geometrically to $<37\%$ at NLO and $<14\%$ at NNLO, thus excluding $x<0.03$ and $ x > 0.88$. 
A list of the above uncertainties at selected $x$ is shown in \tbl{error}. See also \app{final}. We neglect the FT uncertainty as it is extremely small.
As for $m_\pi$ dependence, our associated calculation of the second PDF moment at $m_\pi=140$ MeV~\cite{Gao:2021ghp} shows consistency within 5\% statistical uncertainty, which will be validated by a direct comparison in the future. Previous studies~\cite{Lin:2019ocg,Liu:2020krc} also suggest that the finite volume correction is less than 1\% for our lattice setup.
 At last, by limiting the estimated power corrections to be less than 10\%, we determine the PDF at $0.03\lesssim x\lesssim 0.80$ with 5--20\% uncertainties.
Our result is in great agreement with the recent global fits by \texttt{xFitter}~\cite{Novikov:2020snp} and JAM21nlo~\cite{Barry:2021osv} for $0.2<x<0.6$, but deviates from the earlier GRVPI1~\cite{Gluck:1991ey} and ASV~\cite{Aicher:2010cb} fits.
When compared to a previous analysis of the same lattice data (BNL20)~\cite{Gao:2020ito}, which used a short-distance factorization of the matrix elements at NLO, and a parameterization of the PDF, our new result has shifted central values and considerably reduced uncertainties at moderate $x$, but still agrees within errors. With finite $P^z$ and statistics, lattice QCD can only make predictions for $x\in [x_{\rm min}, x_{\rm max}]$. The PDF parameterization correlates the information at all $x\in[0,1]$, so the larger uncertainties at moderate $x$ in BNL20 could be propagated from the uncontrolled errors in the end-point regions. Besides, there is no practical estimate of the model uncertainty in the parameterization. Therefore, the LaMET calculation for $x\in[x_{\rm min}, x_{\rm max}]$ is more reliable as it does the power expansion and matching directly in $x$-space.

\begin{table}
	\centering
	\begin{tabular}{|c|c|c|c|c|c|}
	\hline
	$x$ & Statistical & Scale & ${\cal O}(\alpha_s^3)$ & Power corrections & ${\cal O}(a^2P_z^2)$ \\
	\hline
	0.03 & 0.10 & 0.04 & $<0.05$ & $<0.01$& $<0.01$\\
	\hline
	0.40	& 0.07 & $<0.01$ & $<0.05$ & 0.04 & $<0.01$\\
	\hline
	0.80 & 0.15 & 0.03 & $<0.05$ & 0.10 & $<0.01$\\
	\hline
	\end{tabular}
\caption{Statistical and systematic uncertainties at given $x$.}
\label{tbl:error}
\end{table}

In summary, we have performed a state-of-the-art lattice QCD calculation of the $x$-dependence of pion valence quark PDF, where we developed a procedure to renormalize the qPDF in the hybrid scheme and match it to the $\MS$ PDF at NNLO. The final results show reduced perturbation theory uncertainty and converge at moderate $x$ with pion momenta greater than $1.45$ GeV, which allows us to reliably estimate the systematic errors. This
calculation can be improved with physical pion mass, continuum extrapolation, and higher statistics for the matrix elements at long distances and at larger boost momenta.

Our renormalization procedure can also be incorporated into the lattice calculations of gluon PDFs, distribution amplitudes, generalized parton distributions and transverse momentum distributions. With the systematics under control, we can expect lattice QCD to provide reliable predictions for these quantities in the future.

\begin{acknowledgments}

We thank Vladimir Braun, Xiangdong Ji, Nikhil Karthik, Yizhuang Liu, Antonio Pineda, Yushan Su and Jianhui Zhang for valuable communications.
This material is based upon work supported by: (i) The U.S. Department of Energy, Office of Science, Office of Nuclear Physics through Contract No.~DE-SC0012704 and No.~DE-AC02-06CH11357; (ii) The U.S. Department of Energy, Office of Science, Office of Nuclear Physics and Office of Advanced Scientific Computing Research within the framework of Scientific Discovery through Advance Computing (SciDAC) award Computing the Properties of Matter with Leadership Computing Resources; (iii) The U.S. Department of Energy, Office of Science, Office of Nuclear Physics, within the framework of the TMD Topical Collaboration. (iv) XG is partially supported by the NSFC under the grant number 11775096 and the Guangdong Major Project of Basic and Applied Basic Research No. 2020B0301030008. (v) SS is supported by the National Science Foundation under CAREER Award PHY-1847893 and by the RHIC Physics Fellow Program of the RIKEN BNL Research Center.. (vi) This research used awards of computer time provided by the INCITE and ALCC programs at Oak Ridge Leadership Computing Facility, a DOE Office of Science User Facility operated under Contract No. DE-AC05-00OR22725. (vii) Computations for this work were carried out in part on facilities of the USQCD Collaboration, which are funded by the Office of Science of the U.S. Department of Energy. viii) YZ is partially supported by an LDRD initiative at Argonne National Laboratory under Project~No.~2020-0020.

\end{acknowledgments}

\appendix

\section{Hybrid scheme renormalization}
\label{app:ren}

\subsection{Definition of scheme}
\label{app:def}

As has been described in the main text, the hybrid scheme renormalization includes two parts:

\begin{itemize}
    \item For $z\le z_S$, we form the ratio of bare matrix elements~\cite{Orginos:2017kos},
    \begin{align}\label{eq:ratios}
        \frac{\tilde h(z,P^z,a)}{\tilde h(z,0,a)}\,,
    \end{align}
    which has a well-defined continuum limit and is renormalization group (RG) invariant.
    
    \item For $z\ge z_S$, the renormalized matrix element is
    \begin{align}
        e^{\delta m(a) |z-z_S|}{\tilde h(z,P^z,a) \over \tilde h(z_S,0,a)} \,,
    \end{align}
which is equal to the ratio in \eq{ratios} at $z=z_S$.
To determine $\delta m(a)$ we use the additive renormalization constant, $c_Q(a)=\delta m(a)$, which is obtained in Ref.~\cite{Bazavov:2018wmo} from the analysis of the free energy of a static quark, $F_Q(T)$, at non-zero temperature $T$ with the normalization condition in \eq{pot}. Recently $F_Q$ has been calculated using one step of HYP smearing \cite{Petreczky:2021mef}, and it was found that HYP smearing does not affect the temperature
dependence of $F_Q(T)$, but only shifts it by an additive constant.
Therefore, we have $F_Q^{B,1}(T)+\delta m(a)=F_Q^{B,0}(T)+c_Q(a)$
with superscripts 0 and 1 referring to the number of HYP smearing
steps in the bare free energy of the static quark. Using the lattice
results for $F_Q^{B,0}(T)$ and $F_Q^{B,1}(T)$ obtained
on $N_{\tau}=12$ lattices 
and temperatures corresponding to $a=0.04$ fm and $a=0.06$ fm (where cutoff effects can be neglected), as well as the values of 
$c_Q$ from Table X of Ref.~\cite{Bazavov:2018wmo} for
$\beta=7.825$ ($a=0.04$ fm) and $\beta=7.373$ ($a=0.06$ fm), we obtain $\delta m(a)$.
The results are $a\delta m(a=0.06{\rm\ fm})=0.1586(8)$ and $a\delta m(a=0.04{\rm\ fm})=0.1508(12)$. 
\end{itemize}

First of all, to test how well the subtraction of $\delta m(a)$ can remove the linear divergences in $\tilde h(z,P^z,a)$, we construct the ratio in \eq{mfit},
\begin{align}\label{eq:ratio}
   \tilde R(z,z_0,a) &\equiv e^{\delta m(a) (z-z_0)} {\tilde h(z,0,a)\over \tilde h(z_0,0,a)}\,,
\end{align}
where $z_0=0.24$ fm for both lattice spacings.
According to \eq{renorm}, the renormalization factor $Z_O(a)$ cancels out in the ratio. Therefore, if $\delta m(a)$ includes all the linear divergences, then $\tilde R(z,z_0,a)$ should have a well-defined continuum limit.

\begin{figure}[htb]
    \centering
    \includegraphics[width=0.8\columnwidth]{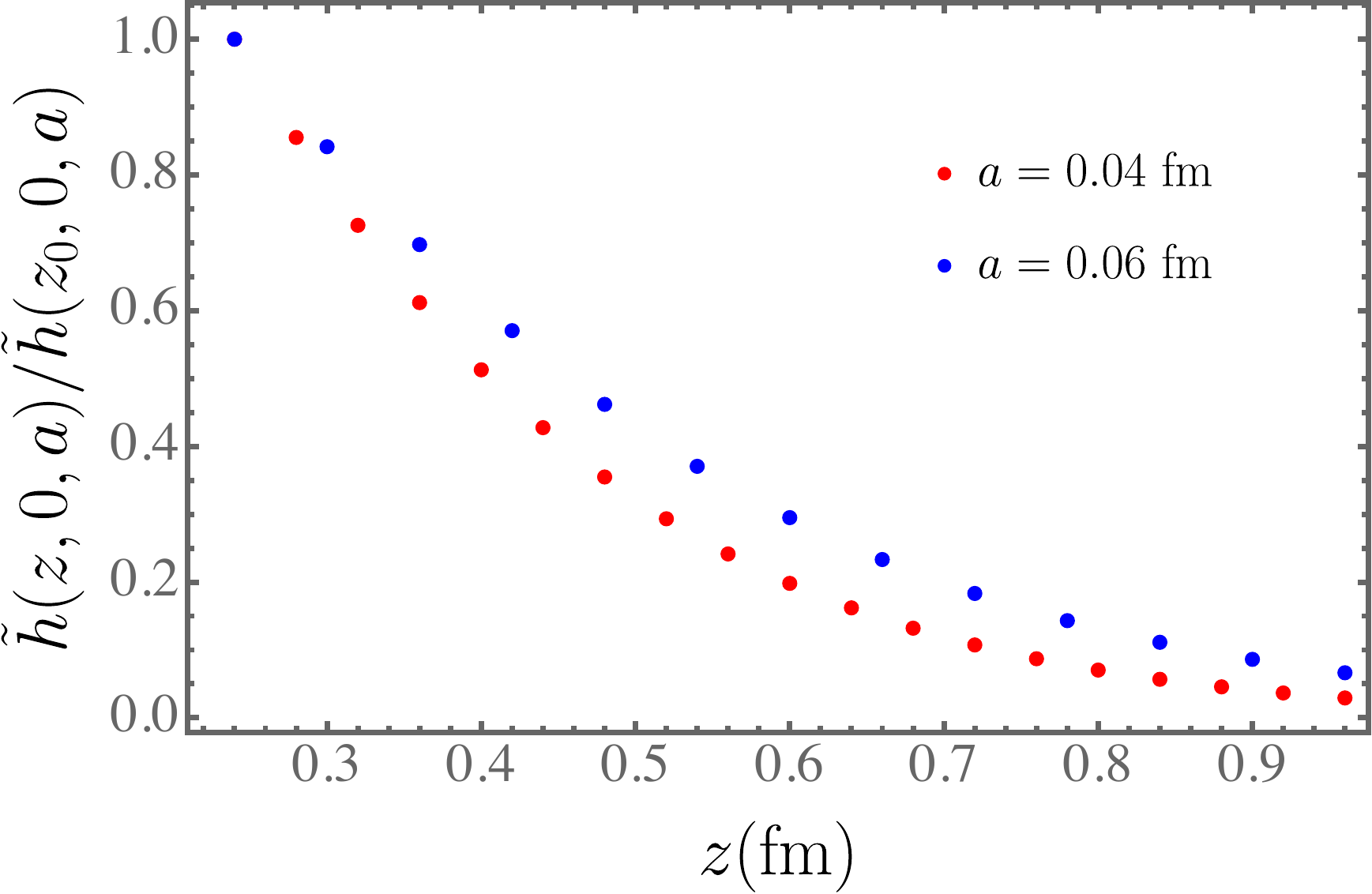}
    \includegraphics[width=0.8\columnwidth]{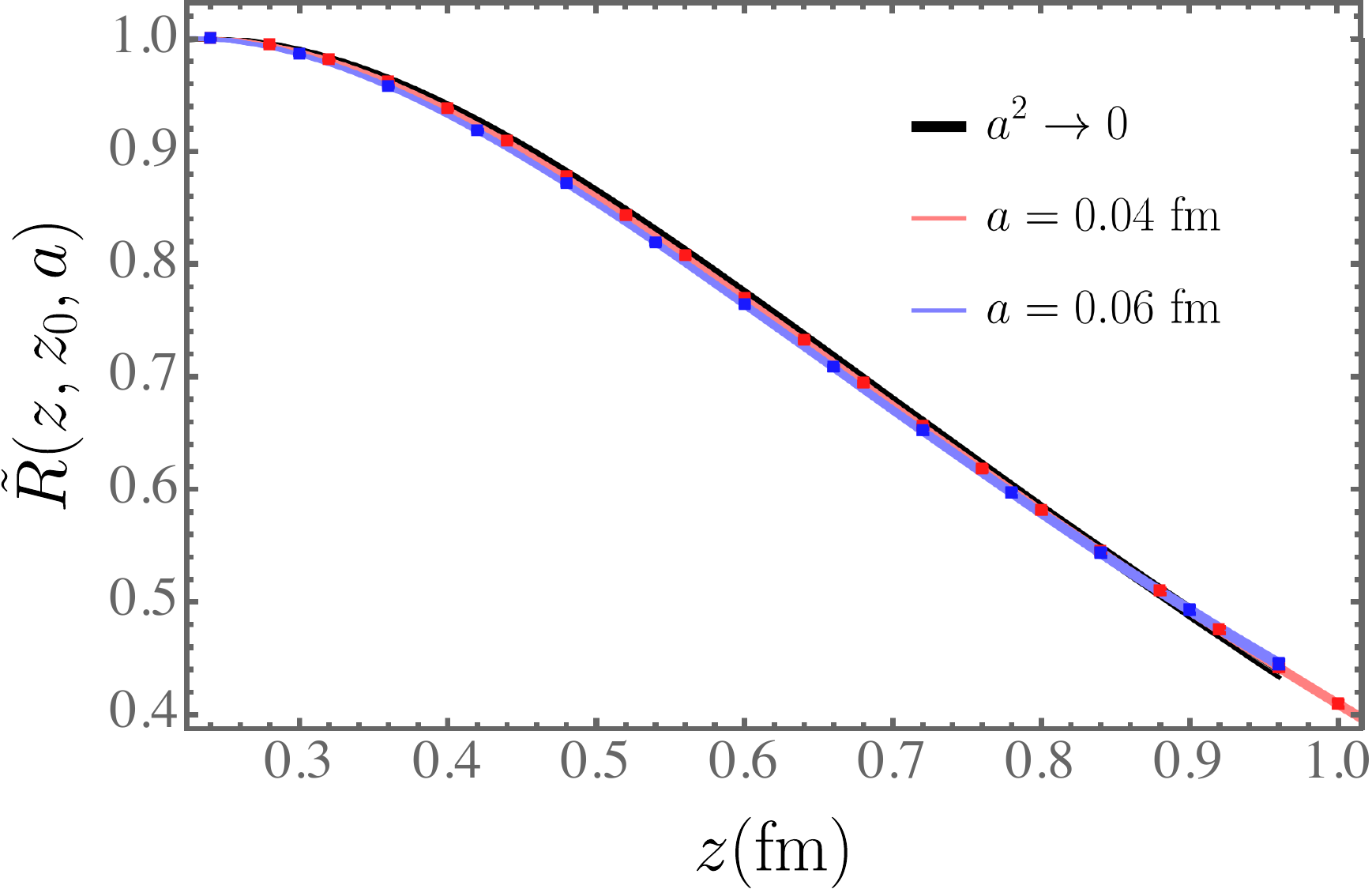}
    \caption{Upper panel: ratios of bare lattice matrix elements without the Wilson-line mass subtraction. Lower panel: the ratio in \eq{ratio} with Wilson-line mass subtraction. The red and blue points are for $a=0.04$ fm and $0.06$ fm. The red and blue bands are interpolations of the points, and the gray band is the continuum extrapolation of them with $a^2$-dependence.}
    \label{fig:ratio}
\end{figure}

Our lattice results for the above ratio with $z_0=0.24$ fm is shown in \fig{ratio}. As one can see, the differences between the ratios at $a=0.04$ fm and $0.06$ fm are at sub-percent level, which clearly shows that the linear divergences have been sufficiently subtracted by $\delta m(a)$.
Therefore, the ratio in \eq{ratio} has a continuum limit 
\begin{align}
   \lim_{a\to0} \tilde R(z,z_0,a) &= \tilde R(z,z_0)\,,
\end{align}
which is RG invariant.

Our next step is to match the lattice subtraction scheme to $\MS$.
When $z,z_0 \ll \Lambda_{\rm QCD}^{-1}$, the $\MS$ matrix element $\tilde h^{\MS}(z,0,\mu)$ has an OPE that goes as
\begin{align}\label{eq:ope}
    \tilde h^{\MS}(z,0,\mu) &= e^{-m^{\MS}_0|z|} \left[C_0(z^2\mu^2) \right.\nn\\
    &\left. + z^2 C_2(z^2\mu^2) \langle P| O_{\rm tw4}(\mu) |P\rangle + \ldots\right]\,,
\end{align}
where $m^{\MS}_0$ is the ${\cal O}(\Lambda_{\rm QCD})$ renormalon ambiguity from the Wilson line self-energy renormalization~\cite{Braun:2018brg}, $O_{\rm tw4}(\mu)$ is a twist-four operator (for example, $\bar{\psi} D^2 \psi$ or $g\bar{\psi} \sigma_{\mu\nu}F^{\mu\nu} \psi$), $C_0$ and $C_2$ are perturbative coefficient functions, and ``$\ldots$'' denotes contributions at higher twists. Since $P^z=0$, $C_0$ is the only Wilson coefficient that contributes at leading-twist.
The leading-twist contribution is proportional to $\langle P|\bar{\psi}\gamma^t \psi|P\rangle/(2P^t)$ which is trivially one due to vector current conservation.
Since $\tilde h^{\MS}(z,0,\mu)$ is multiplicatively renormalizable, both $C_0(z^2\mu^2)$ and $C_2(z^2\mu^2) \langle P| O_{\rm tw4}(\mu) |P\rangle$ must satisfy RG equations with the same anomalous dimension, which is known to next-to-next-to-next-to-leading order (N$^3$LO)~\cite{Braun:2020ymy}. Due to the ambiguity in summing the perturbative series in $C_0(z^2\mu^2)$, there are $O(\Lambda_{\rm QCD}^{2n})$ IR renormalons in the leading-twist contribution that should be cancelled by those from higher-twist condensates, along with the ${\cal O}(\Lambda_{\rm QCD})$ UV renormalon to be cancelled by $m^{\MS}_0$~\cite{Braun:2018brg,Beneke:1994sw}. Both the UV and IR renormalon contributions cannot be well defined unless one specifies how to sum the perturbative series in $C_0(z^2\mu^2)$ to all orders, which, however, is unknown as $C_0(z^2\mu^2)$ has been calculated to only NNLO so far~\cite{Li:2020xml}.

Note that $m^{\MS}_0$ is analogous to the mass renormalization in heavy-quark effective theory (HQET)~\cite{Beneke:1994sw}, which is of UV origin and cannot be attributed to any short-distance condensate. Instead, it appears as a residual mass term in the HQET Lagrangian and exists in $\tilde h^{\MS}(z,0,\mu)$ at all $z$, i.e.,
\begin{align}\label{eq:ms}
    \tilde h^{\MS}(z,0,\mu) &= e^{-m^{\MS}_0|z|}\tilde h^{\MS}_0(z,0,\mu)\,,
\end{align}
where $\tilde h^{\MS}_0(z,0,\mu)$ at short distance reduces to the OPE series in the brackets of \eq{ope}.

The renormalons have been studied extensively for the Polyakov loop and plaquette in lattice QCD~\cite{Bauer:2011ws,Bali:2013pla,Bali:2014fea,Bali:2014sja,Bali:2014gha}. In lattice perturbation theory, one has to compute the perturbative series to very high orders of $\alpha_s$ in order to see the renormalon effects. Nevertheless, in the $\MS$ scheme, the OPE with Wilson coefficient at a few loop orders and the condensate term, turns out to be successful in describing the static potential at short distance up to $\sim0.25$ fm~\cite{Pineda:2002se}. One explanation is that $\alpha_s$ in the $\MS$ scheme is larger than that in lattice perturbation theory, so the renormalon effect which is of ${\cal O}(\alpha_s^n)$ with $n\sim (2\pi)/(\beta_0\alpha_s)$ becomes significant at lower orders. This situation is similar to the OPE in QCD sum rules~\cite{Shifman:1978bx,Shifman:1978by,Novikov:1984ac,Novikov:1984rf,David:1985xj}, which works well in phenomenology.
The reason behind such success is probably due to a proper choice of the renormalization scale $\mu$ so that $\alpha_s(\mu)$ is small enough for the perturbative series to converge, while the $\mu$-dependent effects in the condensate remain insignificant as they should be of the same magnitude of highest order in the truncated perturbative series~\cite{Novikov:1984rf,David:1985xj}.

Therefore, we approximate \eq{ope} as
\begin{align}\label{eq:renormalon}
    \tilde h^{\MS}(z,0,\mu) &\approx e^{-m^{\MS}_0(\mu)|z|} \left[C_0^{\rm FO}(z^2\mu^2) + \Lambda(\mu) z^2  \right]\,,
\end{align}
where ``FO'' stands for fixed order, $\Lambda(\mu)$ is a parameter of ${\cal O}(\Lambda_{\rm QCD}^2)$, and we ignore the higher power corrections by working at not too large $z$. The $\mu$ dependence of the parameters $m^{\MS}_0$ and $\Lambda$ is understandable because this approximation is valid for a small window of $\mu$, and they also depend on the perturbative orders in $C_0^{\rm FO}$ if the latter does not converge fast. Note that the although the model in \eq{renormalon} is not guaranteed to satisfy the RG equation for $\tilde h^{\MS}(z,0,\mu)$, we argue that within the range of $\mu$ where it can describe the physical results, the $\mu$-dependence in the power correction term, which is already suppressed, is weak and can be ignored.

Based on the above approximation, we fit our lattice results of the ratio in \eq{ratio} to the following \textit{ansatz},
\begin{align}\label{eq:ansatz}
    \tilde R(z,z_0) &= e^{-\bar{m}_0(\mu)(z-z_0)}  \frac{C_0^{\rm FO}(z^2\mu^2)  + \Lambda(\mu) z^2}{C_0^{\rm FO}(z_0^2\mu^2)  + \Lambda(\mu) z_0^2}\,,
\end{align}
where the mass shift
\begin{align}
    \bar{m}_0(\mu) &=  -m_0 + m^{\MS}_0(\mu)\,,
\end{align}
cancels the lattice scheme dependence of $m_0$ in \eq{mass} and introduces the renormalon ambiguity of the $\MS$ scheme.
Effectively, $\bar{m}_0$ matches the hybrid-scheme matrix elements at $z\ge z_S$ to the ratio of $\tilde h_0^{\MS}$ as
\begin{align}\label{eq:matching}
 \lim_{a\to0} e^{ (\delta m(a)+\bar{m}_0(\mu))(z-z_S)}{\tilde h(z, P^z, a)\over \tilde h(z_S,0, a)} &= {\tilde h^{\MS}_0(z,P^z,\mu)\over \tilde h^{\MS}_0(z_S,0,\mu)}\,.
\end{align}
Moreover, since the \textit{ansatz} in \eq{ansatz} can describe the short-distance matrix elements well, we can correct the $\Lambda z^2$ term in $\tilde h^{\MS}_0(z,0,\mu)$ at $z\le z_S$ as
\begin{align}
    {\tilde h^{\MS}_0(z,0,\mu)}  \frac{C_0^{\rm FO}(z^2\mu^2)}{C_0^{\rm FO}(z^2\mu^2)  + \Lambda(\mu) z^2}\,,
\end{align}
which is equivalent to replacing ${\tilde h^{\MS}_0(z,0,\mu)}$ by the perturbative $C_0$, as in \eq{hybren}. Eventually, the continuum limit of the matched matrix element in \eq{hybren} is
\begin{align}\label{eq:fullhyb}
    \tilde h(z, z_S,P^z,\mu) &= {\tilde h^{\MS}_0(z,P^z,\mu)\over C_0^{\rm FO}(z^2\mu^2)} \theta(z_S-|z|) \nn\\
    &\qquad + { \tilde h^{\MS}_0(z,P^z,\mu)\over C_0^{\rm FO}(z^2_S\mu^2)} \theta(|z|-z_S)\,,
\end{align}
which is different from $\MS$ through a perturbative matching for all $z$ as long as $z_S\ll \Lambda_{\rm QCD}^{-1}$. Therefore, the qPDF defined as FT of $\tilde h(z, z_S,P^z)$ is still factorizable.

Note that $\bar{m}_0(\mu)$ introduces the ambiguity $m^{\MS}_0(\mu)$ to the matched matrix elements. Nevertheless, we argue that $C^{\rm FO}_0(\mu^2 z^2)$ at NNLO is different from a particular summation prescription by ${\cal O}(\alpha_s^3)$ contributions, which cannot be smaller than the ambiguity in $m^{\MS}_0(\mu)$ as the latter reflects the uncertainty in summing divergent perturbative series at sufficiently high orders. Therefore, we can attribute the renormalon ambiguity in $\bar{m}_0(\mu)$ to higher loop-order effects, and estimate the latter by varying $\mu$ by a factor of $\sqrt{2}$ and $1/\sqrt{2}$. The range of $\mu$ we vary from cannot be too large. If $\mu$ is too small, then $\alpha_s(\mu)$ becomes too large; if $\mu$ is too large, then we need to resum the large $\ln(z^2\mu^2)$ in $C_0(z^2\mu^2)$. In both cases the perturbative series converges slowly. 
In our analysis, we scan $\mu$ within $[0.9, 2.0]$ GeV for $C_0^{\rm NLO}$ and $[1.4, 3.2]$ GeV for $C_0^{\rm NNLO}$ to study the scale dependence and uncertainty from renormalon ambiguity.

\subsection{Fitting of $\bar{m}_0$ and $\Lambda(\mu)$}
\label{app:m0}

Currently, the Wilson coefficient $C_0(\mu^2z^2)$ is known to NNLO~\cite{Chen:2020ody,Li:2020xml} and its anomalous dimension has been calculated at three-loop order~\cite{Braun:2020ymy},
\begin{align}
 &C_0\big(\mu^2z^2,\alpha_s(\mu)\big) = 1+ a_s\left(2 L+\frac{10}{3}\right)\nn\\
    & +\! a_s^2 \left[\frac{13}{2} L^2 \!+\! \frac{1461\!+\!28 \pi ^2}{54}  L  \!+\! \frac{38127 \!-\! 824 \pi^2 \!-\! 4032 \zeta (3)}{648}\right]\nn\\
    & + a_s^3\left[ \frac{143}{6} L^3 + \Big(\frac{6127}{36}+\frac{91 \pi ^2}{27}\Big) L^2 \right. \nn\\
    & \left.\qquad + \frac{690939+760 \pi ^4-8976 \pi^2-94068 \zeta (3)}{972}L + 400 \right]\nn\\
    & + O(a_s^4)\,,
\end{align}
where $a_s=\alpha_s/(2\pi)$, $L=\ln(\mu^2 z^2/b_0^2)$, and $b_0=2e^{-\gamma_E}$,. The factor $400$ in the last square bracket is a simple guess by assuming that the constant part of the perturbative correction grows as a geometric series in the order of $a_s$.

We also consider the RG improved (RGI) Wilson coefficient~\cite{Gao:2021hxl}
\begin{align}
  C_0^{\rm RGI}\big(\mu^2,z^2\big) &= C_0\big(1,\alpha_s(b_0/z)\big)\\
  &\quad \times \exp\Big[\int_{b_0/z}^{\mu} {d\alpha_s(\mu')}\ {\gamma_{\cal O}(\alpha(\mu'))\over \beta(\alpha_s(\mu'))}\Big]\,,\nn
\end{align}
where $\gamma_{\cal O}$ is the anomalous dimension of the operator $O_{\Gamma}(z,\mu)$, and $\beta(\alpha_s(\mu))=d\alpha_s(\mu) / d\ln \mu^2$. In this way, we can first factor out the evolution factor in \eq{renormalon} as it must be satisfied by the full matrix element $\tilde h^{\MS}(z,0,\mu)$, and therefore construct the ratio $\tilde R(z,z_0)$ in an explicitly $\mu$-independent way.

\begin{figure}[htb]
    \centering
    \includegraphics[width=0.8\columnwidth]{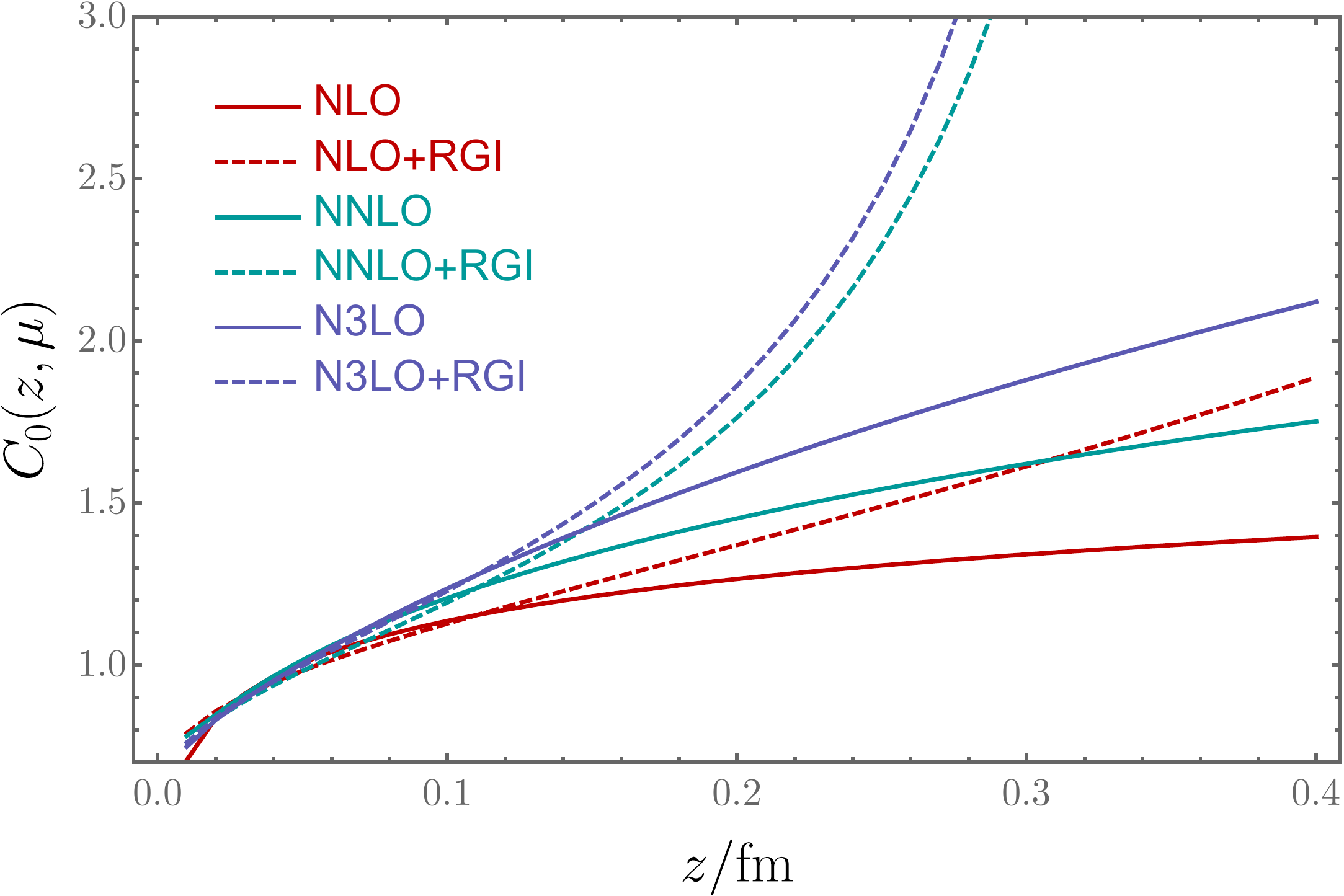}
    \caption{The fixed-order and RGI Wilson coefficients $C_0(z^2\mu^2)$ up to N$^3$LO.}
    \label{fig:c0}
\end{figure}

We compare $C_0$ and $C_0^{\rm RGI}$ at NLO, NNLO and N$^3$LO at $\mu=2.0$ GeV in \fig{c0}. The strong coupling constants at each perturbative order are defined by the corresponding $\Lambda_{\rm QCD}^{\MS}$ with one-, two- and three- loop $\beta$ functions and $n_f=3$, which are fixed by matching to $\alpha_s(\mu=2\ {\rm GeV})=0.293$. The latter is obtained from $\Lambda_{\rm QCD}^{\MS}=332$ MeV with five-loop $\beta$-function and $n_f=3$, as has been calculated using the same lattice ensembles~\cite{Petreczky:2020tky}. As one can see, at $z > 0.2$ fm the RGI Wilson coefficients start to deviate significantly from the fixed-order ones, which is mainly due to the large value of $\alpha_s$ as in RGI Wilson coefficients as we evolve from $\mu$ to $1/\pmb{z}$. This indicates that at $z>0.2$ fm, the scale uncertainty in the perturbative series is significant due to the enhancement of non-perturbative effects, and to use OPE we should work at very short distances ($z<0.2$ fm). However, there will not be enough room for varying $z$ to satisfy $z\gg a$ so that discretization effects are suppressed. Therefore, in our analysis we loosen our requirement for very small $z$ by only using the \textit{ansatz} in \eq{ansatz} and not considering the RGI Wilson coefficients.

\begin{figure}[t]
\centering
    \subfloat[]{
        \centering
        \includegraphics[width=0.4\textwidth]{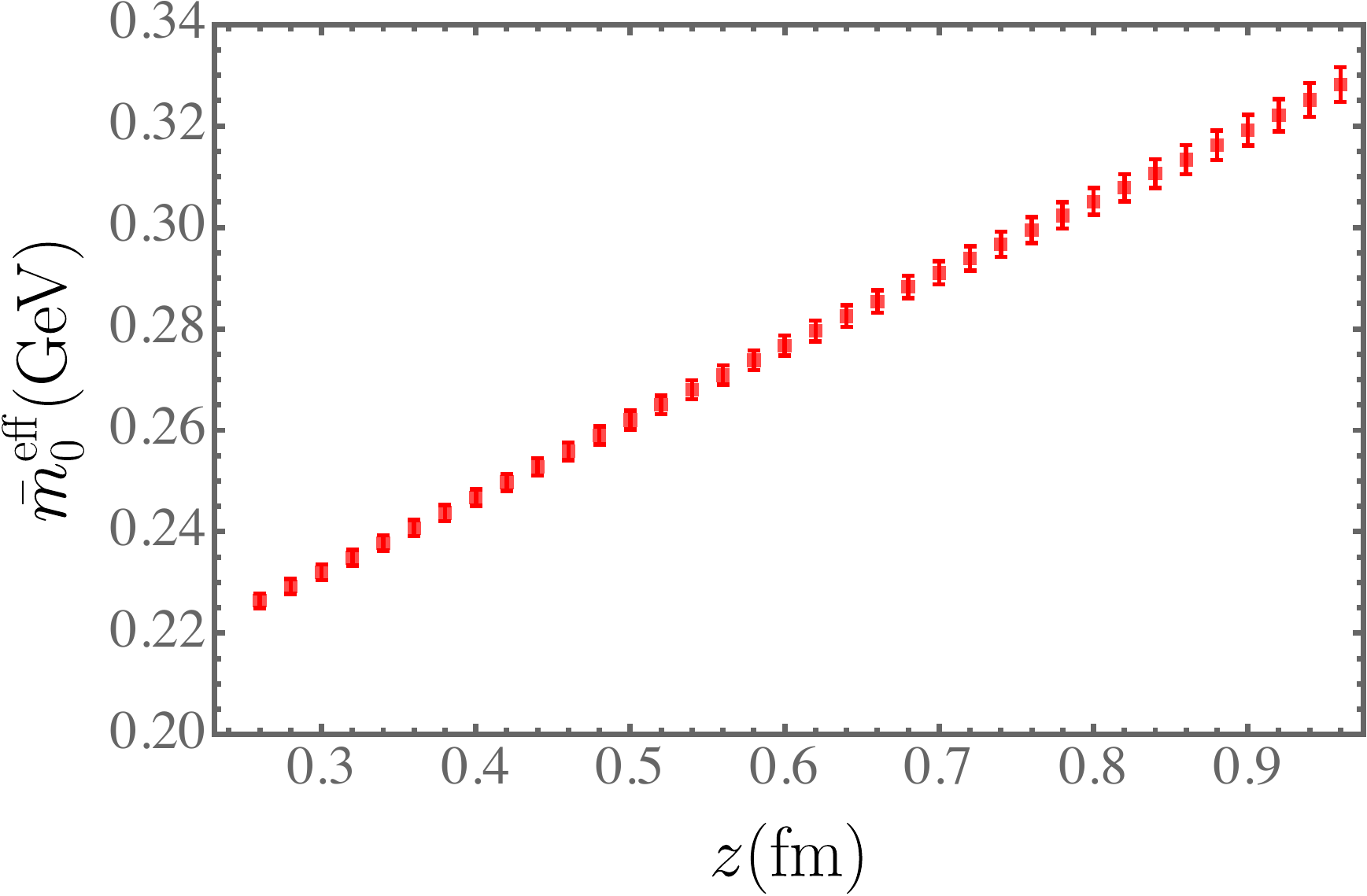}
        \label{fig:meff0}
        }\quad
    \subfloat[]{
        \centering
        \includegraphics[width=0.41\textwidth]{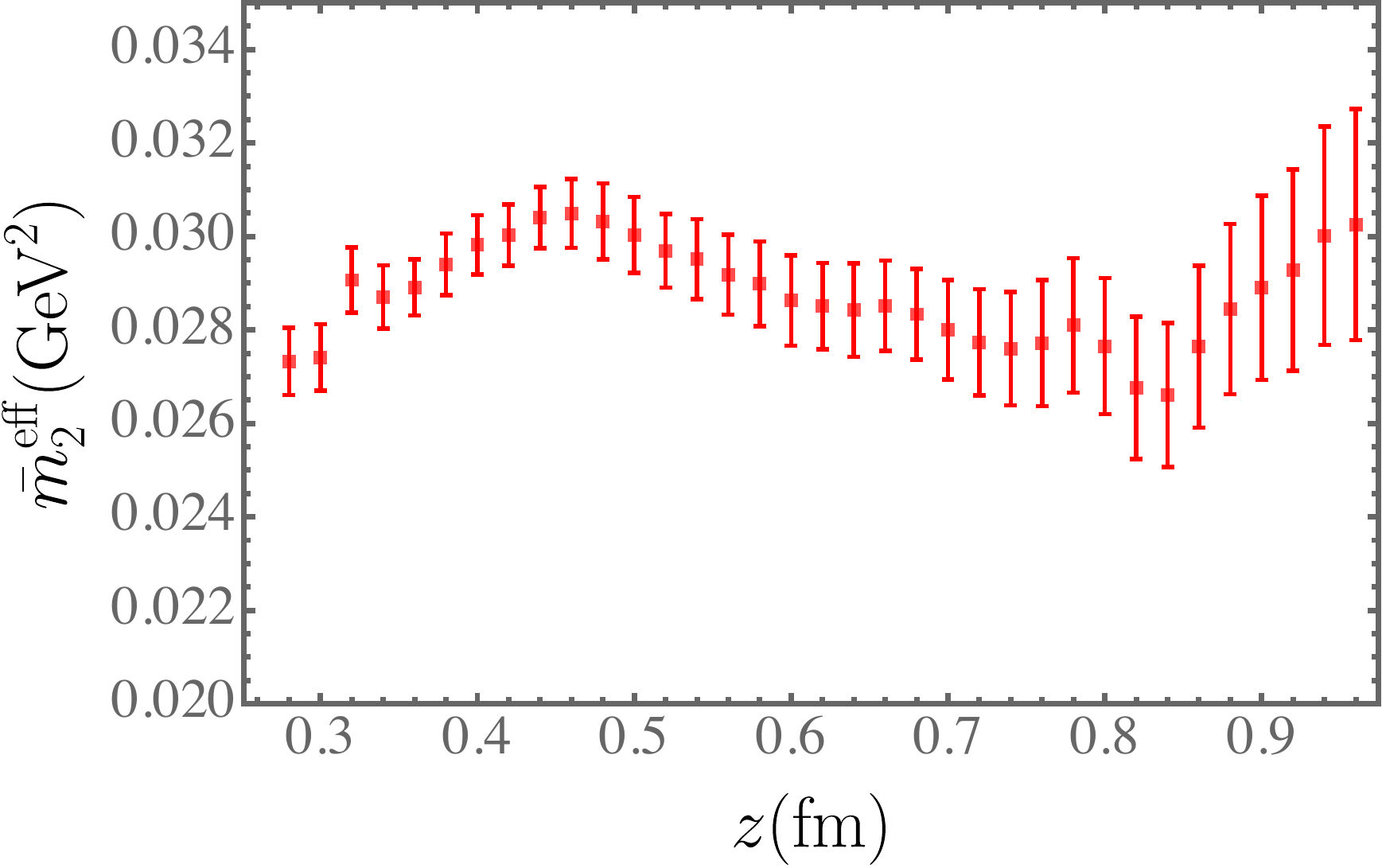}
        \label{fig:meff2} 
        }
        \caption{\label{fig:meff}
           Effective mass $\bar{m}_0^{\rm eff}(z)$ (a) and its slope $\bar{m}_2^{\rm eff}(z)$ (b) vs $z$.}
\end{figure}

In \fig{meff0}, we plot an effective mass $\bar{m}_0^{\rm eff}(z)$ which is defined as
\begin{align}
    \bar{m}_0^{\rm eff}(z)(z-z_0) &\equiv -\ln {\tilde h(z,0,a)\over \tilde h(z_0,0,a)} + \ln {C_0^{\rm NNLO}(z^2\mu^2)\over C_0^{\rm NNLO}(z^2_0\mu^2)}\,,
\end{align}
where $\mu=2.0$ GeV.
If the twist-four condensate is negligible, then we should expect a plateau in $z$, but \fig{meff0} shows that it has an almost constant nonzero slope at $z$ from $0.24$ fm up to $1.0$ fm. In \fig{meff2} we plot its slope
\begin{align}
    \bar{m}_2^{\rm eff}(z) &= {\bar{m}_0^{\rm eff}(z) - \bar{m}_0^{\rm eff}(z-a)\over a}\,,
\end{align}
which is consistent with being constant for a wide range of $z$. This suggests that there is considerable quadratic $z$-dependence from the twist-four condensate, as inclued in the \textit{ansatz} in \eq{ansatz}.

\begin{figure*}[t]
\centering
    \subfloat[]{
        \centering
        \includegraphics[width=0.45\textwidth]{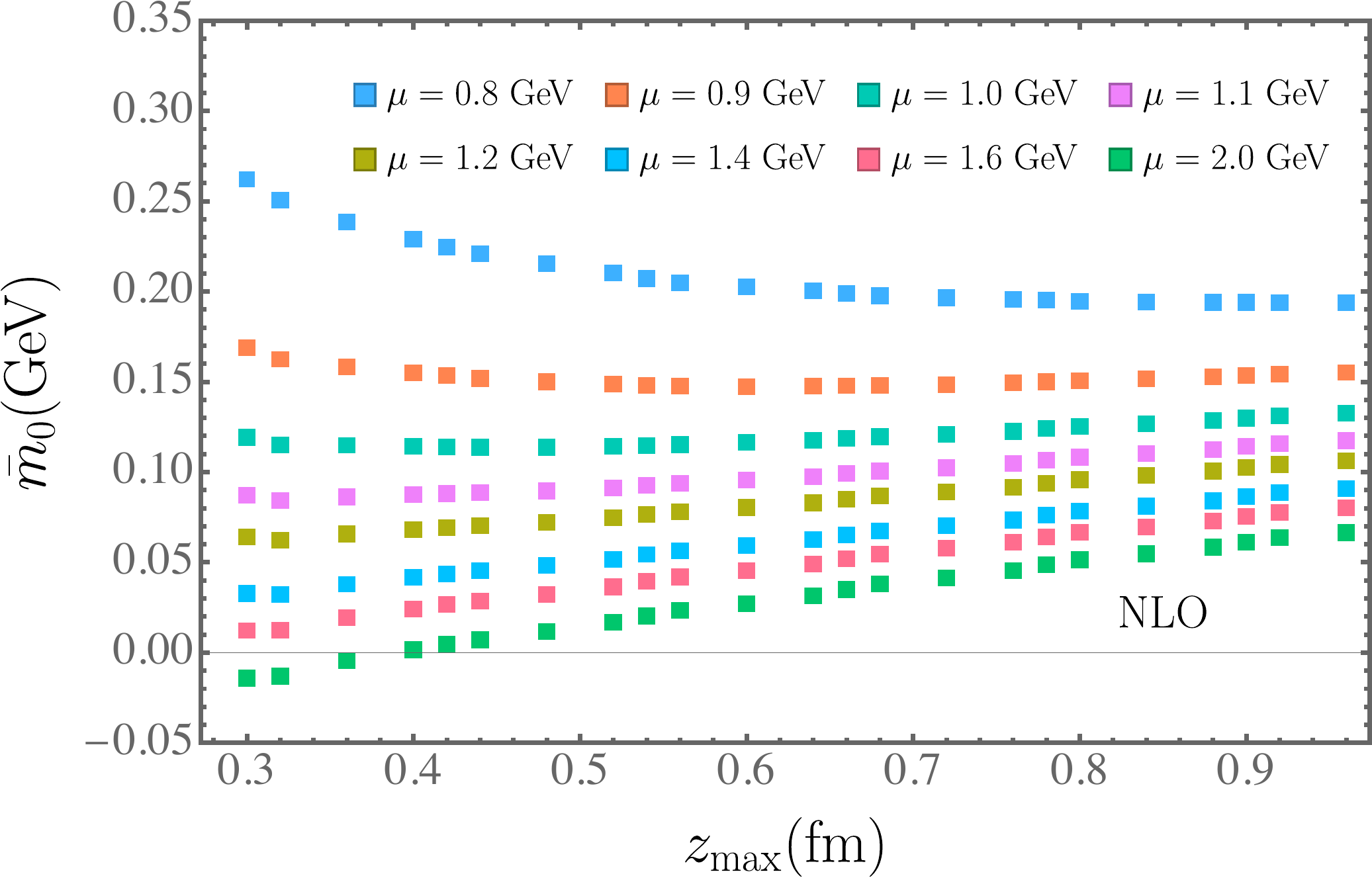}
        \label{fig:m0-nlo}
        }\quad
    \subfloat[]{
        \centering
        \includegraphics[width=0.45\textwidth]{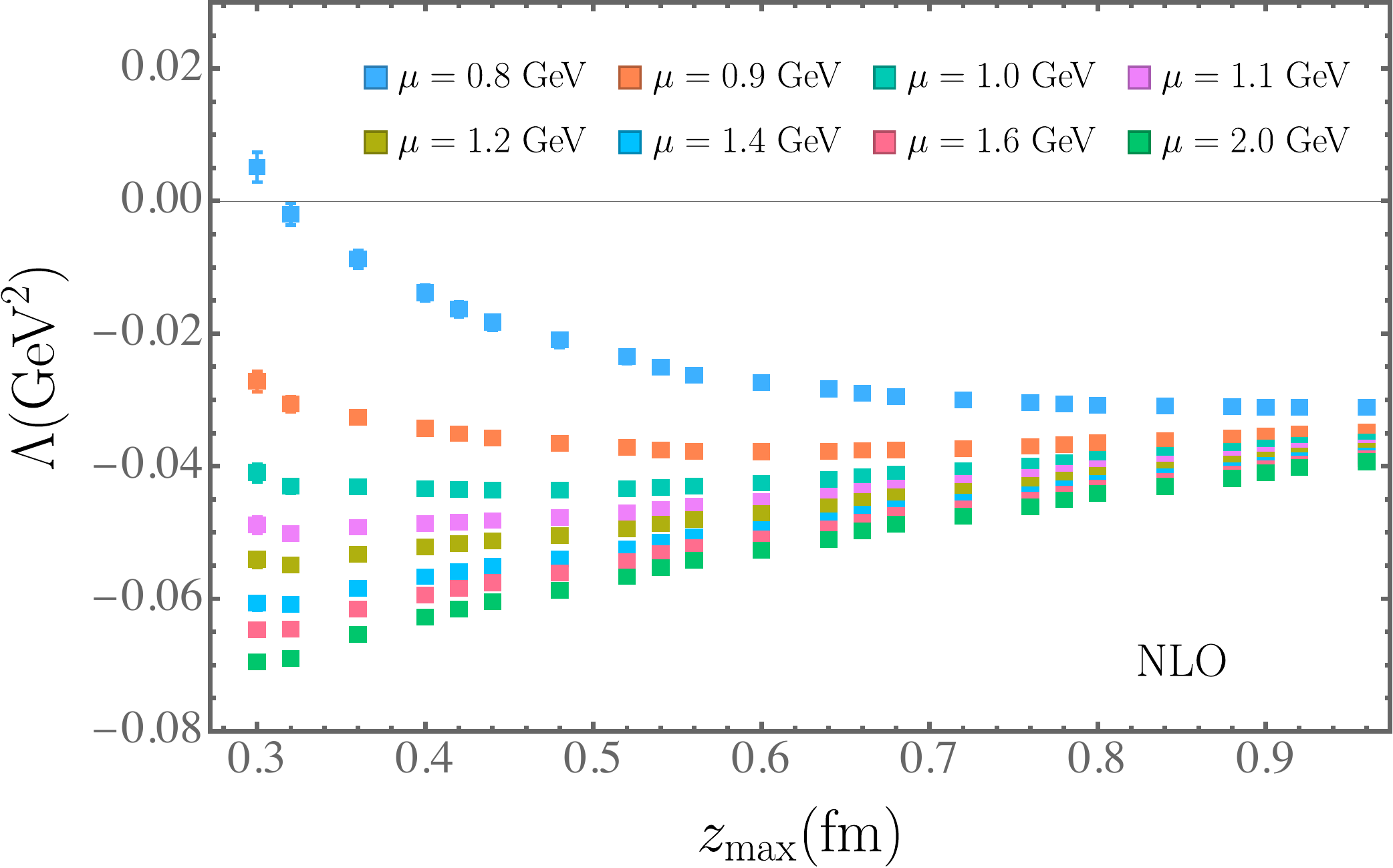}
        \label{fig:lambda-nlo} 
        }\\
    \subfloat[]{
        \centering
        \includegraphics[width=0.45\textwidth]{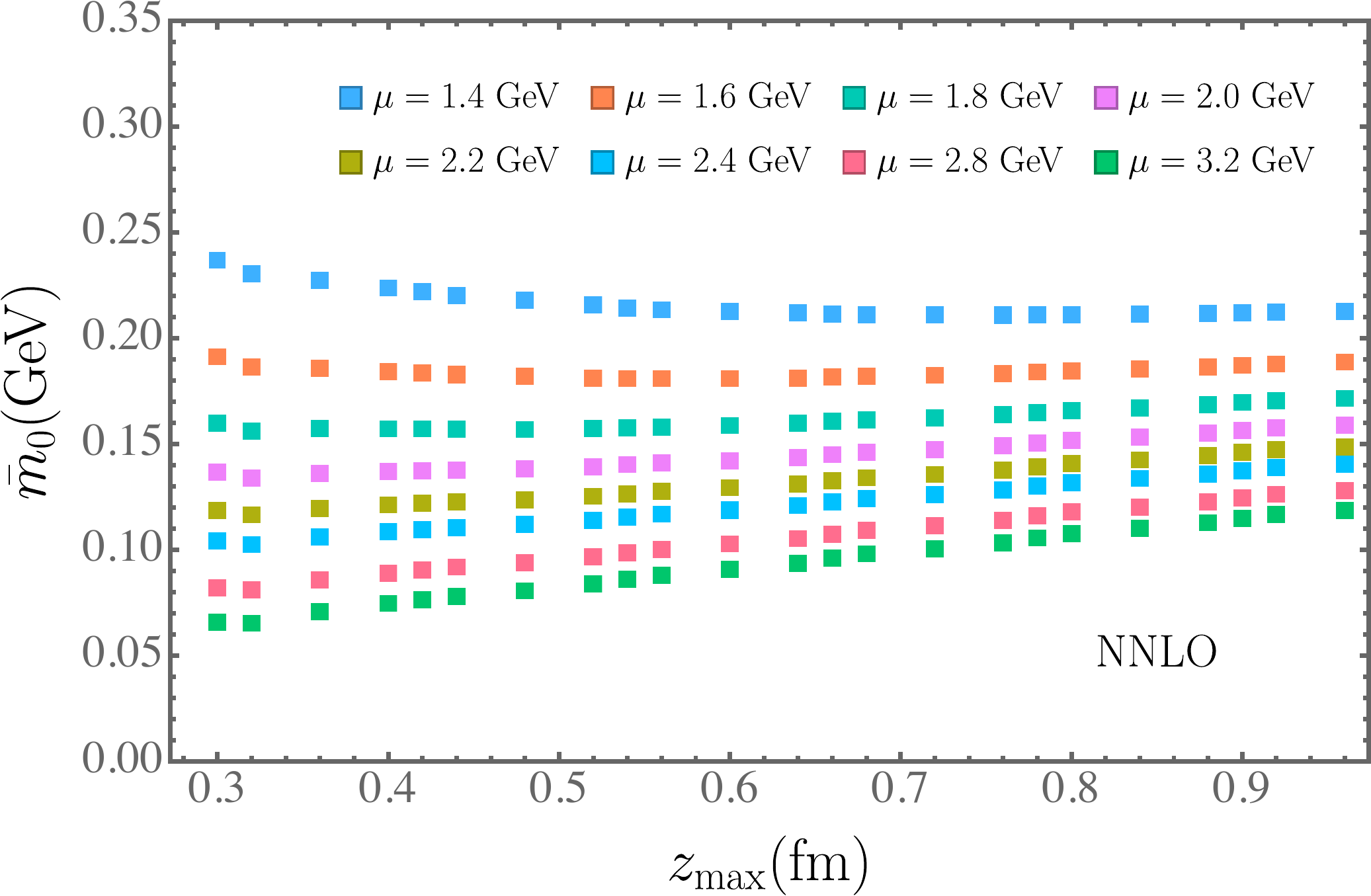}
        \label{fig:m0-nnlo} 
        }\quad
    \subfloat[]{
        \centering
        \includegraphics[width=0.45\textwidth]{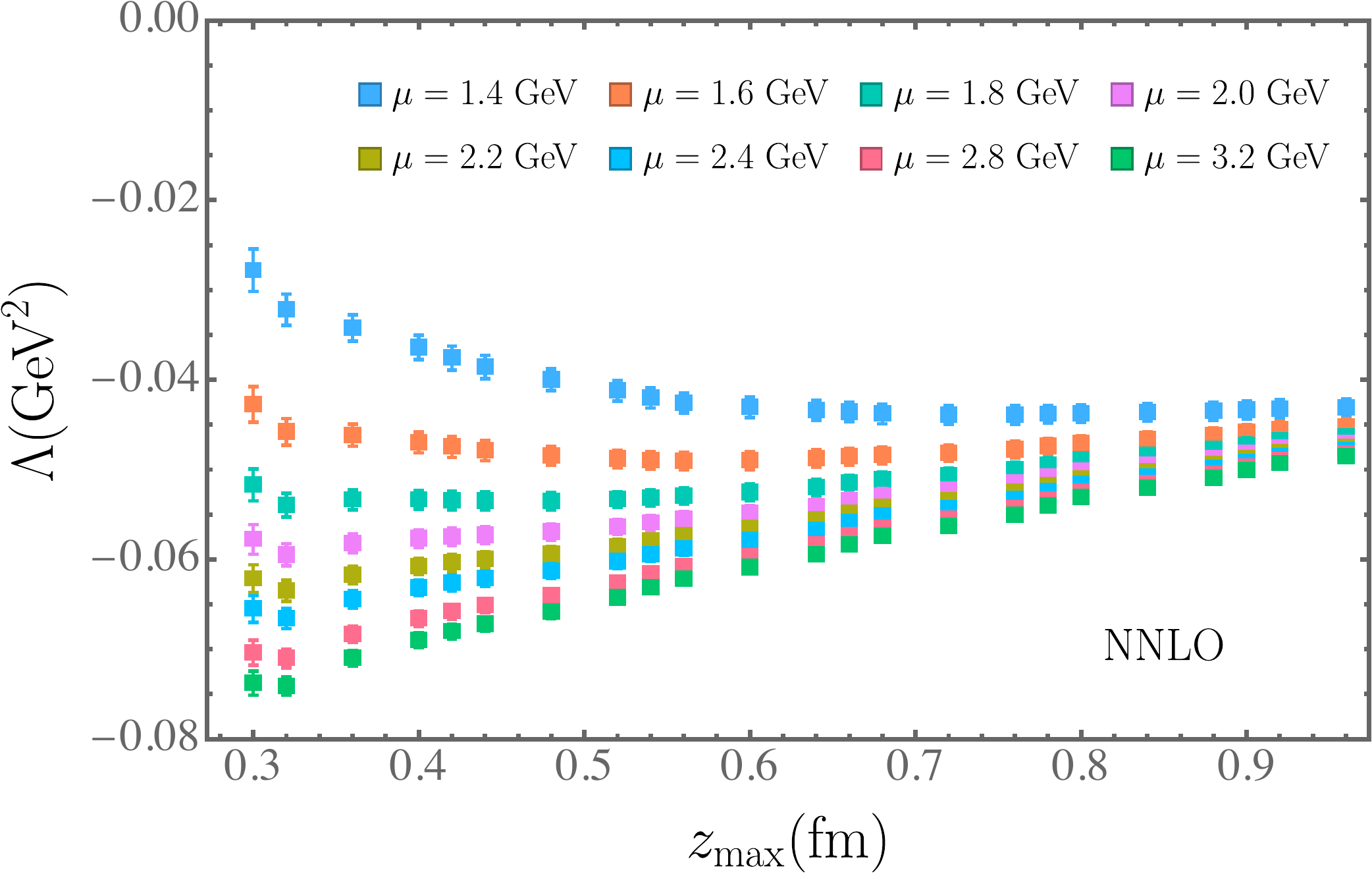}
        \label{fig:lambda-nnlo} 
        }
        \caption{\label{fig:mfit}
           Results for $\bar{m}_0(\mu)$ (a) and $\Lambda(\mu)$ (b) fitted from $\tilde R(z,z_0)$ for $z_0<z<z_{\rm max}$ and $z_0=0.24$ fm, with NLO and NNLO Wilson coefficients at various values of $\mu$.}
\end{figure*}

Our results for $\bar{m}_0$ and $\Lambda$ fitted from $\tilde R(z,z_0)$ for $z_0<z<z_{\rm max}$ with $z_0=0.24$ fm are shown in \fig{mfit}. As one can see, the two parameters remain constant in $z_{\rm max}$ up to around $0.5$ fm within a small window of $\mu$, which is different with the NLO and NNLO Wilson coefficients.
At larger $z$, the higher-twist and $\alpha_s\ln(z^2\mu^2)$ effects become significant, which can no longer be described by the simple \textit{ansatz} in \eq{ansatz}. In this work, we use $\tilde R(z,z_0)$ at $0.24$ fm $<z<0.4$ fm to fit the parameters at all $\mu$ as input for the hybrid scheme renormalization and matching. To estimate the uncertainty from the choice of $\mu$, we will match the qPDFs obtained at different $\mu$ to the corresponding PDFs, and then evolve the final results to $\mu=2.0$ GeV for comparison.

\section{Fourier transform (FT)}
\label{app:ft}

The qPDF is defined as the FT of $\tilde h(z,z_S,P^z)$ or $\tilde h(\lambda,\lambda_S,P^z)$,
\begin{align}
    \tilde f(x, z_S,P^z) &= \int{d\lambda\over 2\pi}\ e^{ix\lambda} \tilde h(\lambda,\lambda_S,P^z)\,.
\end{align}

Since $\tilde h(\lambda,\lambda_S,P^z)$ is perturbatively matched from the $\MS$ scheme, the factorization formula should still be valid for the corresponding qPDF $\tilde f(x, z_S,P^z)$~\cite{Ji:2020brr}.
Therefore, we should integrate over all $z$ in the FT to obtain the $x$-dependence of the qPDF. However, due to finite lattice size effects, worsening signal-to-noise ratio and other systematics at large $z$, we have to truncate $\tilde h(z,z_S,P^z)$ at $z=z_L$ and extrapolate to $z\to\infty$ to complete the FT. 
As a result, the small-$x$ ($x\lesssim 1/\lambda_L$) region is the most sensitive to the extrapolation model, and the corresponding systematic uncertainty cannot be well controlled. On the other hand, the reliability of the $x\gtrsim 1/\lambda_L$ region depends on the premises that the $\tilde h(z)$ is small at $z=z_L$ and exhibits an exponential decay when $z_L$ is large enough. The first condition is easy to understand as a truncated FT will lead to an unphysical oscillation in the $x$-space with amplitude proportional to $|\tilde h(z_L)|$, while the exponential decay guarantees that the FT converges fast and the qPDF at $x\gtrsim 1/\lambda_L$ has very little dependence on the specific model used in the extrapolation.

In this section, we first derive that the equal-time correlator in a hadron state does exhibit an exponential decay at large distances, then we demonstrate that including this constraint in the extrapolation will lead to a reliable FT in the moderate-to-large $x$ region. Finally, we perform the extrapolated FT on our lattice results.

\subsection{Matrix elements at large $z$}
\label{app:exp}

To begin with, let us consider a current-current correlation in the vacuum, $\langle \Omega| J_5(x) J_5(0)| \Omega\rangle$, where $J_5=\bar{q}\gamma_5 q$ and $x^2<0$. If we ignore the existence of zero modes and only consider gapped vacuum excitations, then 
\begin{align}
    &\langle \Omega| J_5(x) J_5(0)| \Omega\rangle\nn\\
    &= \sum_n \int {d^3k_n\over (2\pi)^3 2E_{k_n}} \langle \Omega| J_5(x)|n\rangle \langle n| J_5(0)| \Omega\rangle\nn\\
    &= \sum_n\int {d^3k_n\over (2\pi)^3 2E_{k}} \langle \Omega| J_5(0)|n\rangle \langle n| J_5(0)| \Omega\rangle e^{-ix\cdot k_n}\nn\\
    &= \sum_n |Z_n|^2\int {d^4k_n\over (2\pi)^4} {e^{-ix\cdot k_n}\over k^2_n -m_n^2 +i0}\nn\\
    &= -{i\over 4\pi^2}\sum_n |Z_n|^2 {m_n\over \sqrt{-x^2}}K_1(m_n\sqrt{-x^2})\,.
\end{align}
where $Z_n$ is the overlap between the operator $J_5(x)$ and intermediate sate $|n\rangle$. Here $m_n$ is the mass of the intermediate state particle, and $K_n$ is the modified Bessel function of the second kind. Then, since
\begin{align}
    \lim_{|x|\to\infty} {m_n\over \sqrt{-x^2}}K_1(m_n\sqrt{-x^2}) &= \sqrt{{\pi\over2}} {\sqrt{m_n}\over |x|^{3\over2}}e^{-m_n|x|}\, ,
\end{align}
The correlation function should, therefore, be dominated by the exponential decay of the lowest-lying state that overlaps with $J_5(x)$.

When the external state is a static hadron, it has also been shown that the spacelike correlations exhibit an exponential decay at large distance~\cite{Burkardt:1994pw}.

We are interested in equal-time quark bilinear correlators in a boosted hadron state, which can be expressed in terms of the product of two ``heavy-light'' currents~\cite{Ji:2017oey,Green:2017xeu}, where the ``heavy quark'' $h_{\hat{x}}$ is an auxiliary field defined along the $\hat{x}$ direction, similar to that in HQET. 

Let us choose the external state to be a pion. According to Lorentz covariance, we can decompose the correlation as
\begin{align}
    &\langle \pi(p)| \bar{q}(x)\gamma^\mu h_{\hat{x}}(x) \bar{h}_{\hat{x}}(0) q(0)| \pi(p) \rangle  \nn\\
    &= p^\mu f_p(p\cdot x, x^2)  + x^\mu f_x(p\cdot x, x^2)\,,
\end{align}
where the scalar functions $f_{p,x}(p\cdot x, x^2)$ are analytic functions of $p\cdot x$ and $x^2$. We can select the index $\mu$ such that $x^\mu=0$.
For example, we can choose $\mu=z$ when $x^\mu=(t,0,0,0)$ or $\mu=t$ when $x^\mu=(0,0,0,z)$. The HQET corresponds to the timelike case, as
\begin{align}
p^z f_p(p\cdot x, x^2)
    &= \sum_n\int {d^3k_n\over (2\pi)^3 2E_{k_n}}  e^{-ix\cdot (k_n-m_Qv-p)}\nn\\
    &\quad  \times \langle \pi(p)| \bar{q}\Gamma h_v|n\rangle  \langle n| \bar{h}_v q| \pi(p)\rangle \,,
\end{align}
where $h_v$ is the effective heavy-quark field moving with velocity $v^\mu$ and related to the QCD heavy quark $Q$ by the projection
\begin{align}
h_v(x)&=e^{im_Qv\cdot x} {1+\slashed v\over 2}Q(x)\,.
\end{align}
The lowest intermediate state $|H(v)\rangle$ is a heavy-light meson with mass $m_H=m_Q + \bar{\Lambda}$ and momentum $k^\mu = m_H v^\mu$, where $m_Q$ is the heavy quark pole mass, and $\bar{\Lambda}$ can be interpreted as the mass of the constituent light quark or binding energy. Both $\bar{\Lambda}$ and $m_Q$ have ${\cal O}(\Lambda_{\rm QCD})$ renormalon ambiguities which cancel between each other. In the $\Lambda_{\rm QCD}/m_Q\to0$ limit, $\bar{\Lambda}$ should be independent of the heavy quark mass, but can depend on the light quark mass.

The matrix element $\langle \pi(p)| \bar{q}\Gamma h_v|H(v)\rangle$ is given by the transition form factors~\cite{Falk:1990yz},
\begin{align}
    &\langle \pi(p)| \bar{q}\Gamma h_v|H(v)\rangle \nn\\
    &= -\Tr\left\{\gamma_5\left[f_1(v\cdot p) + f_2(v\cdot p) {\slashed p\over v\cdot p}\right] \Gamma {\cal M}(v)\right\}\,,
\end{align}
where the form factors $f_1$ and $f_2$ only depend on $v\cdot p$ in HQET, and the projetion operator ${\cal M}(v)$ depends on the spin of the heavy-light meson $H(v)$,
\begin{align}
    {\cal M}(v) &= {1+\slashed v\over 2}
    \Bigg\{\begin{array}{cc}
        - \gamma_5\,, & {\rm for}\ J^P=0^-\,, \\
        \slashed \eps\,, & {\rm for}\ J^P=1^-\,,
    \end{array}
\end{align}
with $\eps^\mu$ being the polarization vector for vector mesons.
Therefore,
\begin{align}
    \langle \pi(p)| \bar{q}\gamma^\mu h_v|H(v)\rangle &= 2f_1(v\cdot p)v^\mu + 2f_2(v\cdot p)p^\mu\,,\\
    \langle \pi(p)| \bar{q} h_v|H(v)\rangle &= 2f_1(v\cdot p)+ 2f_2(v\cdot p)\,.
\end{align}

Then, the correlation function becomes
\begin{align}\label{eq:transit}
    & p^z f_p(p\cdot x, x^2)
    \approx 4m_Q^2\sum_n\int {d^3\vec{v}\over (2\pi)^3 2\sqrt{1+\vec{v}^2}}\nn\\
    &\qquad \times e^{-i\left(\bar{\Lambda}\sqrt{1+\vec{v}^2}-p^0\right)x^0 } (f_1+f_2)(f_1 v^z + f_2 p^z)\,.
\end{align}

Note that when $x^0\to\infty$, $x^0 \bar{\Lambda}\sqrt{1+\vec{v}^2}\ge x^0\bar{\Lambda}$ constitutes a large phase, so the integrand is quickly oscillating and should be suppressed. To have a naive estimate, let us assume $f_1$ and $f_2$ are constant in $v\cdot p$, and the remaining integral is simply
\begin{align}
    &\int {d^3\vec{v}\over (2\pi)^3 2\sqrt{1+\vec{v}^2}} e^{-i\left(\bar{\Lambda}\sqrt{1+\vec{v}^2}-p^0\right)x^0 } \nn\\
    &\qquad = {1\over 4\pi^2} K_1
    \left(\bar{\Lambda}\sqrt{-x_0^2}\right){e^{ip^0x^0}\over \bar{\Lambda} \sqrt{-x_0^2}}\nn\\
    &\qquad ={1\over 4\pi^2} K_1
    \left(\bar{\Lambda}\sqrt{-x^2}\right){e^{ip\cdot x}\over \bar{\Lambda} \sqrt{-x^2}}\,,
\end{align}
where we first obtained the result for imaginary $x^0$ and then analytically continued back to the real axis.

Then, using Lorentz invariance and analyticity, we can obtain the result for $x^2<0$, which corresponds to the equal-time correlator that we calculate in this work. At large separation, we have
\begin{align}\label{eq:largez}
    \lim_{|x|\to\infty}f_p(p\cdot x, x^2) &\propto  m_Q^2 {e^{-\bar{\Lambda}|x|}\over (\bar{\Lambda}|x|)^{3\over2}} e^{ip\cdot x}\,,
\end{align}
which also exhibits an exponential behavior with decay constant $\bar{\Lambda}$. Moreover, the correlation also includes a phase $e^{ip\cdot x}$ which becomes $\cos(p\cdot x)$ in the case of the valence quark distribution. Another important takeaway is that $\bar{\Lambda}$ is a Lorentz-invariant quantity and should be independent of the external momentum.

However, it must be pointed out that the conclusion in \eq{largez} is based on a rather crude approximation that $f_1$ and $f_2$ are constant in $v\cdot p$. In practice, the transition form factors could have a pole at the mass of a heavy-light meson created by the current $\bar{q}\gamma^\mu h_v$ or $\bar{h}_v q$, which is different from $m_H$ for the intermediate state $|H(v)\rangle$. As a result, the binding energy $\bar{\Lambda}$ would also be different. If we take this into account in \eq{transit}, then the result will exhibit a more complicated asymptotic behavior at large distance,
\begin{align}\label{eq:largez2}
    \lim_{|x|\to\infty}f_p(p\cdot x, x^2) &\propto  {e^{-\bar{\Lambda}|x|}\over |x|^{d}}\ g[p\cdot x, \cos(p\cdot x), \sin(p\cdot x)]\,,
\end{align}
where the decay constant $\bar{\Lambda}$ should vary among the different binding energies for the heavy-light mesons, which is similar to the observation in Ref.~\cite{Burkardt:1994pw}, and $g$ is a function that can have both oscillating and non-oscillating dependence on $p\cdot x$. For large enough $|x|$, the exponential decay should suppress the correlation and make it or its extremes decrease monotonically in magnitude.\\

Note that after we match the hybrid scheme matrix elements to $\MS$, the renormalon ambiguity in the Wilson line mass, $ m_0^{\MS}$, is subtracted out, so the matched result should exhibit an asymptotic behavior that goes as $e^{-(\bar{\Lambda}-m_0^{\MS})|z|}$ at large $z$. Therefore, the sign of $(\bar{\Lambda}-m_0^{\MS})$ becomes crucial in determining whether it is exponentially decaying or growing.

In QCD sum rule calculations, the result is $\bar{\Lambda} = 0.4-0.6$ GeV from phenomenology, while $m^\MS_0$ is expected to be $0.1-0.2$ GeV~\cite{Beneke:1994sw}, so $\bar{\Lambda}-m^\MS_0=0.2-0.5$ GeV. Since the quarks have heavier-than-physical masses in our lattice calculation, one should expect a larger $\bar{\Lambda}$, so it is very likely that $\bar{\Lambda}-m^\MS_0$ still remains positive. After all, this can be always put to test on the $P^z=0$ matrix elements since $\bar{\Lambda}-m^\MS_0$ is a Lorentz-invariant quantity.

\subsection{Extrapolation and FT}
\label{app:ext}

If $z_L$ is large enough for the correlation $\tilde h(z)$ to reach the asymptotic region, then an extrapolation that encodes the exponential decay behavior we derived in \app{exp} should lead to reliable FT for moderate-to-large $x$. To be more precise, there is a rigorous upper bound for the uncertainty of FT which decreases with $x$.

To prove the above statement, let us consider extrapolation based on the general model
\begin{align}
  \tilde h(\lambda) &= e^{-c|\lambda-\lambda_L|}g(\lambda)\,,
\end{align}
where $g(\lambda_L)= \tilde h(\lambda_L)$, and $c=m_{\rm eff}/P^z$ with $m_{\rm eff}$ being the effective mass for the exponential decay.
Motivated by QCD sum rule results, we expect $m_{\rm eff}\sim 0.2-0.5$ GeV, which can be larger since we have used heavier-than-physical quark masses. Therefore, for $P^z\sim 2.0$ GeV in the current work, we should have $c\sim 0.10-0.25$ or higher.

Now let us compare two extrapolations $h_1$ and $h_2$ with different $g_1$ and $g_2$.
The difference between the two extrapolations,
\begin{align}
    \delta \tilde h(\lambda) &\equiv \tilde h_1(\lambda) - \tilde h_2(\lambda)\,,
\end{align}
should satisfy $\delta \tilde h(\lambda_L)=0$ and $\delta \tilde h(\infty)=0$. 
The difference in the FT with extrapolation is therefore
\begin{align}
    \delta \tilde f(x) &= \int_{\lambda_L}^\infty {d\lambda\over \pi} \delta \tilde h(\lambda) \cos(x\lambda)\,.
\end{align}

If we can approximate $\delta \tilde h(\lambda)$ as a flat curve within one period of the oscillatory function $\cos(x\lambda)$, then the integral in that region vanishes. This condition can be satisfied if $|\delta \tilde h'(\lambda)| \ll x$, which should be reached very quickly due to the exponential suppression at large $\lambda$. For each $x$, there should be a minimal integer $N_x$ which satisfies $|\delta \tilde h'(\lambda_L+N_x 2\pi/x)| \ll x$, so that we can approximate $\delta \tilde f(x)$ as
\begin{align}\label{eq:df}
    \delta \tilde f(x) &\approx \int_{\lambda_L}^{\lambda_L+N_x {2\pi\over x}} {d\lambda\over \pi} \delta \tilde h(\lambda) \cos(x\lambda)\,.
\end{align}

Since $\delta \tilde h(\lambda_L)=0$ and $\delta \tilde h(\infty)=0$, there must be at least one extremum of $\delta \tilde h(\lambda)$ for $\lambda_L<\lambda <\infty$,
so we have the inequality
\begin{align}\label{eq:bound}
    |\delta \tilde f(x)| & < \int_{\lambda_L}^{\lambda_L+N_x {2\pi\over x}}{d\lambda\over \pi}\ |\delta \tilde h(\lambda)| |\cos(x\lambda)| \nn\\
    & < N_x|\delta \tilde h(\lambda)|_{\rm max} \int_{\lambda_L}^{\lambda_L+ {2\pi\over x}} {d\lambda\over \pi} |\cos(x\lambda)|\nn\\
    &= {4N_x|\delta \tilde h(\lambda)|_{\rm max}\over \pi x} \lesssim {4N_x|\tilde h(\lambda_L)|\over \pi x} \,.
\end{align}

According to our estimate of $\bar{\Lambda} - m_0^{\MS}$, $c\gtrsim 0.1$ at $P^z\sim 2$ GeV, so 
\begin{align}
    e^{-cN_x (2\pi)/x} \lesssim  e^{-0.6 N_x/x} \,,
\end{align}
and $N_x\sim {\cal O}(1)$ should be sufficient to satisfy $|\delta \tilde h'(\lambda_L+N_x 2\pi/x)| \ll x$ with $0<x<1$.
Therefore, in \eq{bound} we demonstrate that there is an upper bound for the model uncertainty in the FT with exponential extrapolation, which decreases in $x$. The error is also proportional to $|\delta \tilde h(\lambda)|_{\rm max}$ which can be much smaller than $|\tilde h(\lambda_L)|$ that is already close to zero. 
If $h(\lambda_L)=0.1$, $|\delta \tilde h(\lambda)|_{\rm max}=0.05$, and $N_x=1$, then we have
\begin{align}
    |\delta \tilde f(x)| < {0.07\over x} \,,
\end{align}
which is less than 0.15 at $x=0.5$ and around 15\% of the central value of the qPDF as we obtain below.
It is worth pointing out that our estimate of the upper bound in \eq{bound} can be highly overestimated, as $\delta \tilde h(\lambda)$ has an oscillation from $\cos(\lambda)$ and $\sin(\lambda)$ which are out of pace with $\cos(x\lambda)$ for $0<x<1$, and $|\delta \tilde h(\lambda)|_{\rm max}$ could be much smaller than $|\tilde h(\lambda_L)|$ and at a sharp peak within $\lambda_L < \lambda < \lambda_L + N_x 2\pi/x$. 

Therefore, the FT with exponential extrapolation is under control for moderate and large $x$. When $\tilde h(\lambda_L)$ is small enough, the model uncertainty from the extrapolation can be controlled to be much smaller than the other systematic uncertainties which are about $10\%-20\%$ in this work.

It is worth to compare with the extrapolation error when the correlation function decreases algebraically as $1/|\lambda|^{d}$, which corresponds to the generic model
\begin{align}
    \tilde h(\lambda) &= \left({\lambda_L\over \lambda}\right)^d g(\lambda)\,.
\end{align}
Suppose we truncate at $\lambda_L=10$, then
\begin{align} \label{eq:algebraic}
    \left({\lambda_L\over \lambda_L + N_x 2\pi/x}\right)^d & \sim (1+0.6N_x/x)^{-d}\,.
\end{align}
The power $d$ is related to the small-$x$ behavior of the PDF. If we parameterize the PDF as $\sim x^a(1-x)^b$, then with LO matching one can derive that $d={\rm min}\{1+a, 1+b\}$~\cite{Ji:2020brr}, which is ${\cal O}(1)$ empirically. Therefore, it will take $N_x\gg 1$ for the factor in \eq{algebraic} to decrease sufficiently to satisfy the condition $|\delta \tilde h'(\lambda_L+N_x2\pi/x)| \ll x$. As a result, the uncertainty in the FT is of orders of magnitude larger than that of extrapolation with exponential decay.\\

To test our claim of controlled FT error with exponential decay, we choose a particular model
\begin{align}\label{eq:extest}
    \tilde h(\lambda) &= \tilde h(\lambda_L) \left({\lambda_L\over \lambda}\right)^de^{- c|\lambda - \lambda_L|}\,.
\end{align}
Suppose that the extrapolation is done at $\lambda_L=10$ with $\tilde h(\lambda_L)=0.15$, and the parameters $c$ and $d$ are fitted with errors $\delta c$ and $\delta d$, then we analytically FT the extrapolated result to the $x$-space, and calculate its error using
\begin{align}
    \delta \tilde f(x,c,d) &= \sqrt{\left({\partial \tilde f \over \partial c}\right)^2\delta c^2 + \left({\partial \tilde f \over \partial d}\right)^2\delta d^2}\,.
\end{align}

\begin{figure}[htb]
    \centering
    \includegraphics[width=0.8\columnwidth]{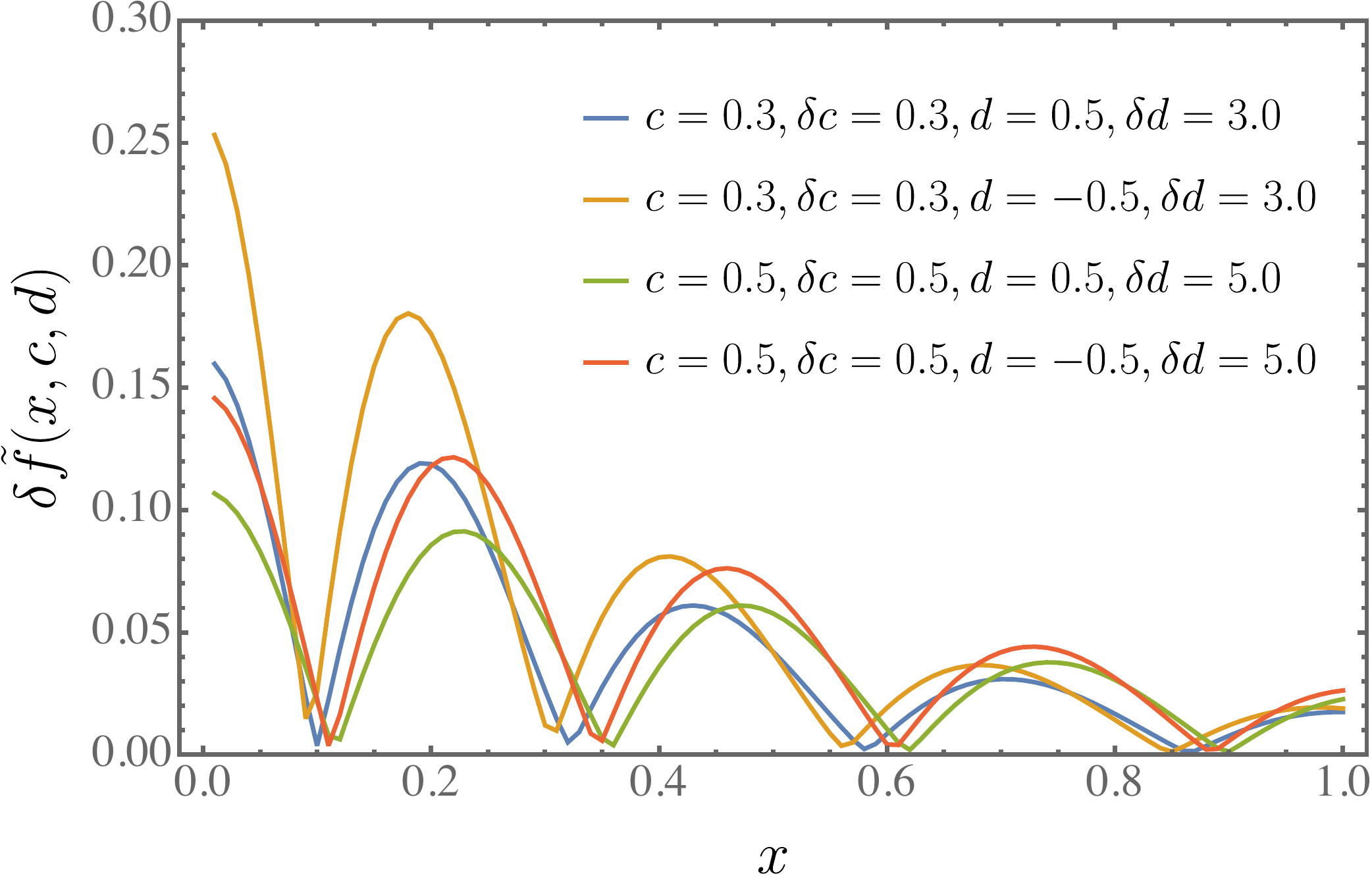}
    \caption{Estimate of error in the FT with extroplation using the model in \eq{extest}.}
    \label{fig:error}
\end{figure}

In \fig{error}, we plot the extrapolation error against $x$. We have chosen different central values of the parameters $c$ and $d$ and fairly large uncertainties in them. The parameter $d$ cannot have a large negative value, otherwise it would make $\tilde h(\lambda)$ grow beyond $\lambda_L$. 
In most of the scenarios considered, the error is $\lesssim 0.1$ for $x>0.1$. As we shall see below, the actual extrapolation error is much smaller than this estimate and thus negligible when compared to the other systematic errors.\\

In the following, we perform the extrapolation with four different models. The extrapolation is carried out on each bootstrap sample by a minimal-square fit. For each $P^z$, we truncate $\tilde h(z)$ at the largest $z$, $z_{>0}$, where the central value of $\tilde h(z)$ remains positive, and choose $z_{\rm max}=\{z_{>0}-2a, z_{>0}-a, z_{>0}\}$ to estimate the truncation error. The range of $z$ used to fit the parameters is $z_{\rm min} \le z \le z_{\rm max}$ where $z_{\rm min}$ satisfies $\tilde h(z_{\rm min})< 0.2$. The continuty condition between data and model was imposed in the middle point of the fit range, namely $z_L$, which is listed in Table~\ref{tab:zL}. The extrapolation models are:

\begin{table}
\centering
\begin{tabular} {|>{\centering\arraybackslash} p{1.0cm}|>{\centering\arraybackslash} p{3.0cm}|>{\centering\arraybackslash} p{3.0cm}|}
\hline
     &  \multicolumn{2}{c|}{$z_L/a$}\\
\hline
$n_z$    & $a=0.04$ fm & $a=0.06$ fm \\
\hline
$1$ & $\{29, 30, 31\}$ & N/A \\
\hline
$2$ & $\{26, 27, 28\}$ & $\{19, 20, 21\}$\\
\hline
$3$ & $\{19, 20, 21\}$ & $\{16, 17, 18\}$\\
\hline
$4$ & $\{24, 25, 26\}$  & $\{14, 15, 16\}$\\
\hline 
$5$ & $\{21, 22, 23\}$ & $\{15, 16, 17\}$ \\
\hline
\end{tabular}
\caption{Choices of $z_L$ for the extrapolations.}
\label{tab:zL}
\end{table}

\paragraph*{Exponential decay model,}  or ``model-exp''. 
The model for extrapolation is
\begin{align}\label{eq:exponential}
    A {e^{-m_{\rm eff}|z|}\over |\lambda|^d}\,.
\end{align}
We have tried to fit $m_{\rm eff}$ from the same range of $z$ for $P^z=0$ matrix elements with a similar form, $A e^{-m_{\rm eff}|z|}/|z|^d$, and found that $m_{\rm eff}$ is around $0.1$ GeV, about the same scale as the phenomenological estimate. For the $P^z\neq0$ matrix elements, we do not fix $m_{\rm eff}$, but constrain it with a prior $m_{\rm eff} \ge m_{\rm min}$. To test the dependence on this prior condition, we have set $m_{\rm min}=\{0, 0.1, 0.2\}$ GeV. Besides, we also impose $A>0$ and $d>0$ to ensure that the extrapolated result is positive and decreases in $\lambda$.

\paragraph*{Power-law decay model,} or ``model-pow''. The model is defined by setting $m_{\rm eff}=0$ in model-exp. As the $P^z\to\infty$ limit of model-exp, model-pow can be used to give a coarse estimate of the significance of higher-twist effects, although its FT error is not well under control as we discussed above. We impose the conditions $A, d >0$ so that the fitted results decrease to zero as $\lambda\to\infty$.

\paragraph*{Two-parameter model with exponential decay,} or ``model-2p-exp''. As we can see from \fig{hybme}, the matrix elements at $\lambda_L\sim 6-10$ do not show a clear exponential decay, although they can be fitted by the latter with $\chi^2/d.o.f < 1$ due to the large errors. This may indicate that there is oscillation in $\tilde h(\lambda)$.
To incorporate such dependence, we ignore the higher-twist contributions and assume that the qPDF is parameterized as
\begin{align}
    f_v(x;a,b)&={\Gamma(2+a+b)\over \Gamma(1+a)\Gamma(1+b)} |x|^a(1-|x|)^b\nn\\
    &\qquad \times \theta(|x|)\theta(1-|x|)\,.
\end{align} 
By doing an inverse FT into the $\lambda$-space, the asymptotic form of $\tilde h_{\rm 2p}(\lambda)$ at large $\lambda$ reads,
\begin{align}\label{eq:asym}
\tilde h_{\rm 2p}(\lambda) & = A\ {\rm Re}\left[\frac{\Gamma(1+a)}{(-i|\lambda|)^{a+1}}+e^{i\lambda}\frac{\Gamma(1+b)}{(i|\lambda|)^{b+1}}\right]\,.
\end{align}
Then we multiply $\tilde h_{\rm 2p}(\lambda)$ with an exponential decay factor as our model for extrapolation,
\begin{align}\label{eq:2pexp}
    \tilde h_{\text{2p-exp}} &=\tilde h_{\rm 2p}(\lambda) e^{-m_{\rm eff}(z-z_L)}\,.
\end{align}

\paragraph*{Two-parameter model,} or ``model-2p''. Again, we ignore the exponential decay and use $\tilde h_{\rm 2p}$ as the extrapolation model, which can help us estimate the significance of higher-twist effects.\\

\begin{figure*}[htb]
    \centering
    \includegraphics[width=0.33\textwidth]{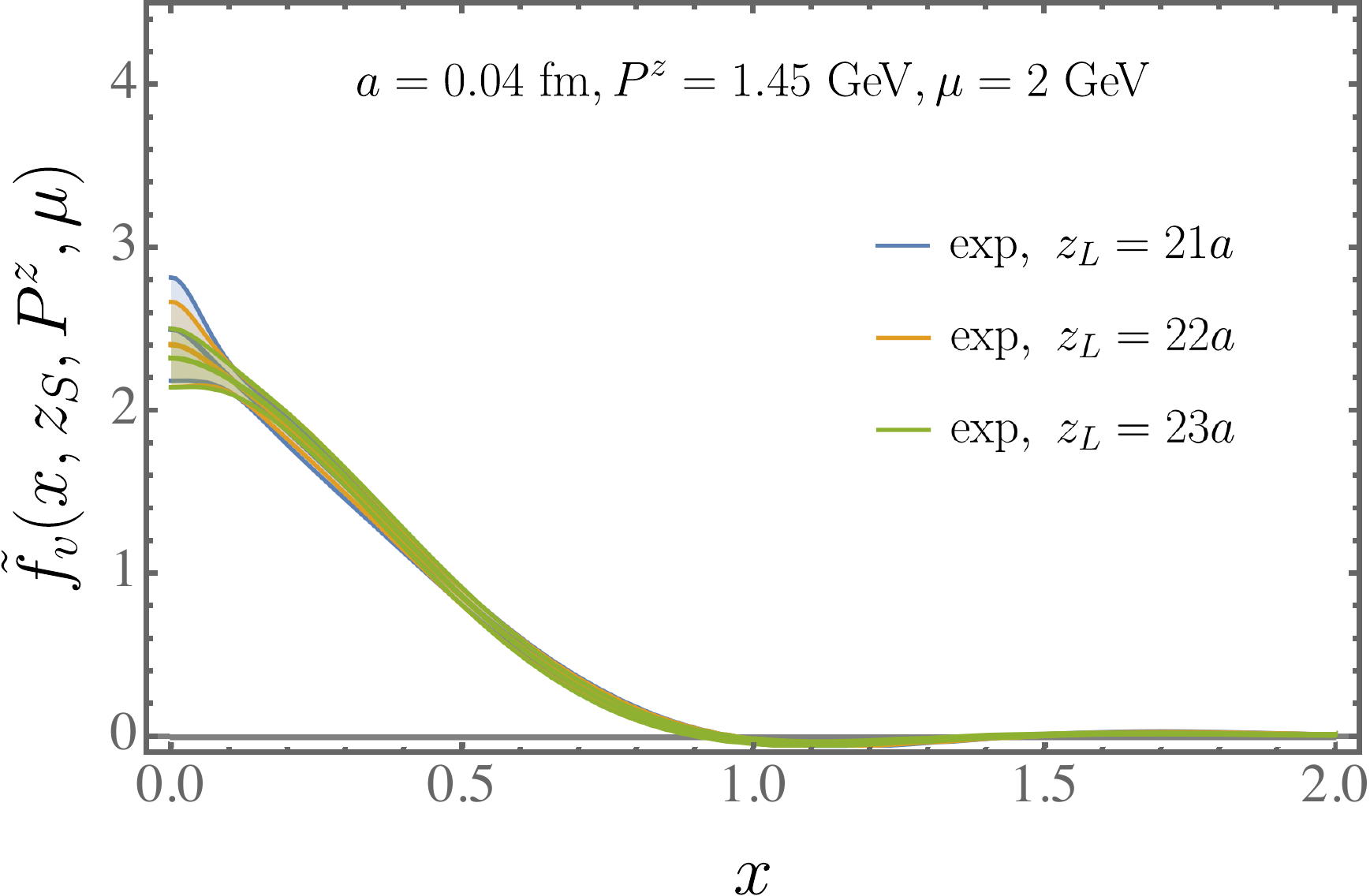}
    \includegraphics[width=0.32\textwidth]{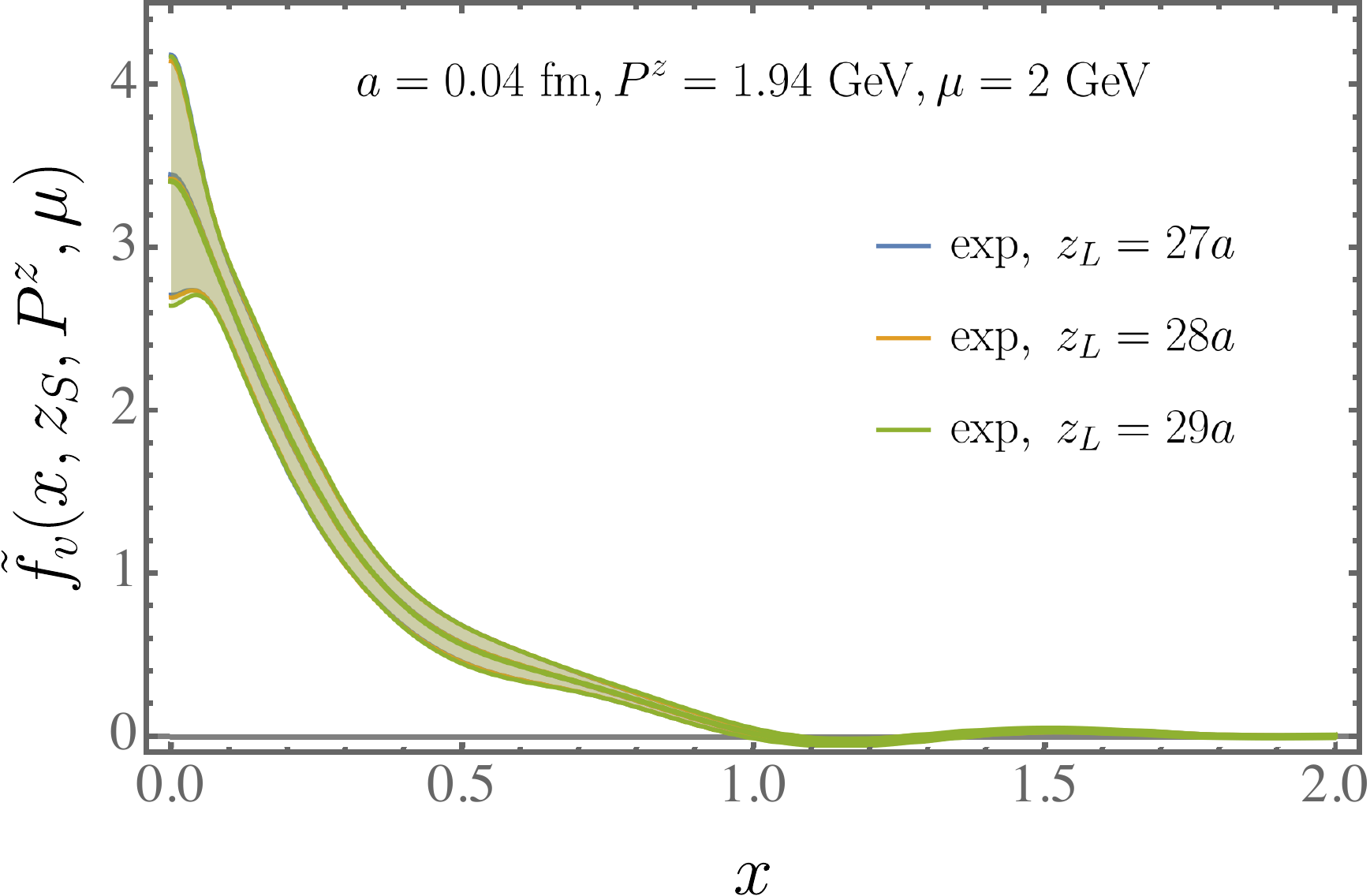}
    \includegraphics[width=0.32\textwidth]{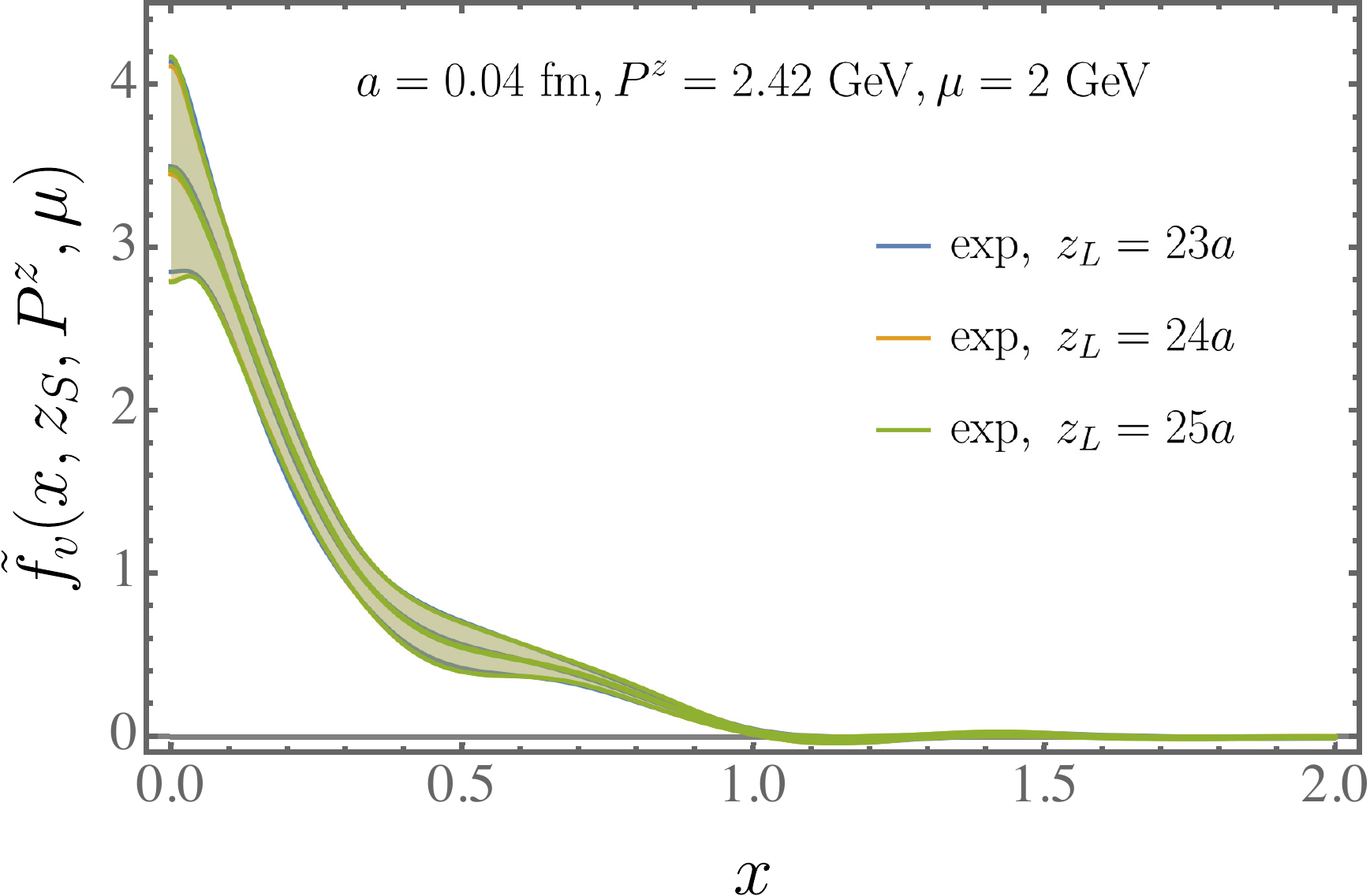}
    \caption{FT with different $z_L$ for model-exp extrapolation (with prior $m_{\rm eff}>0.1$ GeV) of the NNLO-matched  $\tilde h(\lambda, \lambda_S, P^z,\mu,a)$ at $z_S=0.24$ fm.}
    \label{fig:zLs}
\end{figure*}

In \fig{zLs} we compare the FT with different $z_L$ for extrapolation with model-exp and condition $m_{\rm eff}>0.1$ GeV. Except for very small $x$, the results are consistent, and those at smaller $z_L$ have smaller errors because the error of the matrix element grows with $z$.
Therefore, for the rest of our analysis, we simply use the largest $z_L$ for each $P^z$.

\begin{figure*}[htb]
    \centering
    \includegraphics[width=0.33\textwidth]{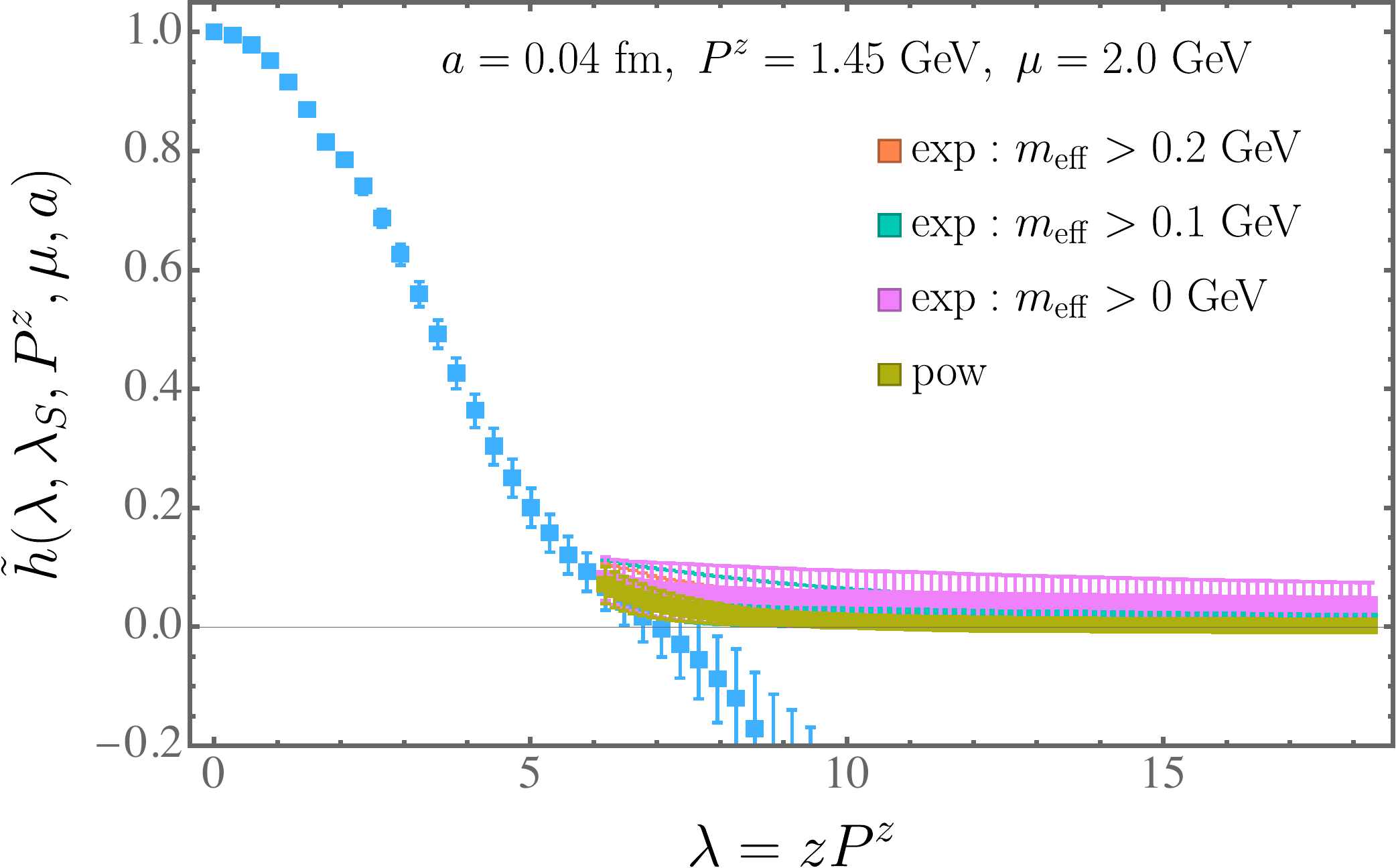}
    \includegraphics[width=0.32\textwidth]{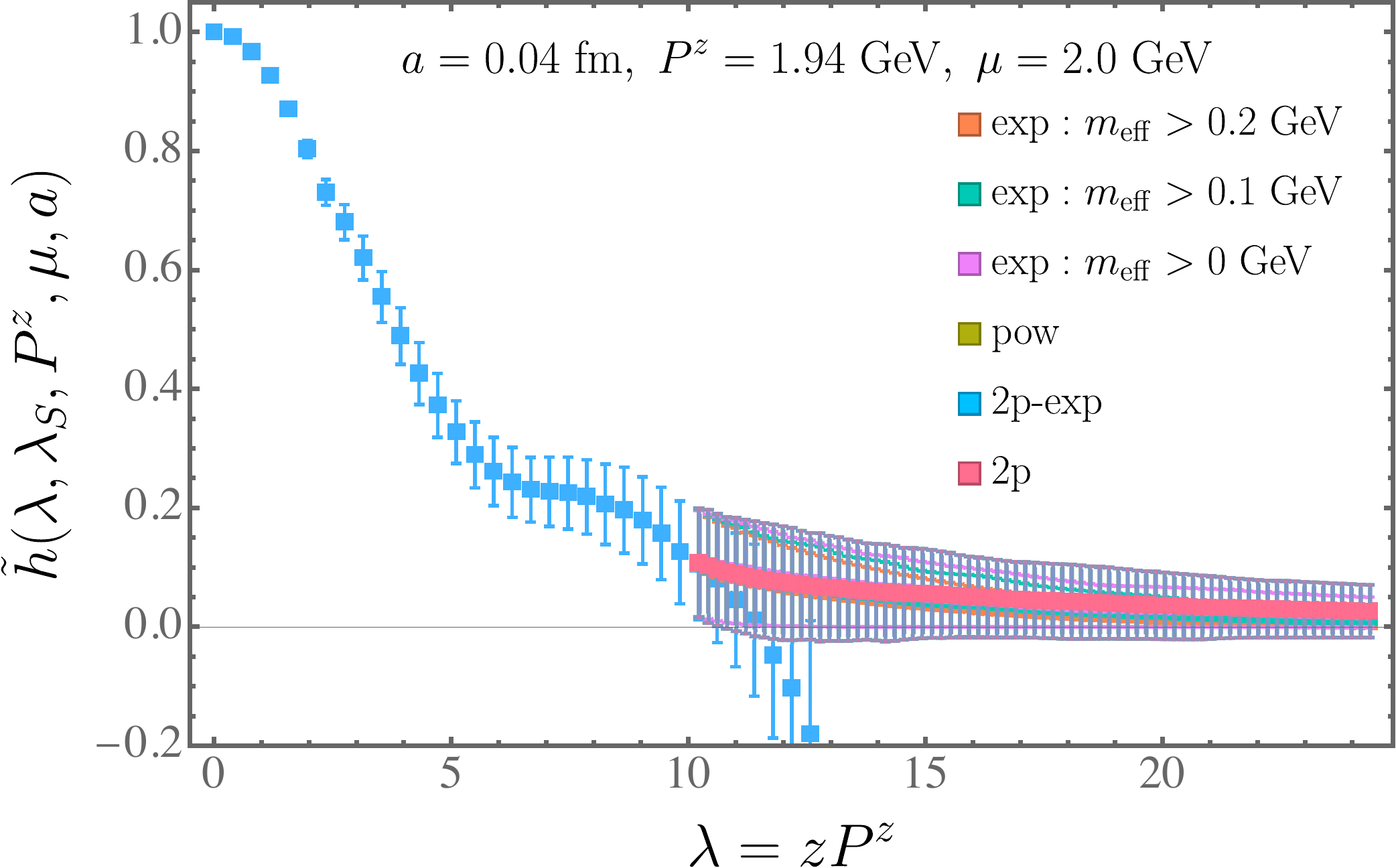}
    \includegraphics[width=0.32\textwidth]{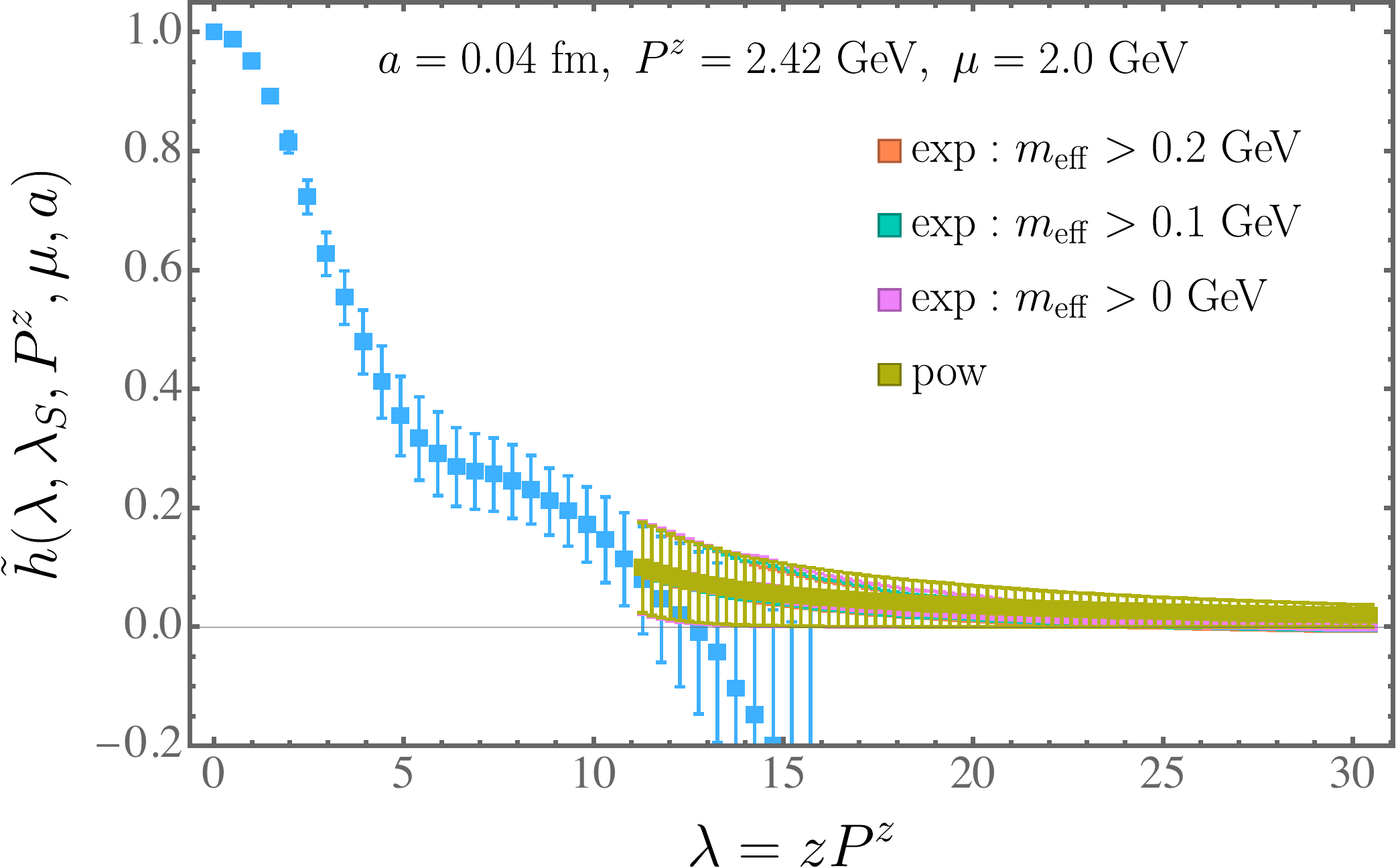}
    \caption{Extrapolation with different models for the NNLO-matched $\tilde h(\lambda, \lambda_S, P^z,\mu,a)$. At $P^z=1.94$ GeV, we have added the comparison with the 2p-exp and 2p models.}
    \label{fig:extrapolation}
\end{figure*}

\begin{figure*}[htb]
    \centering
    \includegraphics[width=0.33\textwidth]{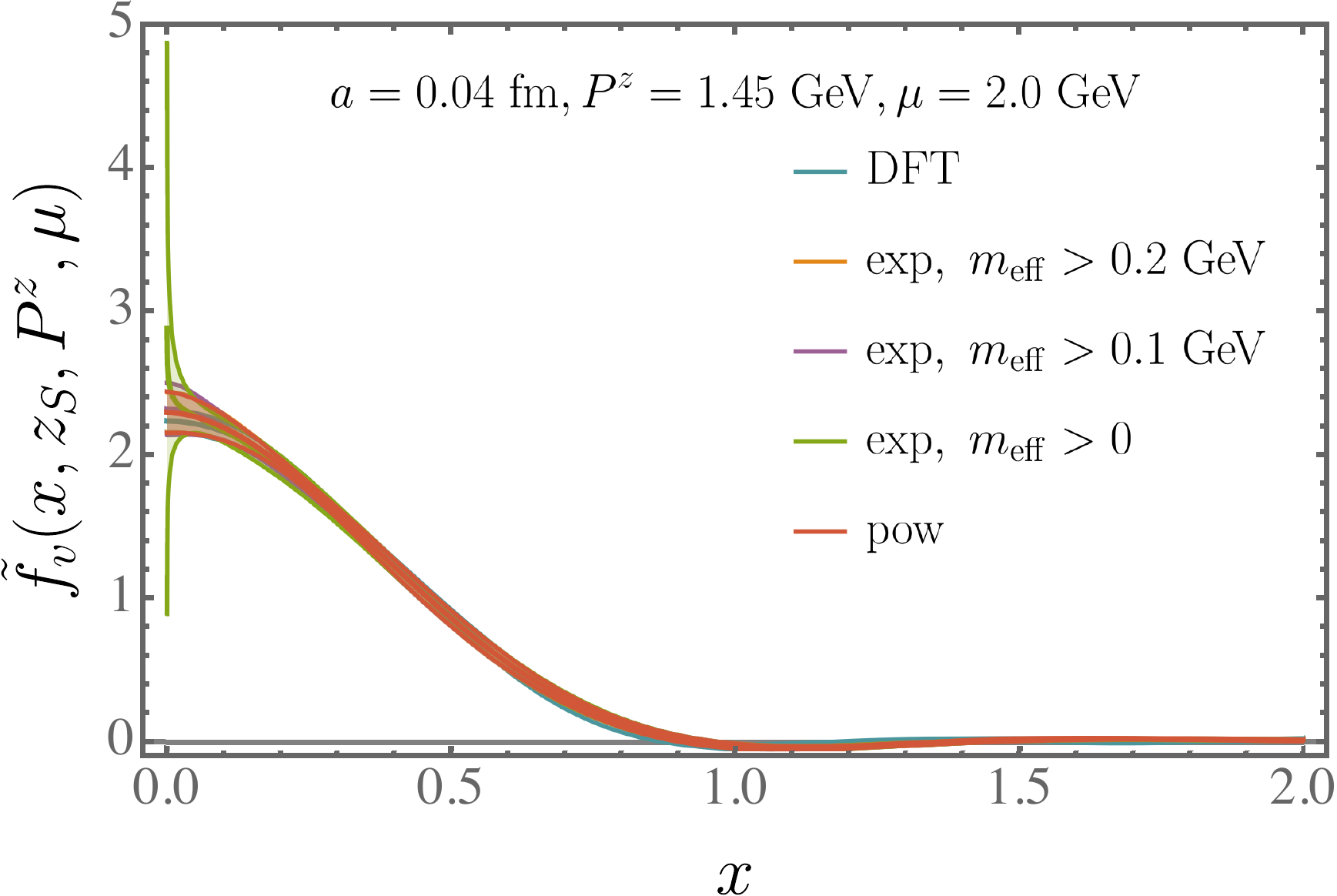}
    \includegraphics[width=0.32\textwidth]{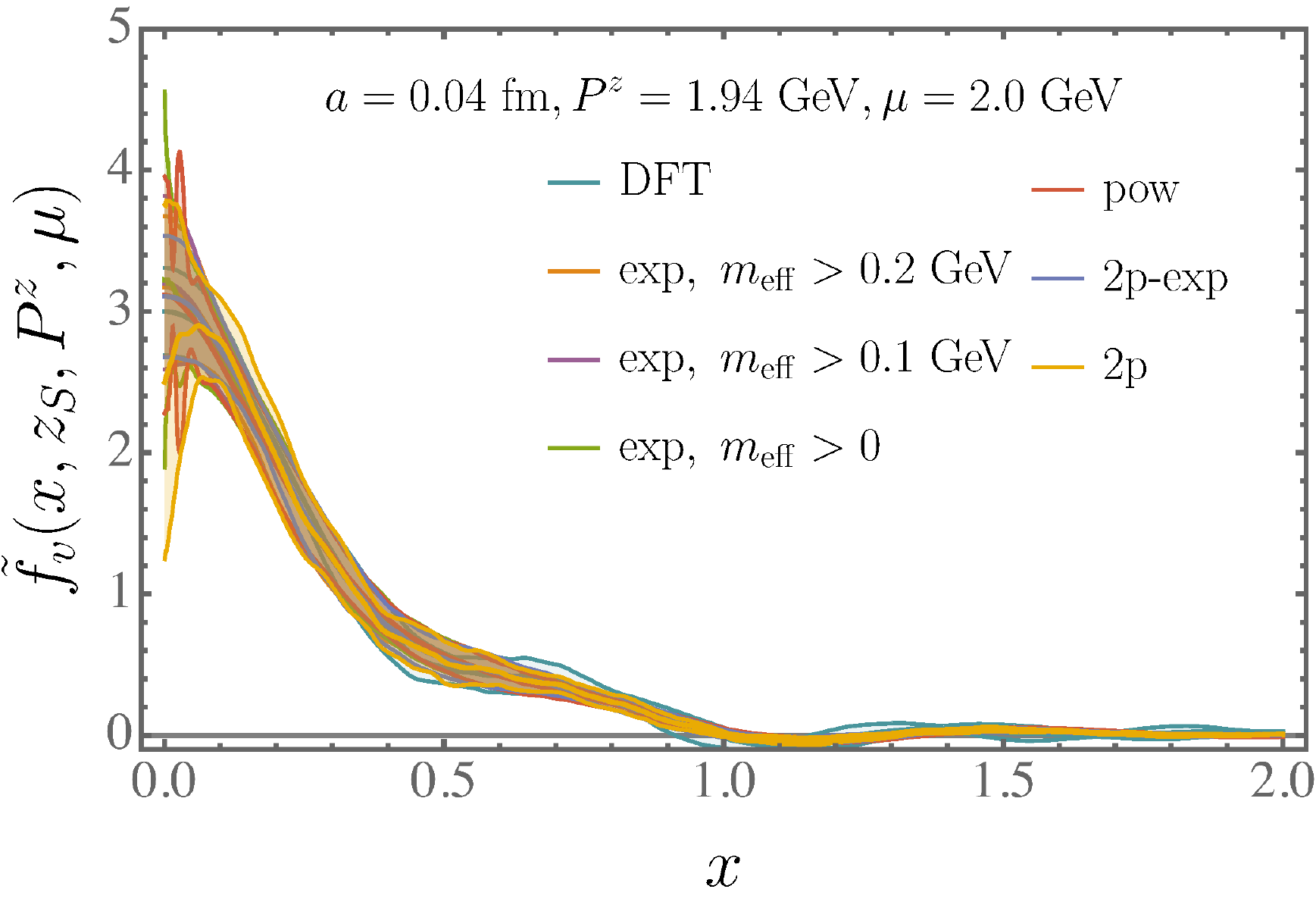}
    \includegraphics[width=0.32\textwidth]{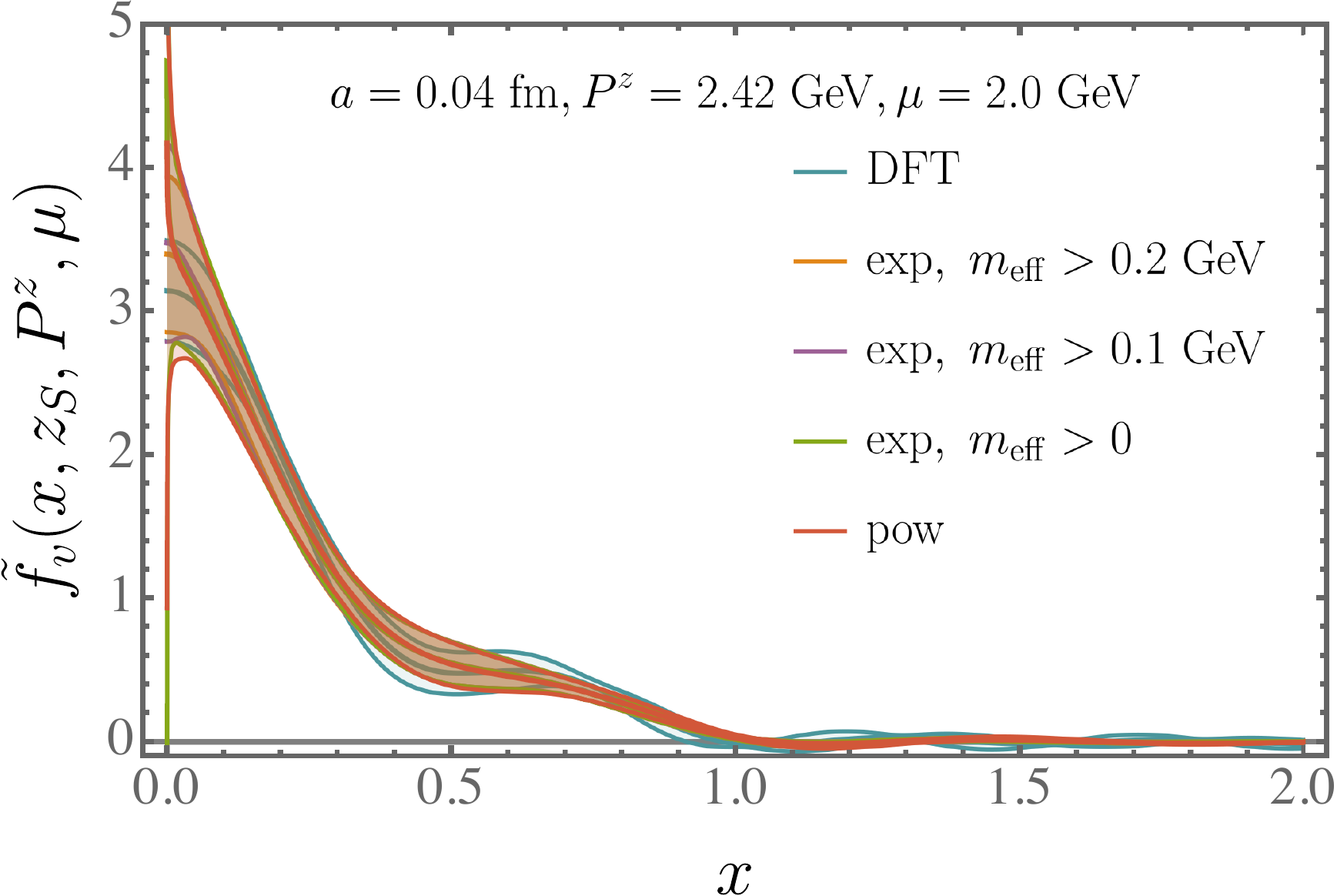}
    \caption{Comparison of DFT and FT with different extrapolation models for the NNLO-matched $\tilde h(\lambda, \lambda_S, P^z,\mu,a)$ at $z_S=0.24$ fm. At $P^z=1.94$ GeV, we have added the comparison with the 2p-exp and 2p models.}
    \label{fig:models}
\end{figure*}

In \fig{extrapolation} we show the extrapolations with different models, which have noticeable differences at $\lambda > \lambda_L$.
In \fig{models} we compare the FT with different extrapolation models as well as with the discrete FT (DFT). As we can see, the DFT introduces unphysical oscillation in the qPDF which is due to the truncation of $\tilde h(\lambda)$ at $\lambda_L$.
In contrast, the extrapolations are free of such oscillation, and different models yield consistent qPDFs at moderate and large $x$, thought they differ significantly at small $x$. We notice that the qPDF from model-2p extrapolation still has slight oscillations despite its agreement with the others, because the extrapolated result decays too slowly at $\lambda > \lambda_L$.
As expected, the models with exponential decay lead to regular qPDFs at $x=0$, whereas model-pow and model-2p give divergent qPDFs as $x\to0$.

Based on the above results, we use model-exp with $m_{\rm eff}>0.1$ GeV for the FT in our following analysis. To have a coarse estimate of the uncertainties from extrapolation model and higher-twist contributions, we look into the difference between final PDFs matched from qPDFs with model-exp and model-pow extrapolations.

\begin{figure}[htb]
	\centering
	\includegraphics[width=0.8\columnwidth]{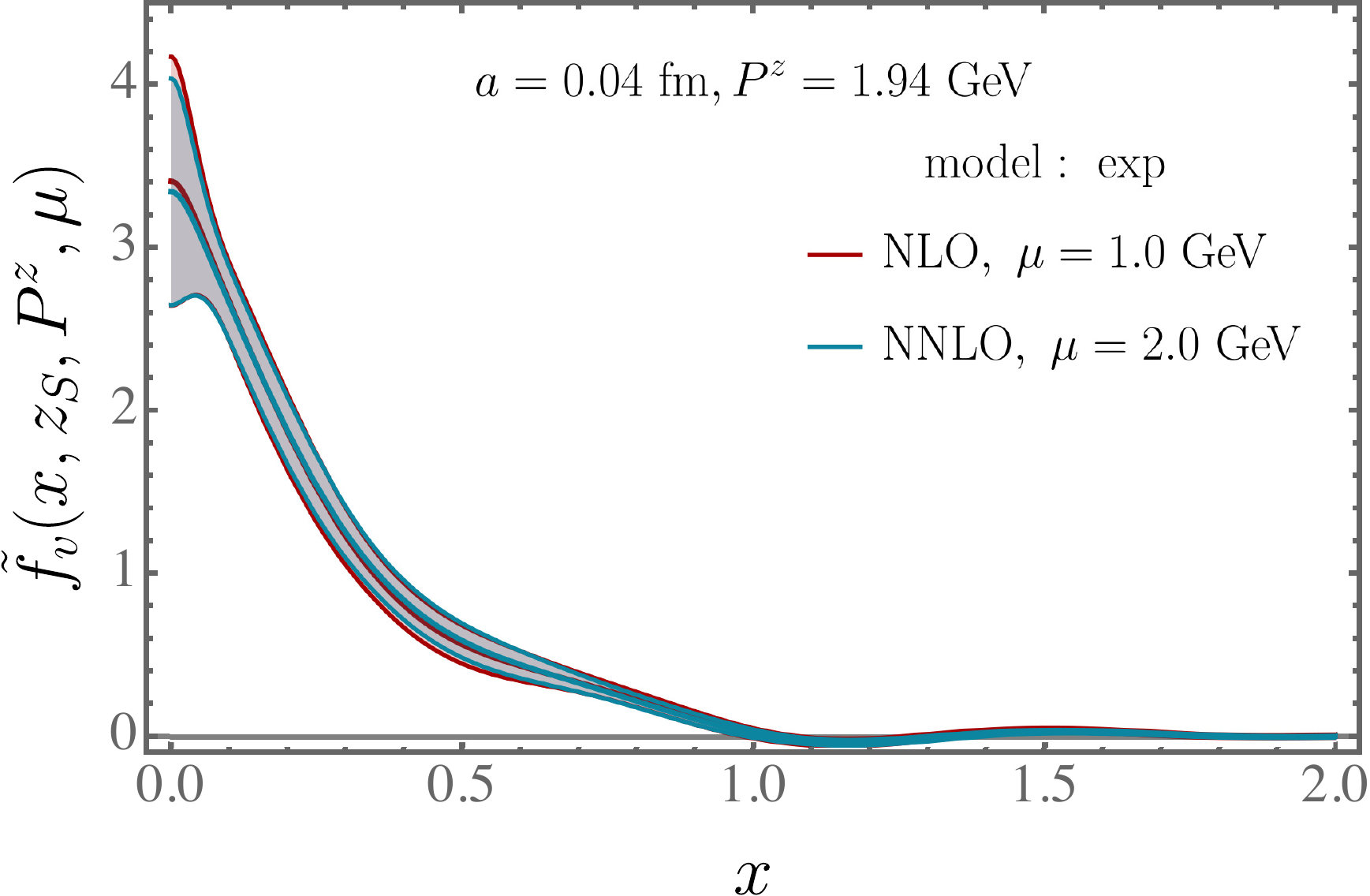}
	\caption{Comparison of the qPDF with model-exp extrapolation (with $m_{\rm eff}>0.1$ GeV) of the NLO- and NNLO-matched $\tilde h(\lambda, \lambda_S, P^z,\mu,a)$ at $z_S=0.24$ fm and $z_L=26a$. The choices of $\mu$ are based on where the renormalon model best describes the matrix elements.}
	\label{fig:nlovsnnlo}
\end{figure}

\begin{figure}[htb]
	\centering
	\includegraphics[width=0.8\columnwidth]{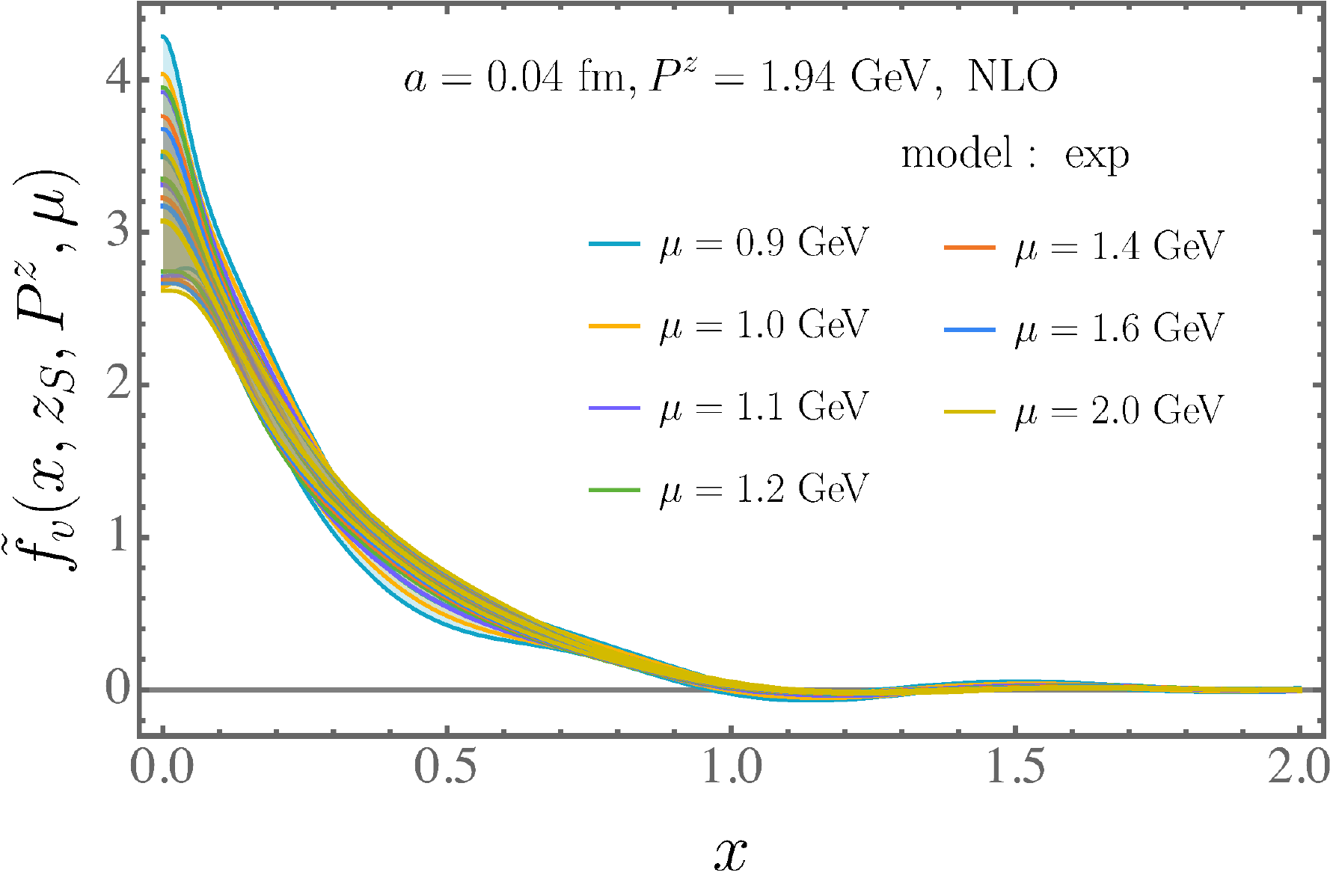}
	\includegraphics[width=0.8\columnwidth]{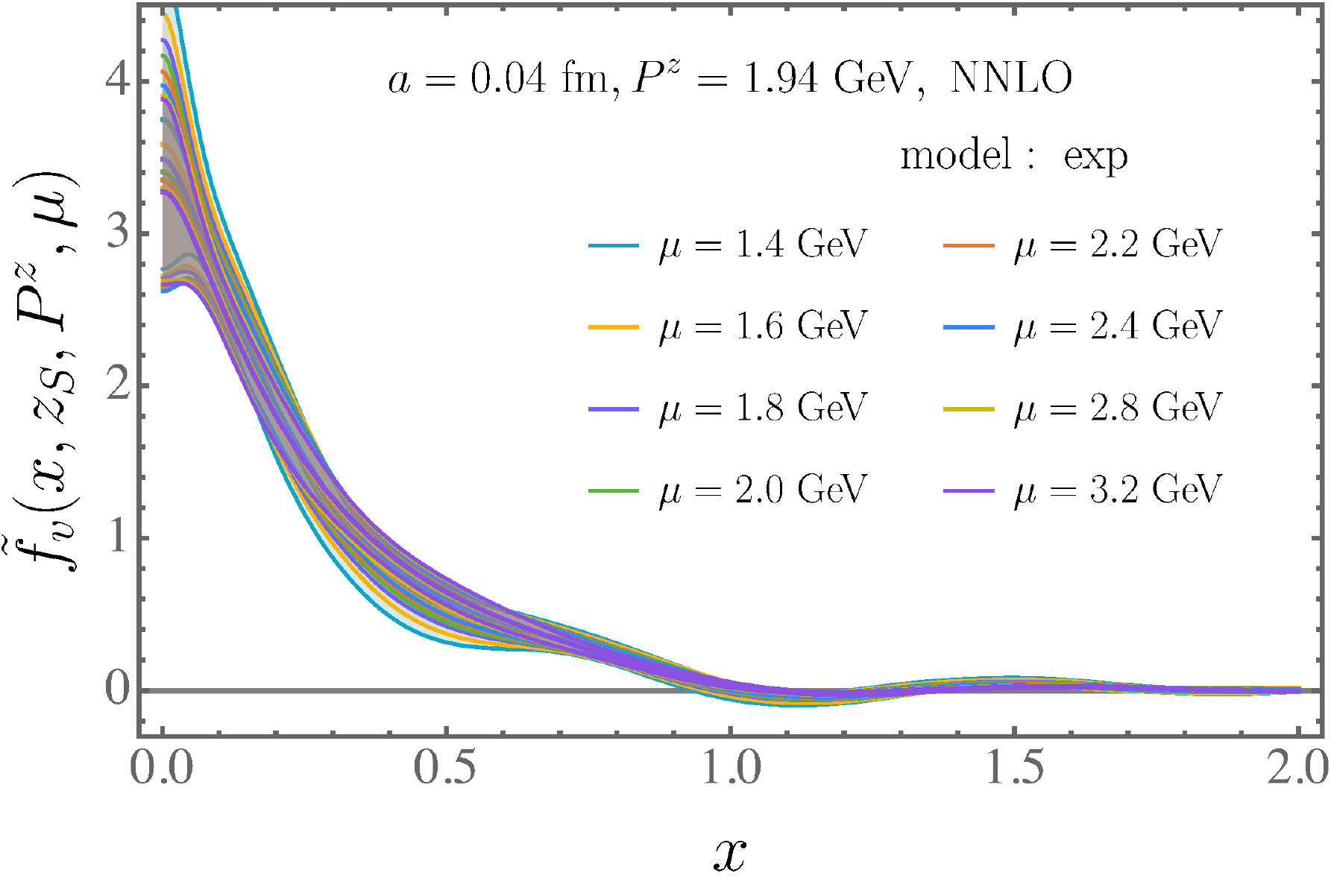}
	\caption{Comparison of the qPDF at different $\mu$ with model-exp extrapolation of the NLO- and NNLO-matched $\tilde h(\lambda, \lambda_S, P^z,\mu,a)$.}
	\label{fig:mu}
\end{figure}

Recall that although the hybrid-scheme matrix elements $\tilde h(\lambda,\lambda_S,P^z)$ should be RG invariant, they can still depend on $\mu$ due to the fixed-order Wilson coefficients used in the matching between lattice and $\MS$ schemes. In \fig{nlovsnnlo}, we compare the qPDFs which are FTs of $\tilde h(\lambda,\lambda_S,P^z,\mu,a)$ obtained at $a=0.04$ fm with $C_0^{\rm NLO}$ and $C_0^{\rm NNLO}$. We choose $\mu=1.0$ GeV for $C_0^{\rm NLO}$ and $\mu=2.0$ GeV for $C_0^{\rm NNLO}$ as the \textit{ansatz} in \eq{ansatz} appear to best describe the lattice matrix elements according to \fig{mfit} at these scales. The results are almost identical to each other, which shows that the renormalon-inspired model with fixed-order Wilson coefficient can indeed describe the data within a specific window of $\mu$. At NLO, smaller $\mu$ is favored as $\alpha_s(\mu)$ is larger so that the renormalon effects become important at lower orders. In \fig{mu} we show the $\mu$-dependence of the qPDFs from NLO- and NNLO-matched $\tilde h(\lambda,\lambda_S,P^z,\mu,a)$. As one can see, the results have mild dependence on $\mu$ which becomes more significant at lower scales. Therefore, the uncertainty from scale variation will also be larger in this region.

\section{Perturbative matching}
\label{app:match}

In this section we perform the perturbative matching to the qPDF. Recall that \eq{fact} relates the qPDF to the PDF, 
\begin{align}\label{eq:fact2}
	f_v(x, \mu)=& \int_{-\infty}^{\infty} \frac{dy}{|y|} \ C^{-1}\!\left(\frac{x}{y}, \frac{\mu}{yP^z},|y|\lambda_S\right) \tilde f_v(y,z_S,P^z) \nn\\
	&\qquad + {\cal O}\Big(\frac{\Lambda_{\text{QCD}}^2}{(xP^z)^2},\frac{\Lambda_{\text{QCD}}^2}{((1-x)P^z)^2}\Big)\,.
\end{align}
The matching kernel $C$ can be expanded to $O(\alpha_s)$ as
\begin{align}
    &C\!\left(\frac{x}{y}, \frac{\mu}{yP^z},|y|\lambda_S\right) \nn\\
    &= \delta\left(\frac{x}{y} - 1\right)  + \alpha_s C^{(1)}\!\left(\frac{x}{y}, \frac{\mu}{yP^z},|y|\lambda_S\right) \nn\\
    & \quad + \alpha_s^2 C^{(2)}\!\left(\frac{x}{y}, \frac{\mu}{yP^z},|y|\lambda_S\right) + {\cal O}(\alpha_s^3)\,.
\end{align}
The inverse matching kernel $C^{-1}$ can obtained by solving
\begin{align}
 \int{dz\over |z|} C^{-1}\!\left(\frac{x}{z}, \frac{\mu}{zP^z},|z|\lambda_S\right)C\!\left(\frac{z}{y}, \frac{\mu}{yP^z},|y|\lambda_S\right) &= \delta\big({x\over y}-1\big)
\end{align}
order by order in $\alpha_s$~\cite{Zhao:2021xxx}, and the result is
\begin{align}\label{eq:invmatch}
    &C^{-1}\!\left(\frac{x}{y}, \frac{\mu}{yP^z},|y|\lambda_S\right) \nn\\
    &= \delta\left(\frac{x}{y} - 1\right)  - \alpha_s C^{(1)}\!\left(\frac{x}{y}, \frac{\mu}{yP^z},|y|\lambda_S\right) \nn\\
    & \quad + \alpha_s^2 \int{dz\over |z|} C^{(1)}\!\left(\frac{x}{z}, \frac{\mu}{zP^z},|z|\lambda_S\right)C^{(1)}\!\left(\frac{z}{y}, \frac{\mu}{yP^z},|y|\lambda_S\right) \nn\\
    & \quad - \alpha_s^2 C^{(2)}\!\left(\frac{x}{y}, \frac{\mu}{yP^z},|y|\lambda_S\right) + {\cal O}(\alpha_s^3)\,.
\end{align}
It has been shown in Ref.~\cite{Zhao:2021xxx} that the inverse matching coefficient satisfies the correct RG and $P^z$-evolution equations.

\subsection{Numerical implementation of matching}
Since in the asymptotic regions,
\begin{align}
	\lim_{y\to\infty}C\Big({x\over y}\Big) \to {\rm\ finite}\,,\quad \quad \lim_{y\to0}C\Big({x\over y}\Big) \propto {y^2\over x^2}\,,
\end{align}
and
\begin{align}
	C\left({x\over y}\right) &\equiv C_r\left({x\over y}\right) - \delta\left({x\over y}-1\right) \int_{-\infty}^\infty dy'\ C_r(y')
\end{align}
is a plus function (with ``$r$'' denotes the $x\neq y$ part) that regulates the singularity at $y=x$, the convolution integral in \eq{fact2} is convergent and insensitive to the cutoffs for $y\to 0, x, \infty$, as long as the qPDF is integrable. Therefore, we are able to evaluate the integral numerically within a finite range of $y$ with a target precision.

\begin{figure}[htb]
	\centering
	\includegraphics[width=0.8\columnwidth]{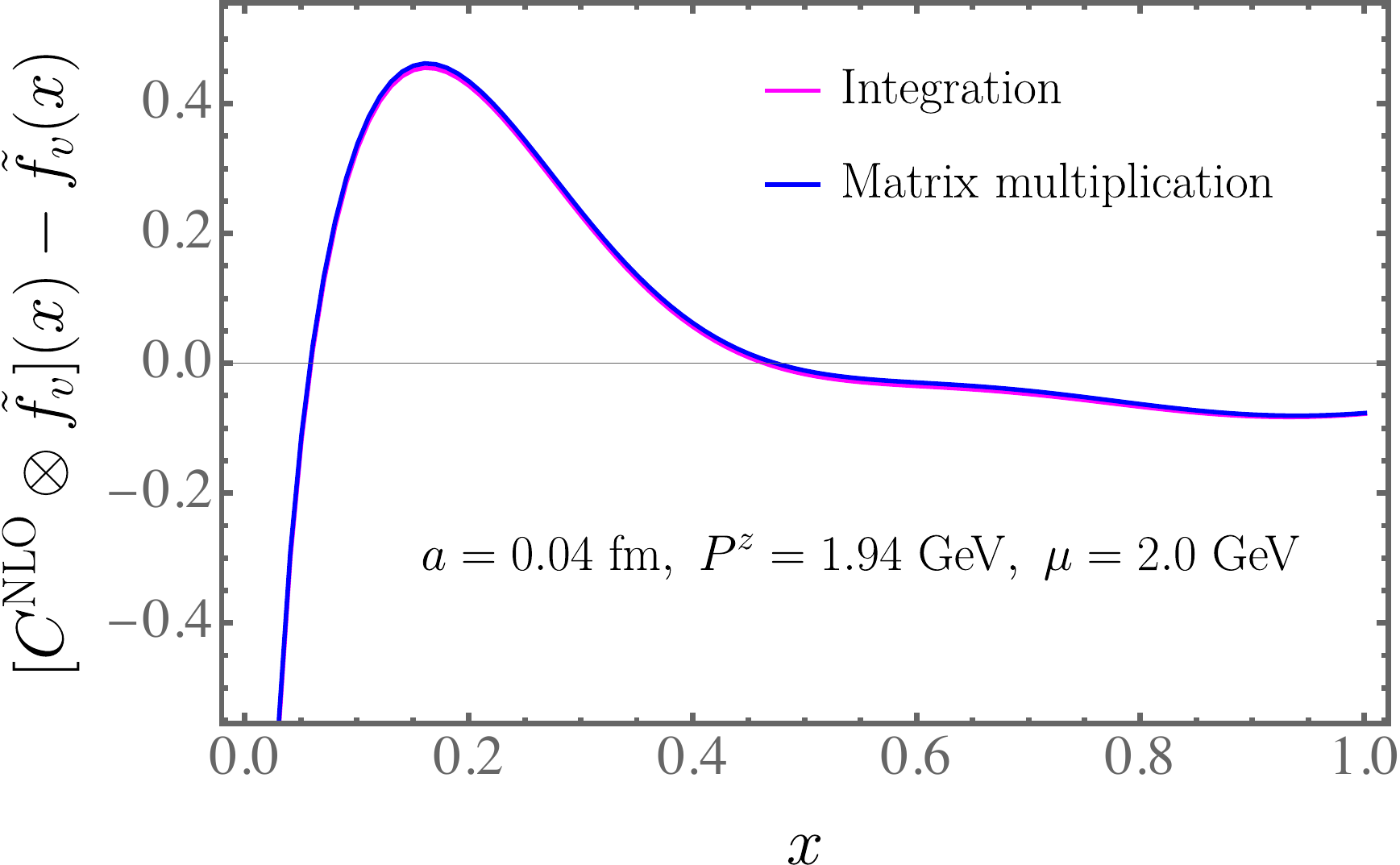}
	\caption{Comparison of matrix multiplication to direct numerical integration for the NLO matching correction to one qPDF sample.}
	\label{fig:matrix}
\end{figure}

The numerical integration in \eq{fact} is time consuming, especially when we have to perform the matching for the qPDF on each bootstrap sample. Therefore, to speed up the matching procedure, we discretize the integral in \eq{fact} and reexpress it as multiplication of a matching matrix and the qPDF vector. In our implementation, our integration domain is $-2.0< y < 2.0$ discretized with a step size $\delta y = 0.001$. Since the qPDF falls very close to zero at $|y|=2.0$, the corresponding uncertainty is negligible as we have varied the truncation point. Note that the matching coefficient is a plus function, the step size $\delta y$ also serves as a soft cutoff for the singularity near $|x/y|=1$ in the plus functions. To test how well the matrix multiplication can reproduce the exact numerical intergration, we compare the NLO corrections to the qPDF from one bootstrap sample using the two methods in \fig{matrix}. With our current step size, the results are almost indentical for $x$ as small as $0.01$.

Moreover, to test the reliability of our inverse matching coefficient, which is obtained through expansion in $\alpha_s$, we compare it to direct matrix inversion. To be specific, we construct a square matching matrix $C$ in $x$ and $y$ with $x,y\in [-2,2]$, which is asymmetric but has dominant diagonal elements, and then invert it to obtain the inverse matching matrix $C^{-1}$. At small $\alpha_s$, the matrix $C$ can be schematically expressed as
\begin{align}
	C & = {\cal I} + {\cal E}\,,
\end{align}
where ${\cal I}$ is an identity matrix, whereas ${\cal E}$ is ${\cal O}(\alpha_s)$, so that its inverse can be expanded as
\begin{align}\label{eq:iter}
	C^{-1} &= {\cal I} - {\cal E} + {\cal E}^2 - {\cal E}^3 + \ldots\,.
\end{align}

\begin{figure}[htb]
	\centering
\subfloat[]{
	\centering
	\includegraphics[width=0.8\columnwidth]{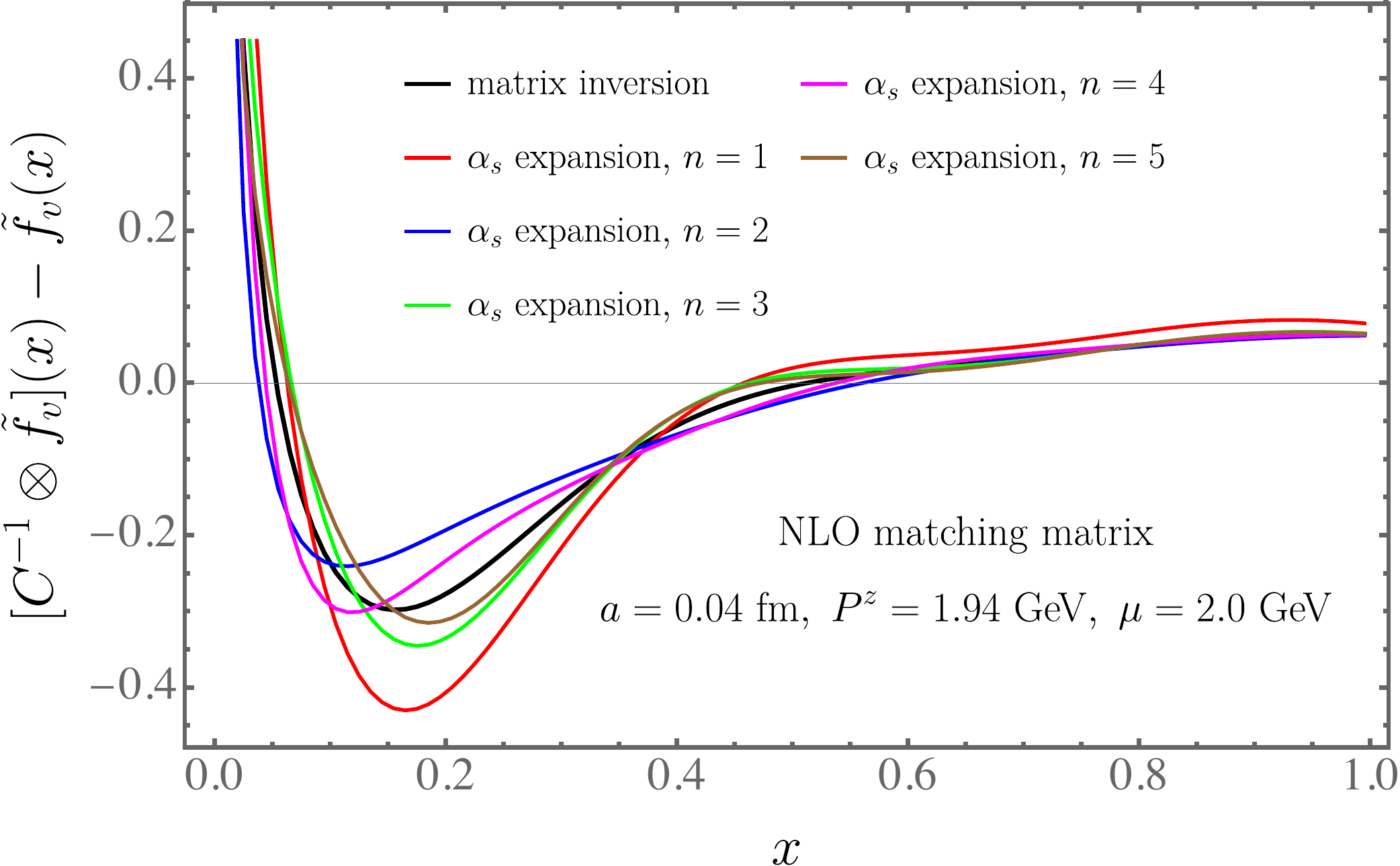}
	\label{fig:matinv1}
}

\subfloat[]{
	\centering
	\includegraphics[width=0.8\columnwidth]{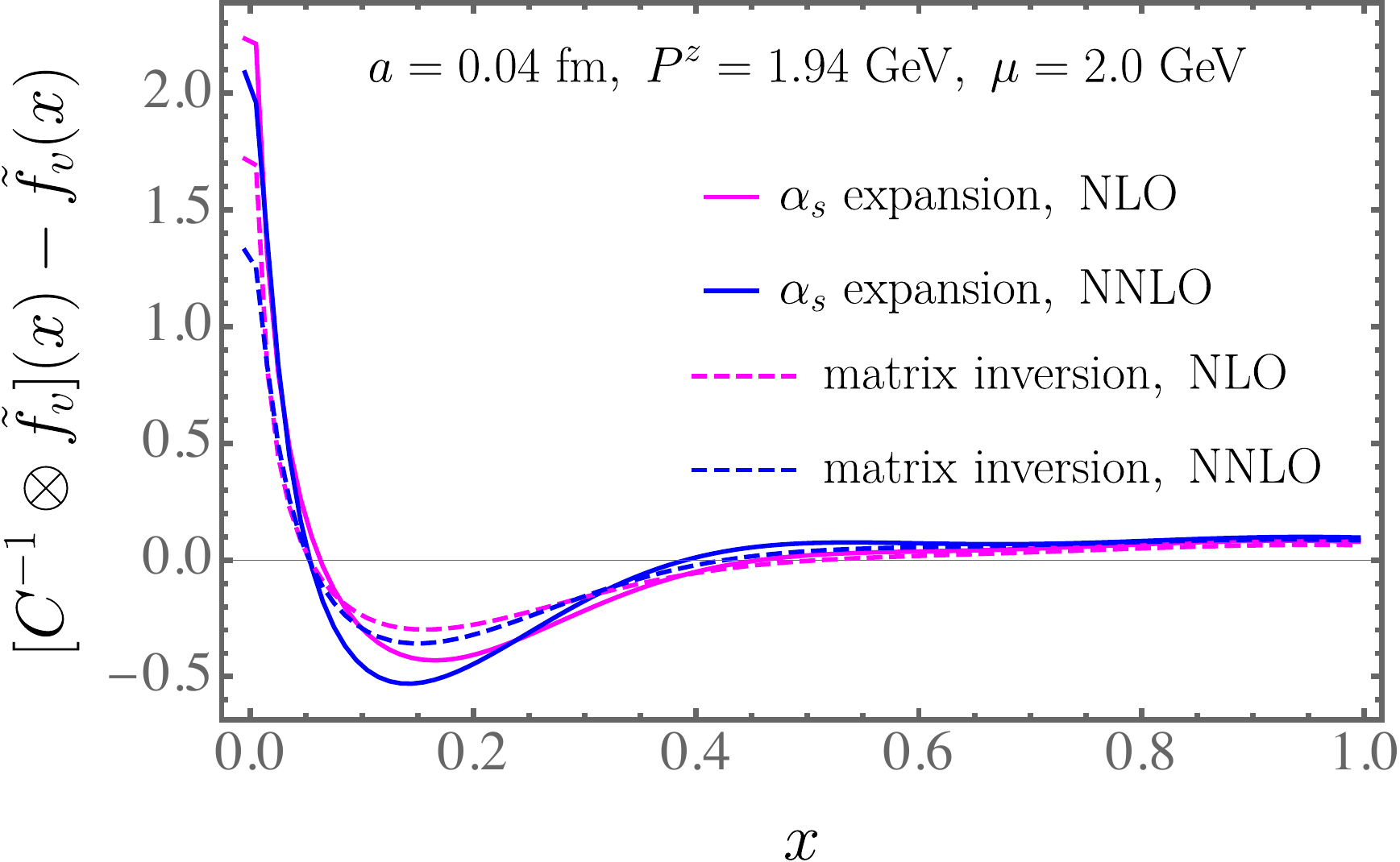}
	\label{fig:matinv2}
}
	\caption{(a) Comparison of the NLO matching correction to the qPDF with matrix inversion and the expansion in \eq{iter} to order $n$. 
		(b) Comparison of NLO and NNLO matching corrections to the qPDF from direct matrix inversion and the $\alpha_s$-expansion in \eq{invmatch}.}
	\label{fig:inv}
\end{figure}

In \fig{matinv1} we first test the convergence of the solution in \eq{iter} for the NLO matching matrix. By expanding the solution to order $n$, we calculate the NLO matching correction to a qPDF sample, and then compare it to the result from direct matrix inversion. Since our main purpose is to compare the two inversion methods, we increase the step size to $\delta y=0.01$ to reduce the computing time regardless the accuracy of numerical integration. We find that by increasing $n$, the expansion method gradually coverges to direct inversion, as expected. Of course, in perturbation theory, we should calculate the matching coefficient to $n$-loop accuracy for consistency, for $\alpha_s$ is the actual power-counting parameter.

In \fig{matinv2} we compare the NLO and NNLO matching corrections to a qPDF sample using direct matrix inversion and the $\alpha_s$-expansion methods. The results are basically consistent with each other for almost the entire range of $x\in(0,1)$, except for small deviations. This is because direct matrix inversion includes all-order terms in $\alpha_s$, and the deviations reflect the size of higher-order effects, whose smallness shows that the perturbation series is convergent. With our current two-loop accuracy, we adopt the $\alpha_s$-expansion method.

\subsection{Perturbative convergence}

\begin{figure*}[htb]
    \centering
    \includegraphics[width=0.32\textwidth]{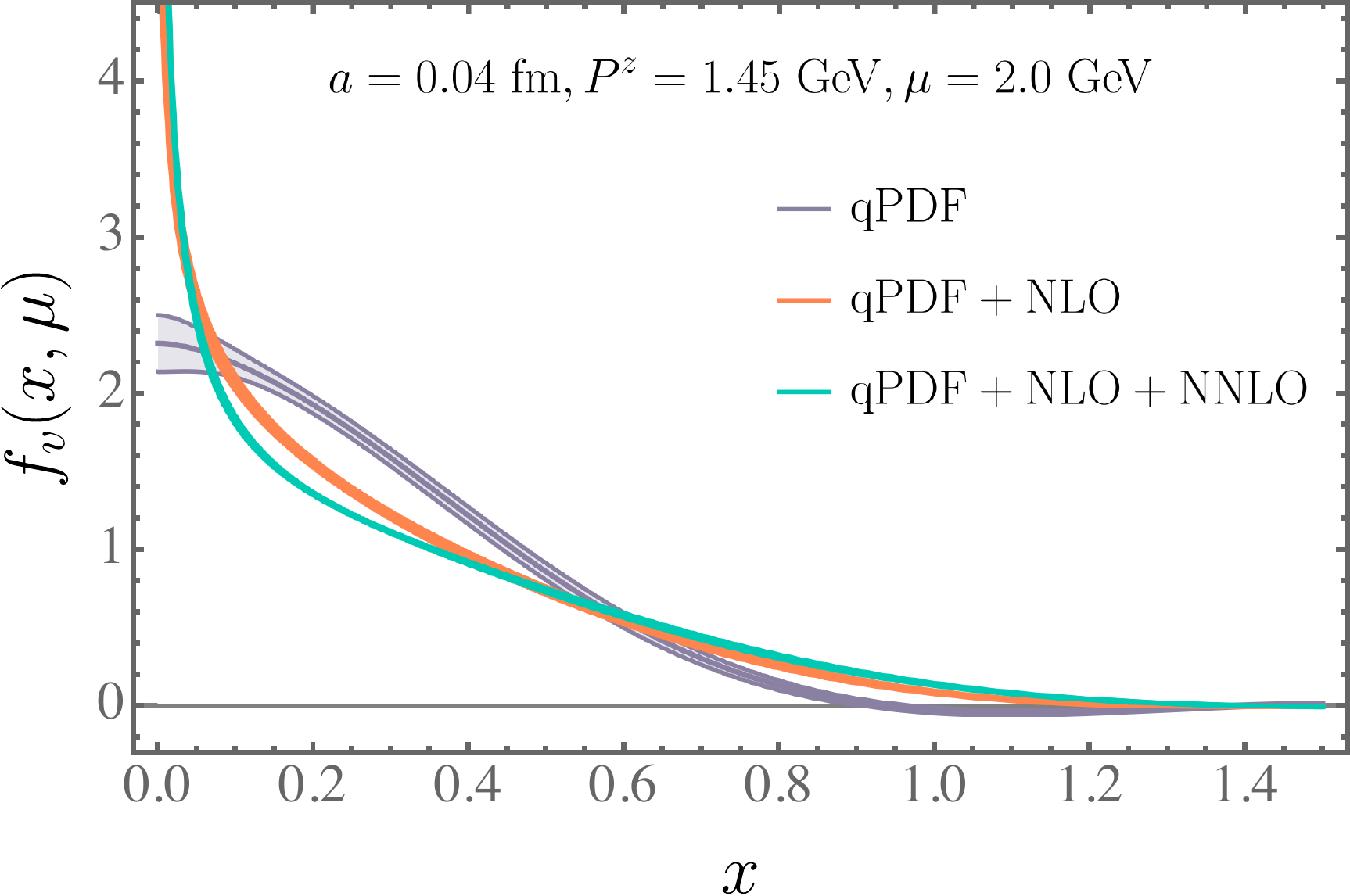}
    \includegraphics[width=0.32\textwidth]{convergence}
    \includegraphics[width=0.32\textwidth]{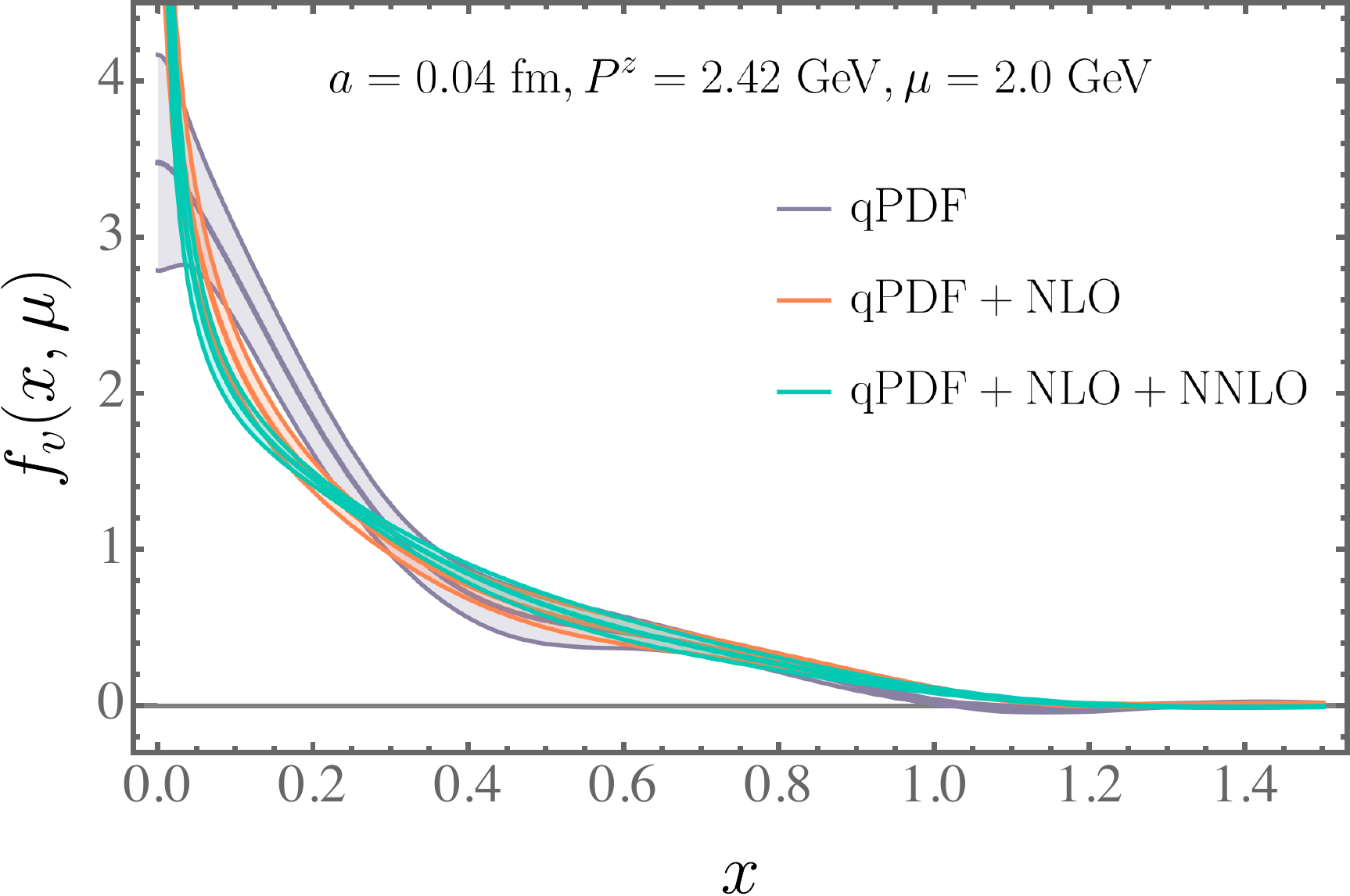}
    \includegraphics[width=0.32\textwidth]{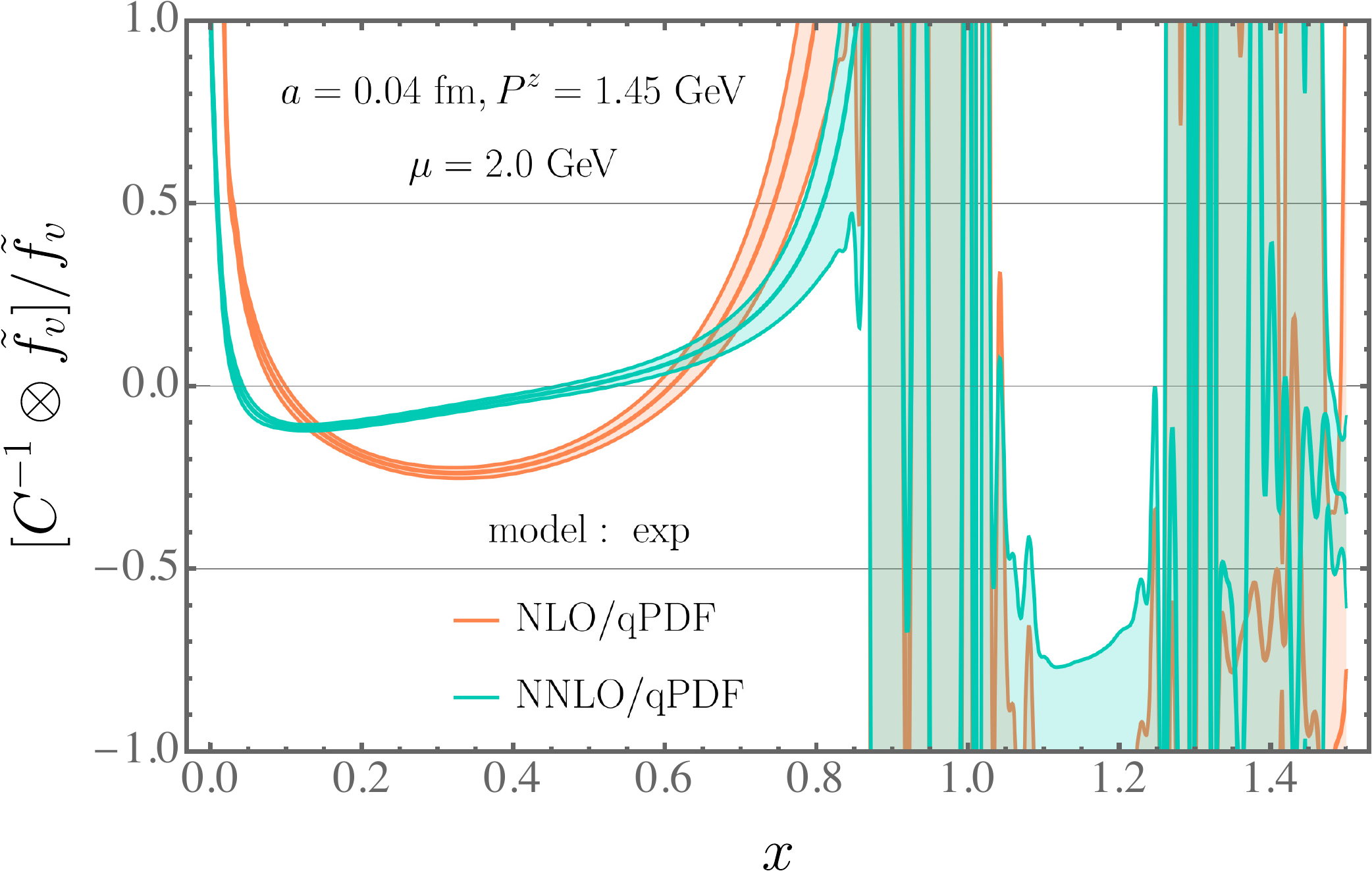}
    \includegraphics[width=0.32\textwidth]{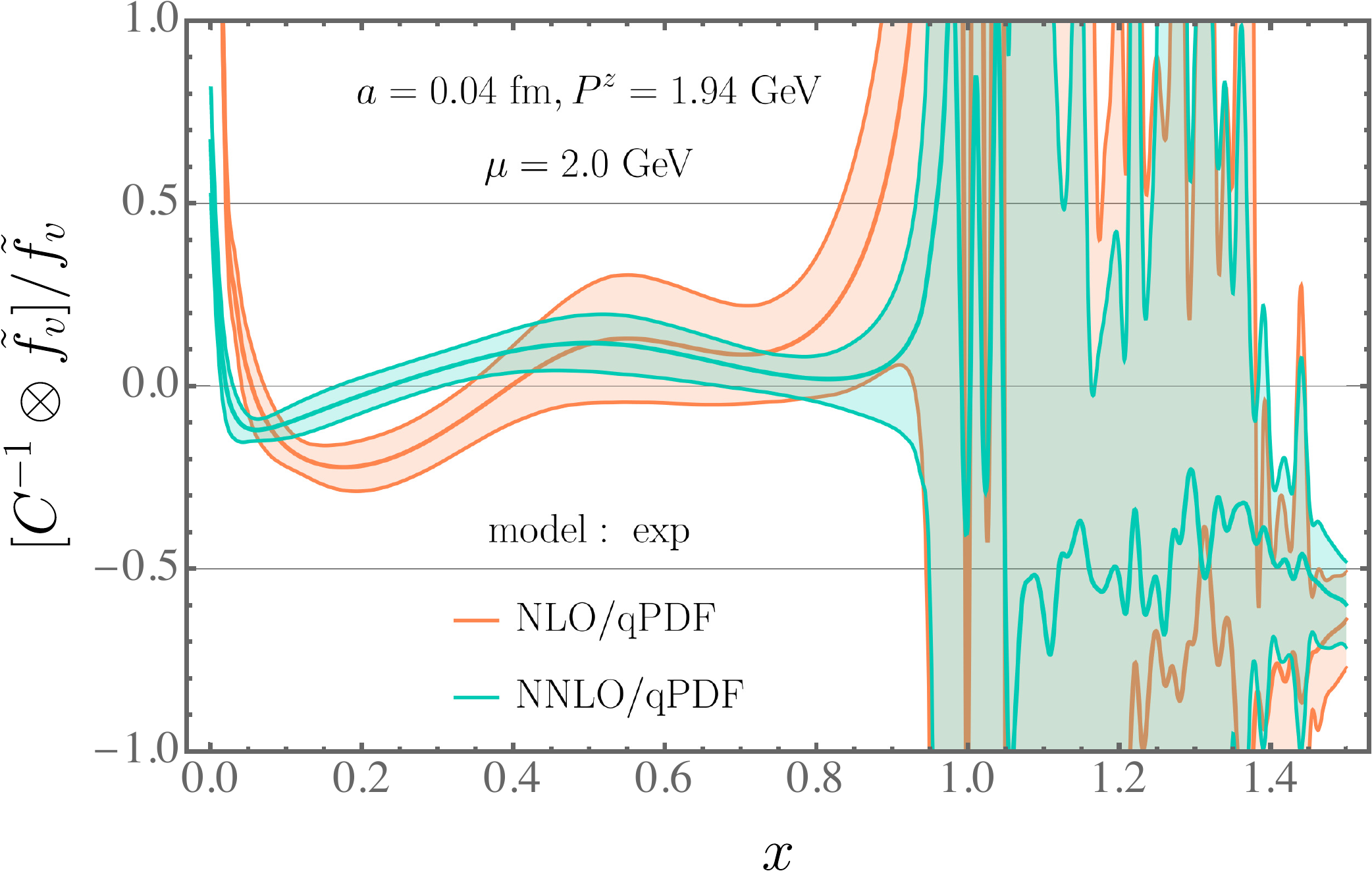}
    \includegraphics[width=0.32\textwidth]{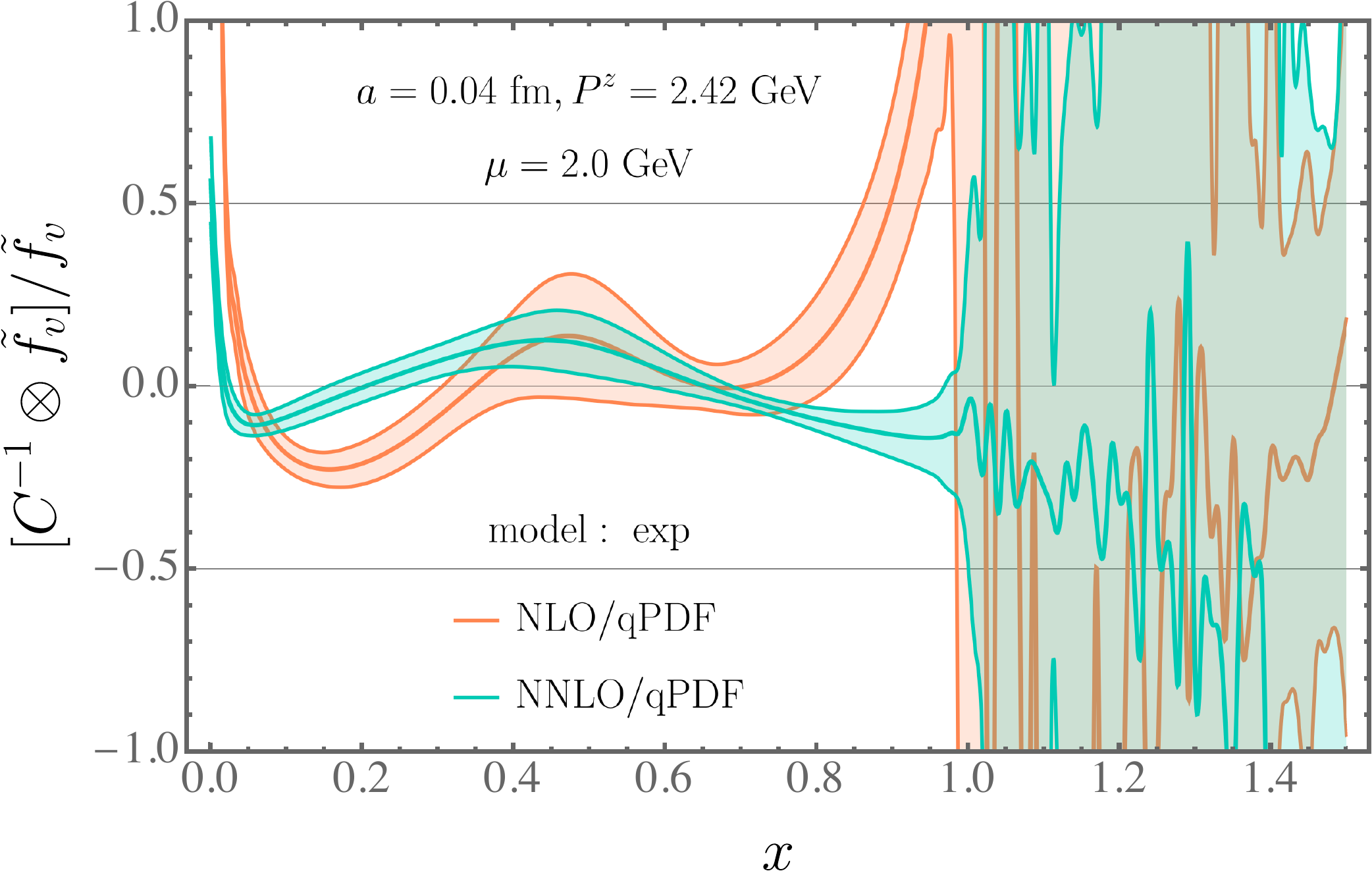}
    \caption{Upper row: the PDFs from NLO and NNLO matching corrections are compared to the qPDF (or LO PDF), which is obtained from model-exp (with $m_{\rm eff}>0.1$ GeV) extrapolation of the NNLO-matched $\tilde h(\lambda, \lambda_S, P^z,\mu,a)$. Lower row: the ratio of NLO and NNLO corrections to the qPDF.}
    \label{fig:nnlo}
\end{figure*}

In \fig{nnlo} we show the matched results for the PDF from the qPDF obtained from model-exp extrapolation (with $m_{\rm eff}>0.1$ GeV) of the NNLO-matched $\tilde h(\lambda, \lambda_S, P^z,\mu,a)$. As one can see, the NNLO correction is generally smaller than the NLO correction for moderate $x$, which indicates good perturbative convergence. 
Near the end-point regions, the NLO and NNLO corrections become larger than 50\%, which suggests that higher-order corrections or resummation effects become important.

\begin{figure}[htb]
	\centering
	\includegraphics[width=0.8\columnwidth]{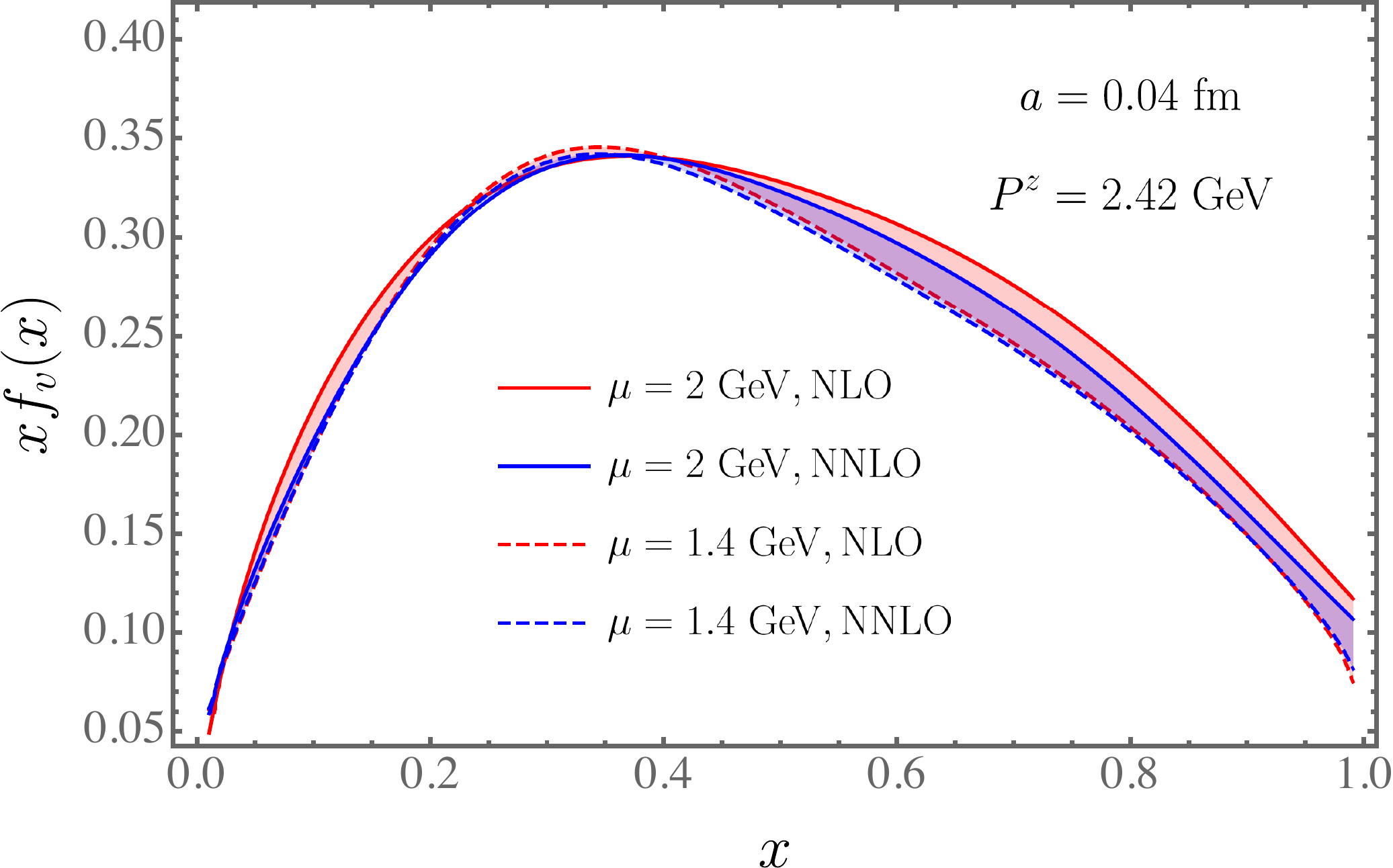}
	\caption{Comparison of the PDFs at different $\mu$ obtained from the NLO- and NNLO-matched $\tilde h(\lambda, \lambda_S, P^z,\mu,a)$.}
	\label{fig:scale_var}
\end{figure}

To see whether the NNLO matching reduces the uncertainty from scale variation, we match qPDFs at different $\mu$ to the corresponding PDFs, and then use DGLAP equation to evolve the results to $\mu=2.0$ GeV. We use NLO matching coefficient and LO DGLAP evolution kernel for the qPDF obtained from the NLO-matched $\tilde h(\lambda,\lambda_s,P^z,\mu,a)$, and NNLO matching coefficient and NLO DGLAP evolution kernel for the qPDF obtained from the NNLO-matched $\tilde h(\lambda,\lambda_s,P^z,\mu,a)$. The NLO DGLAP evolution formula takes the following form,
\begin{align}
	f_v(x,\mu) &= f_v(x,\mu_0) \\
	&+ {\alpha_s(\mu_0)t\over 2\pi} \int_x^1{dy\over |y|}P_{qq}^{(0)}\left({x\over y}\right) f_v(y,\mu_0) \nn\\
	&+ \left({\alpha_s(\mu_0)t\over 2\pi}\right)^2 \int_x^1{dy\over |y|}\left[P_{qq}^{V(1)} + {1\over2}P_{qq}^{(0)}\otimes P_{qq}^{(0)} \right.\nn\\
	&\qquad\qquad\qquad\left. - {\beta_0\over2}P_{qq}^{(0)}\right]\left({x\over y}\right) f_v(y,\mu_0)\,,
\end{align}
where $t=\ln(\mu^2/\mu_0^2)$, $\beta_0=(11C_A-2n_f)/6$, $P_{qq}^{(0)}$ is the LO splitting kernel, and $P_{qq}^{V(1)}$ is the NLO splitting kernel~\cite{Curci:1980uw} for the valence quark PDF.

\begin{figure}[htb]
	\centering
	\includegraphics[width=0.8\columnwidth]{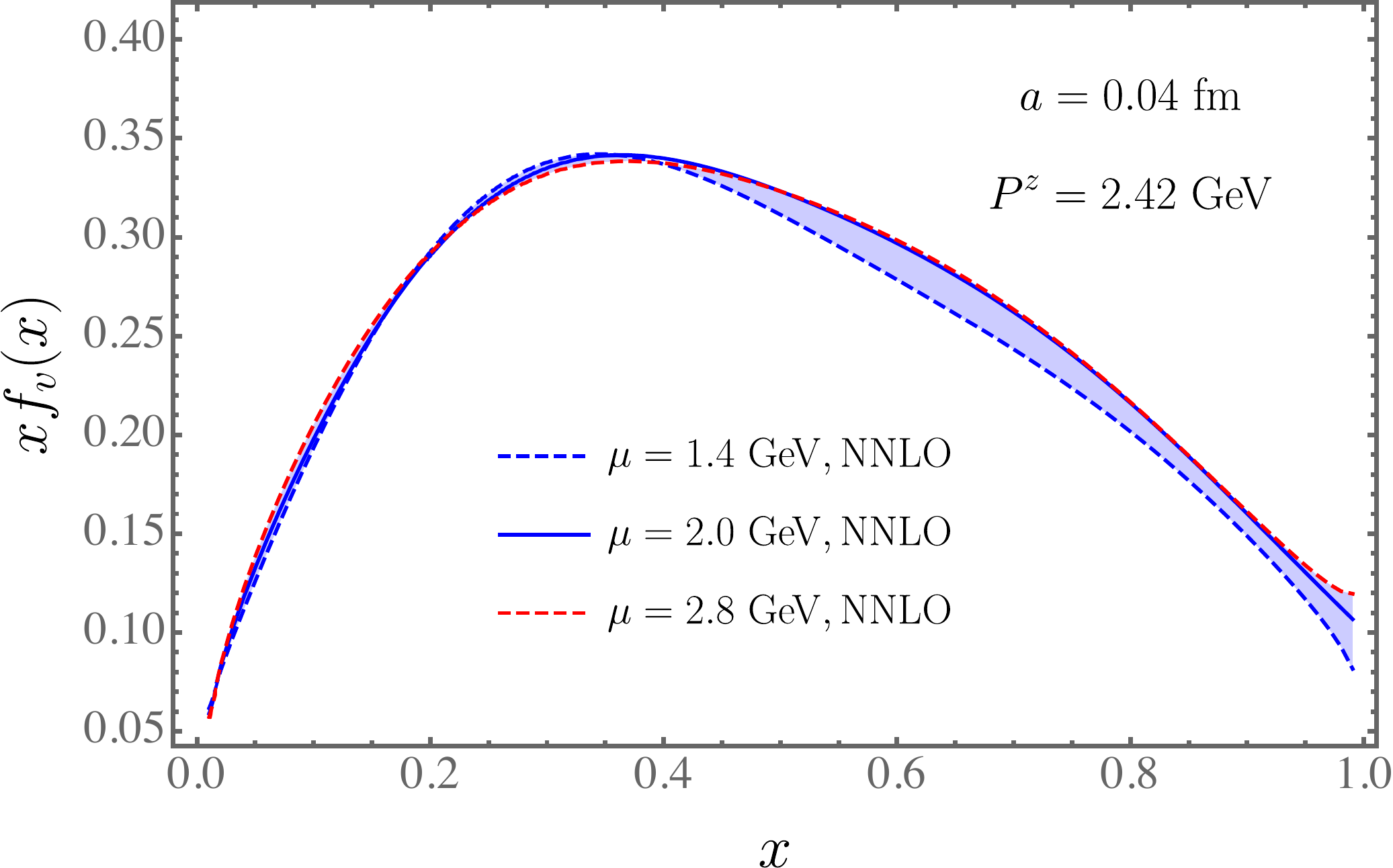}
	\caption{Comparison of the PDFs obtained from NNLO matching of the qPDFs at different $\mu$ and NLO DGLAP evolution to $\mu=2.0$ GeV.}
	\label{fig:scale_var2}
\end{figure}

Since there are only a few common $\mu$ values for the NLO- and NNLO- matched $\tilde h(\lambda,\lambda_s,P^z,\mu,a)$, we choose $\mu=1.4$ and $2.0$ GeV for our comparison. In \fig{scale_var} we show the scale variation of the PDFs from NLO and NNLO matching, where only the central values are plotted for our purpose. As one can see, the NNLO matching correction significantly reduces the uncertainty for $x\lesssim 0.4$ at NLO, while for $x\gtrsim 0.4$ the NNLO uncertainty band is still about a factor of one half of the NLO case. Therefore, the NNLO matching does indeed improve the perturbation theory uncertainty.

Finally, for the NNLO matching we vary $\mu=2.0$ GeV by a factor of $\sqrt{2}$ and $1/\sqrt{2}$, and then use NLO DGLAP equation to evolve the matched results to $\mu=2.0$ GeV, whose central vavlues are shown in \fig{scale_var2}. As one can see, there is virtually no difference between choosing $\mu=2.0$ and $2.8$ GeV as the factorization scale, but the lower choice of $\mu=1.4$ GeV does introduce larger uncertainty mainly because $\alpha_s$ becomes too large. Nevertheless, such uncertainty is still quite small compared to the other systematics.\\

\subsection{Dependence on $P^z$, $a$ and extrapolation model}
\label{app:pzdependence}

	\begin{figure*}[htb]
		\centering
		\includegraphics[width=0.32\textwidth]{pz_dependence}
		\includegraphics[width=0.32\textwidth]{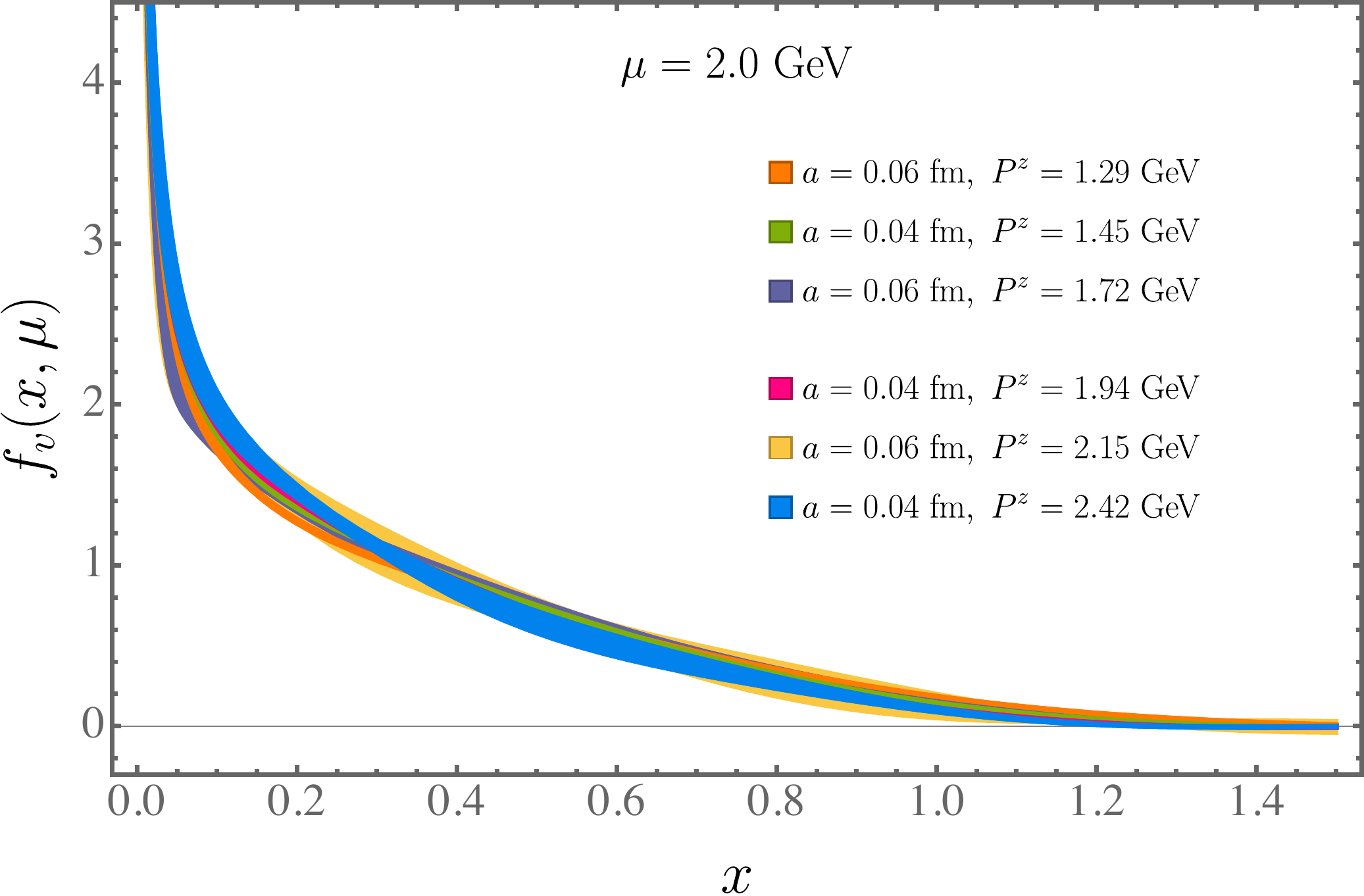}
		\includegraphics[width=0.34\textwidth]{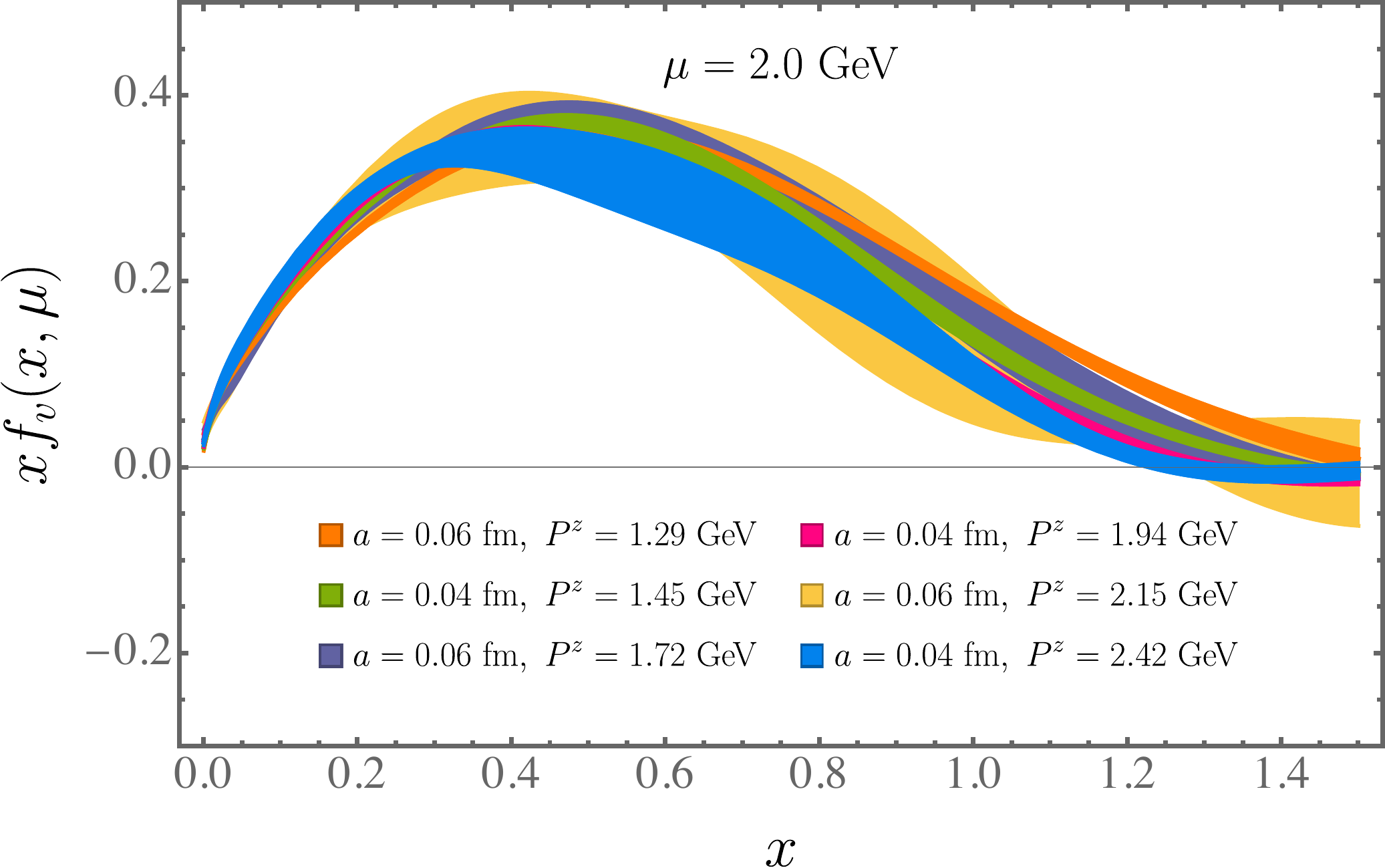}
		\caption{The PDFs from NNLO matching of the qPDFs at different $P^z$, which is obtained from model-exp extrapolation of the NNLO-matched $\tilde h(\lambda, \lambda_S, P^z,\mu,a)$.}
		\label{fig:pzs}
	\end{figure*}
	
	\begin{figure*}[htb]
		\centering
		\includegraphics[width=0.32\textwidth]{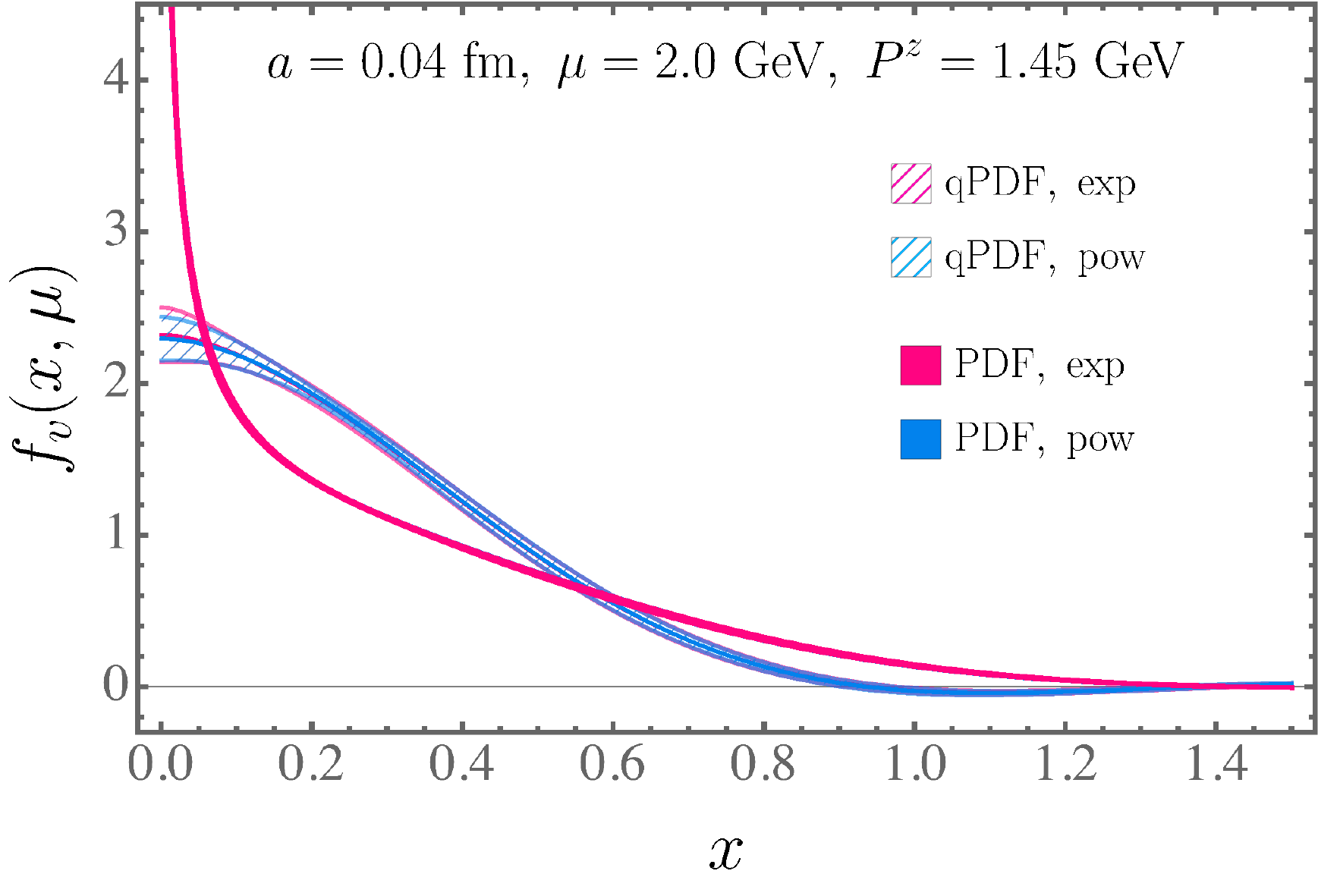}
		\includegraphics[width=0.32\textwidth]{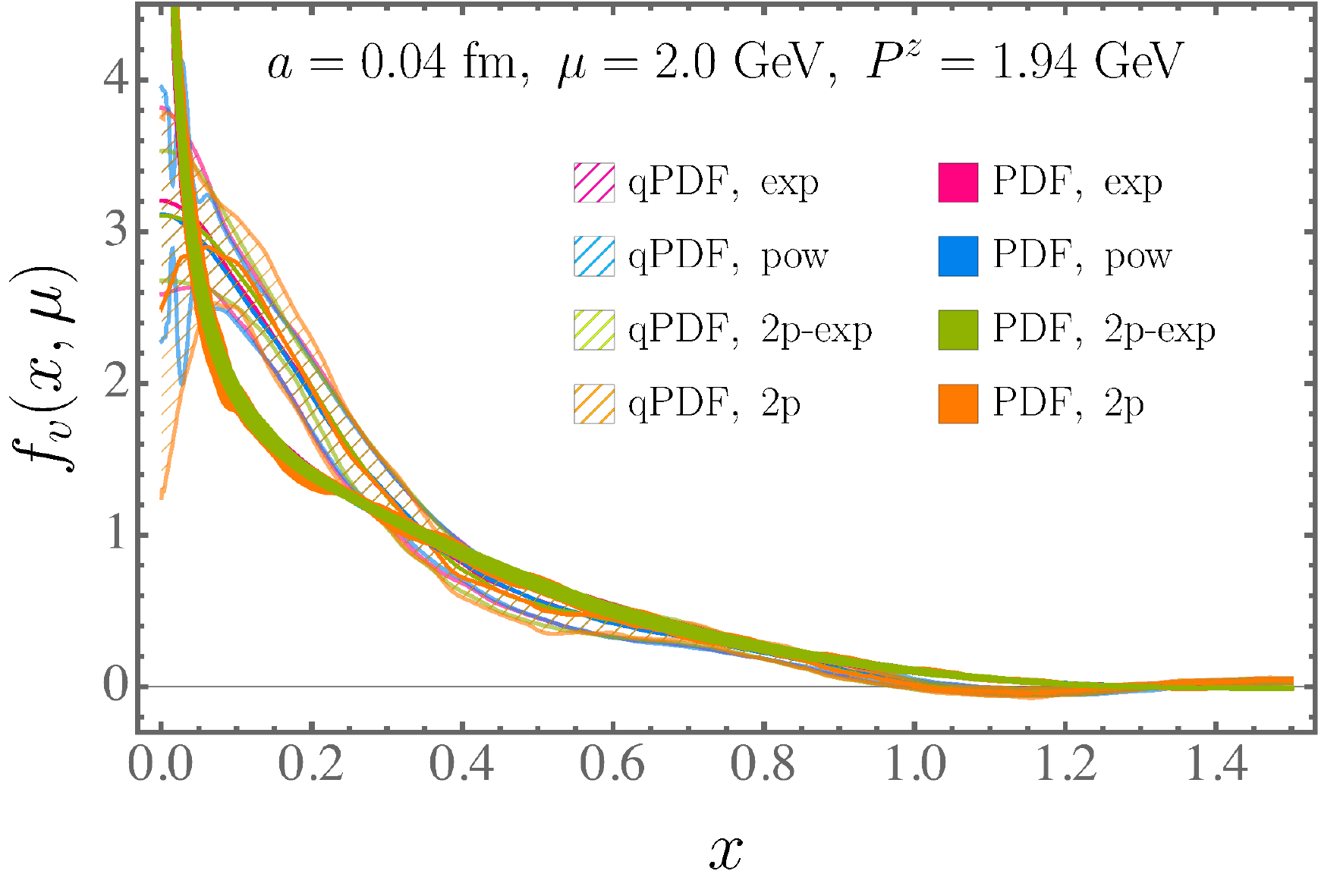}
		\includegraphics[width=0.32\textwidth]{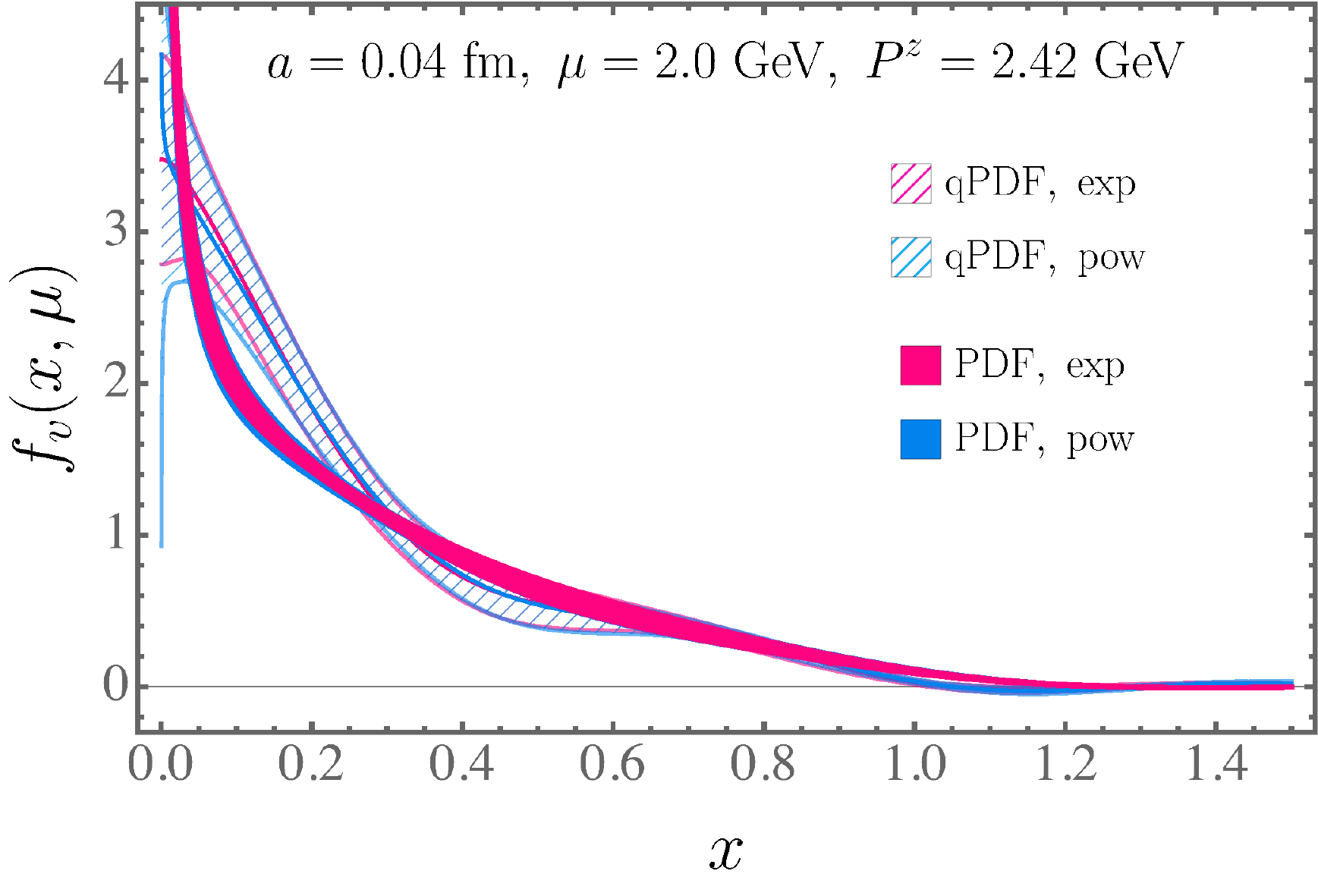}
		\caption{Comparison of the final results from qPDFs obtained by different model extrapolations for the FT.}
		\label{fig:exp_vs_pow}
	\end{figure*}

In \fig{pzs} we show the $P^z$-dependence of the PDF with NNLO matching correction. We find that despite the considerable differences between the qPDFs at $P^z\le 1.45$ GeV and those at $P^z\ge 1.94$ GeV, the matching corrections bring the final results closer, which shows the effectiveness of LaMET. Note that the matching drives the qPDF closer to the smaller $x$ region, so the error bands of the PDFs also shrink after matching as they are contributed from the larger $x$ region.
Moreover, we find that the PDFs start to converge at $P^z\ge 1.29$ GeV, which corresponds to a boost factor of $\sim 4$. As $P^z$ increases, the results becomes smaller as $x\to1$, which agrees with our expectation that large momentum suppresses the higher-twist contributions. It is worth mentioning that both the $P^z$-dependence and matching correction appear to be small for $x$ as low as $0.05$, which hints that the power correction and resummation effects are less severe than our naive estimate through power counting.

In \fig{exp_vs_pow} we compare the PDFs matched from the qPDFs with model-exp (with $m_{\rm eff}>0.1$ GeV) and model-pow extrapolations. For $a=0.04$ fm and $P^z=1.94$ GeV, we also added comparison to the model-2p-exp and model-2p extrapolations. Despite the differences between the qPDFs at small $x$, the matched results are almost identical even at the smallest $x$ shown in the plot.
Again, this is the outcome of the PDF receiving contributions from the qPDF at larger $x$ through matching, which suggests that the extrapolation error can still be under control for $x$ as small as $\sim 0.01$.  Note that the result from model-2p also shows agreement, but it includes slight oscillations in the $x$-space, because the extrapolated $\tilde h(\lambda)$ decays too slowly in the coordinate space. Therefore, in the region where other systematic errors are under control, the difference between model-exp and other extrapolations is negligible, and we will use the model-exp extrapolation to obtain the final results.

\section{Final results}
\label{app:final}

The central value of our final result is obtained from the qPDF at $a=0.04$ fm, $z_S=0.24$ fm, $z_L=0.92$ fm, $\mu=2.0$ GeV and $P^z=2.42$ GeV with exponential extrapolation ($m_{\rm eff}>0.1$ GeV) and NNLO matching. The error from variation of the factorization scale is obtained by repeating the same procedure for $\mu=1.4$ and $2.8$ GeV and evolving the matched results to $\mu=2.0$ GeV with the NLO DGLAP equation, as shown in \fig{scale_var2}, where let the error band cover all the data sets from the three different factorization scales.

In order to obtain a target precision of 10\%, we aim to control the relative ${\cal O}(\alpha_s^3)$ matching correction at $\mu=2.0$ GeV be smaller than 5\%. By assuming that the perturbation series grows geometrically, it means that the relative NLO correction should be less than $^3\sqrt{5\%}=37\%$ and the relative NNLO correction less than 14\%. By comparing to \fig{nnlo}, it means that we should exclude the regions $x<0.03$ and $ x > 0.88$.

 \begin{figure}
	\centering
	\includegraphics[width=0.8\columnwidth]{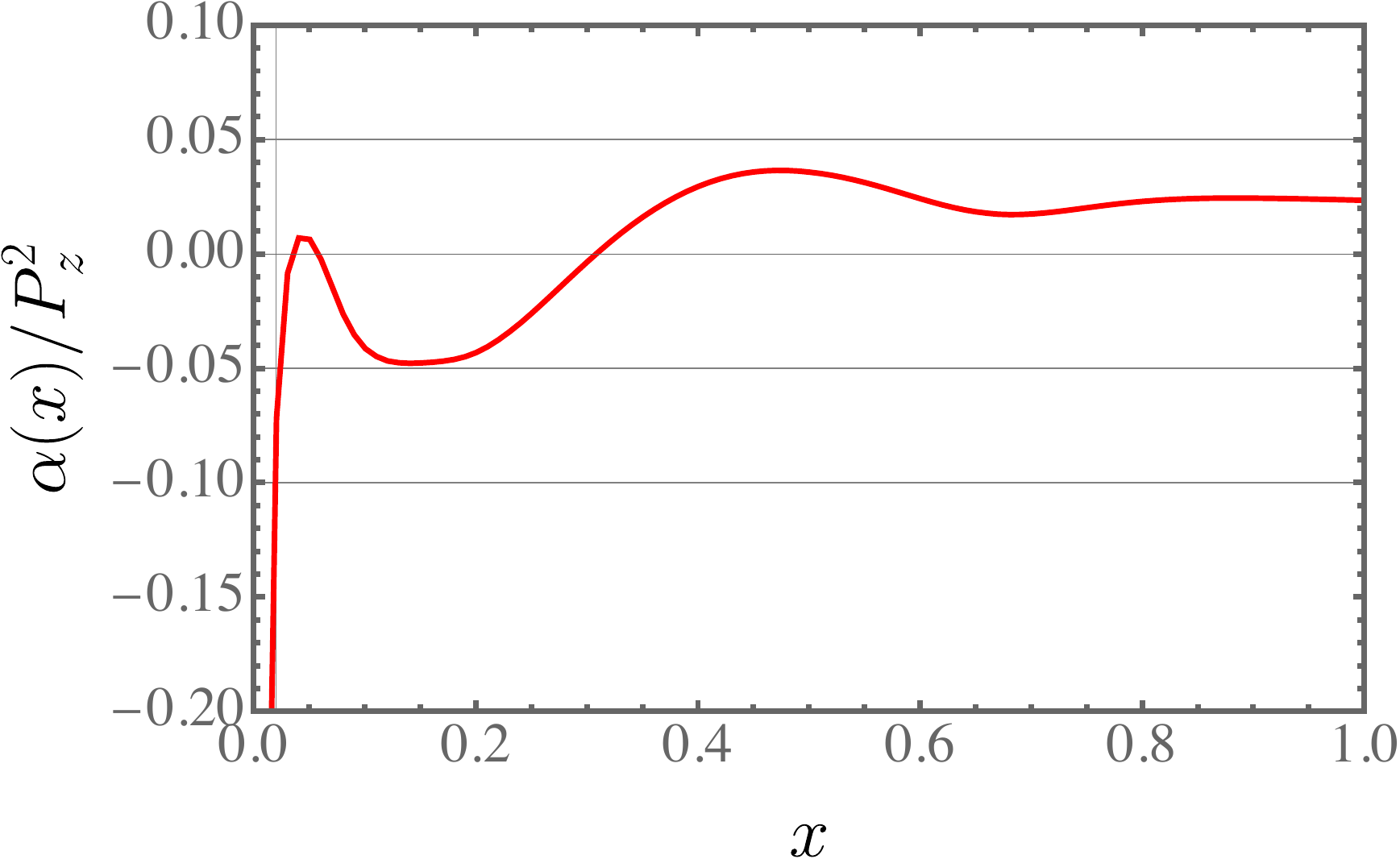}
	\includegraphics[width=0.8\columnwidth]{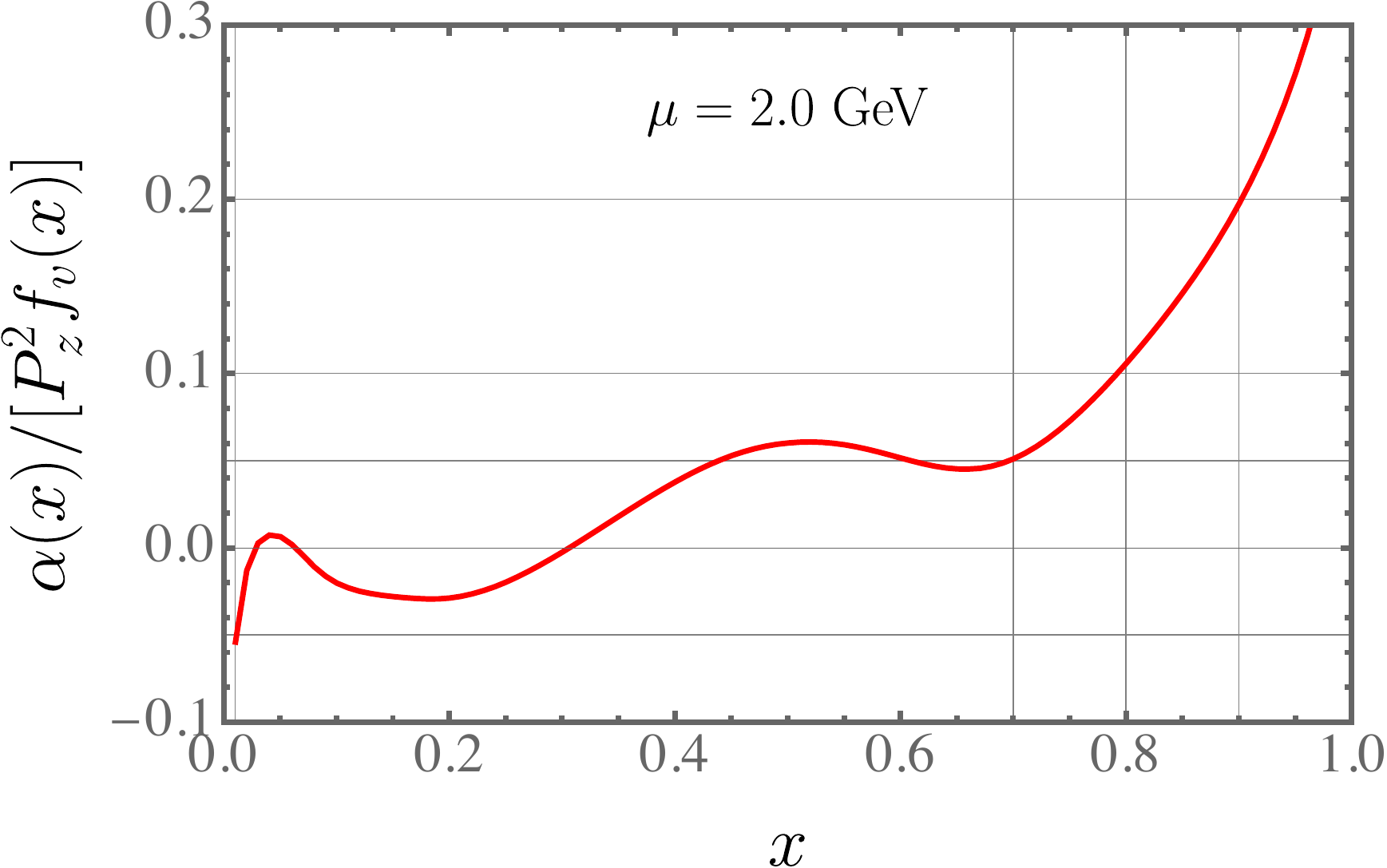}
	\caption{Estimate of the size of power correction $\alpha(x)/P_z^2$ (upper panel) and its relative size to the qPDF $\tilde f_v(x, P^z=2.42\ {\rm GeV})$ (lower panel).}
	\label{fig:power}
\end{figure}
 
To estimate the size of the power corrections, we fit the PDFs obtained at $a=0.04$ fm, $P^z=\{1.45,1.94,2.42\}$ GeV and $a=0.06$ fm, $P^z=\{1.72,2.15\}$ GeV to the \textit{ansatz} $f_v(x) + \alpha(x) / P_z^2$ for each fixed $x$, and show the size of the power correction term in \fig{power}.
At $P^z=2.42$ GeV, we find that the absolute value of the power correction diverges at very small $x$, as expected, but its relative size $ \alpha(x) / [P_z^2 f_v(x)]$ remains finite because the PDF also diverges. On the contrary, $ \alpha(x) / [P_z^2 f_v(x)]$ starts to grow as $x\to1$. According to our estimate, $ \alpha(x) / [P_z^2 f_v(x)] \lesssim 0.1$ for $0.01<x<0.80$ and $ \alpha(x) / [P_z^2 f_v(x)] \lesssim 0.05$ for $0.01<x<0.70$.
 According to \fig{exp_vs_pow}, the qPDF from power-law extrapolation leads to almost identical PDF after the matching correction for $x$ as small as 0.01.
 Our explanation is that the matching correction drives the qPDF to smaller $x$, so the PDF at a given $x$ receives contributions from the larger-$x$ region of the qPDF which has less $P^z$ dependence. Although there are logarithms of $\mu/(xP^z)$ in the matching coefficient which become large at small $x$, they are always multiplied by the DGLAP splitting function, which when convoluted with the qPDF always drives the result to smaller $x$, thus the perturbative correction remains small even at $x=0.03$.
 
 \begin{figure}
	\centering
	\includegraphics[width=0.8\columnwidth]{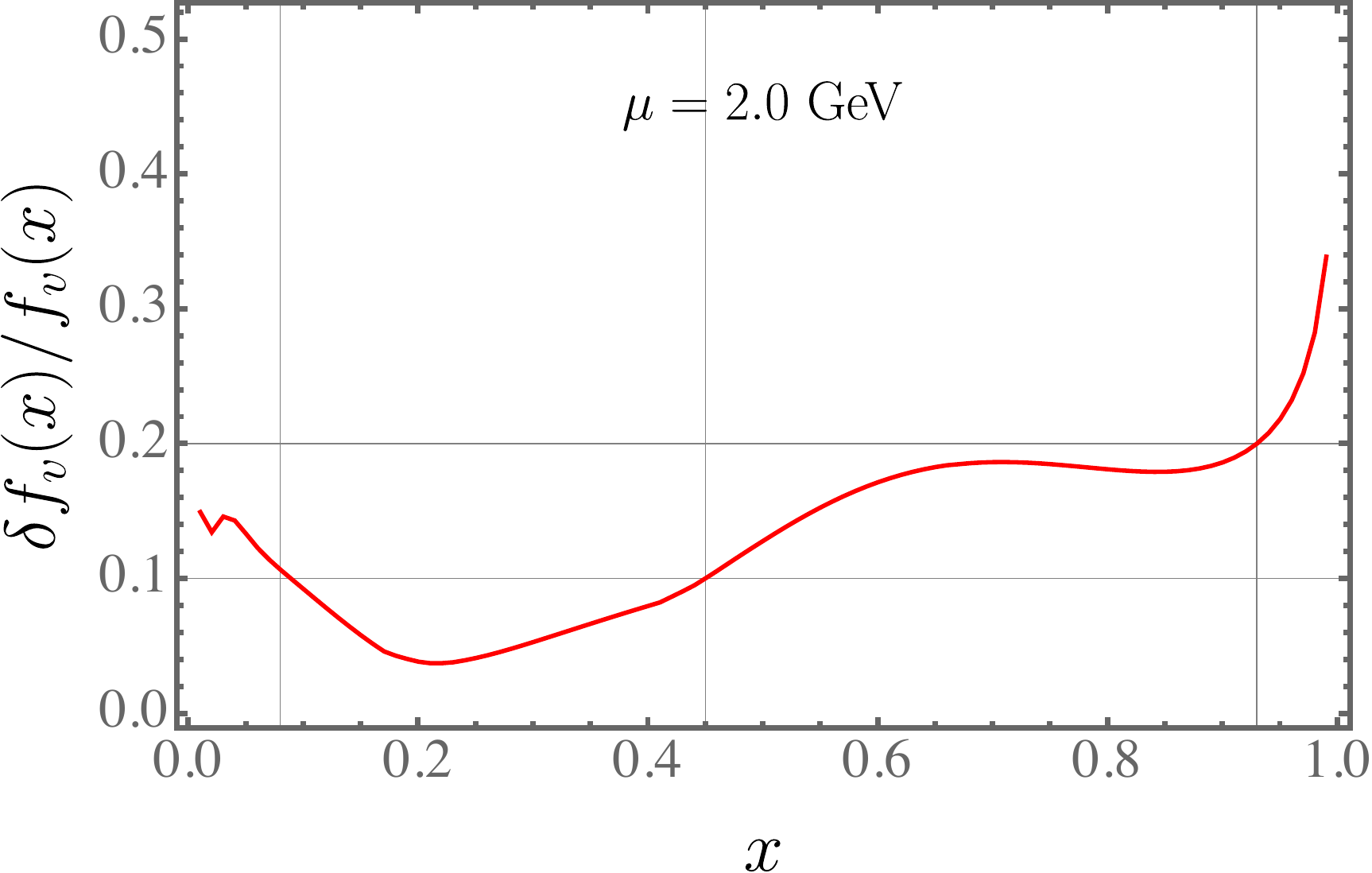}
	\caption{Statistical and scale-variation uncertainty of the PDF obtained from the qPDF at $a=0.04$ fm and $P^z=2.42$ GeV.}
	\label{fig:sigma}
\end{figure}
 
In \fig{sigma} we show uncertainty of the PDF, $\delta f_v(x)/f_v(x)$, where $\delta f_v(x)$ includes both statistical and scale-variation errors. The uncertainty is $\le 20\%$ for $0.01\le x \le 0.93$, as $x=0.01$ is the smallest $x$ that we show in the plot, and $\le 10\%$ for $0.08\le x \le 0.45$.
 
Therefore, by combining the estimates of power correction, higher-order perturbative correction, statistical and scale-variation errors, we determine the PDF at $0.03\lesssim x\lesssim 0.80$ with $\le 20\%$ uncertainty and at $0.08\lesssim x \lesssim 0.45$ with $\le 10\%$ uncertainty, which is shown in \fig{comp}.

\newpage
\bibliography{xpdf}

\end{document}